\documentclass[aps,prl,nofootinbib,twocolumn,superscriptaddress,preprintnumbers,balancelastpage,longbibliography]{revtex4-1}

\usepackage{amsmath,amssymb,mathtools,bm}
\usepackage{graphicx, color, hepunits}
\usepackage[dvipsnames]{xcolor}
\usepackage{float}
\usepackage{filecontents}
\usepackage{multirow}
 \usepackage{hyperref} 
\hypersetup{
    colorlinks=true,       
    linkcolor=blue,        
    citecolor=blue,        
    filecolor=magenta,     
    urlcolor=blue          
}
\usepackage[utf8]{inputenc}
\usepackage[english]{babel}

\newcommand{\gagg}{g_{a \gamma \gamma}}

\newcommand{\new}[1]{\textcolor{black}{#1}}

\newcommand{\es}[2] {\begin{equation} \label{#1} \begin{split} #2 \end{split} \end{equation}}

\begin{document}

\title{$X$-ray Searches for Axions from Super Star Clusters}

\author{Christopher Dessert}
\affiliation{Leinweber Center for Theoretical Physics, Department of Physics, University of Michigan, Ann Arbor, MI 48109 U.S.A.}
\affiliation{Berkeley Center for Theoretical Physics, University of California, Berkeley, CA 94720, U.S.A.}
\affiliation{Theoretical Physics Group, Lawrence Berkeley National Laboratory, Berkeley, CA 94720, U.S.A.}

\author{Joshua W. Foster}
\affiliation{Leinweber Center for Theoretical Physics, Department of Physics, University of Michigan, Ann Arbor, MI 48109 U.S.A.}
\affiliation{Berkeley Center for Theoretical Physics, University of California, Berkeley, CA 94720, U.S.A.}
\affiliation{Theoretical Physics Group, Lawrence Berkeley National Laboratory, Berkeley, CA 94720, U.S.A.}

\author{Benjamin R. Safdi}
\affiliation{Leinweber Center for Theoretical Physics, Department of Physics, University of Michigan, Ann Arbor, MI 48109 U.S.A.}
\affiliation{Berkeley Center for Theoretical Physics, University of California, Berkeley, CA 94720, U.S.A.}
\affiliation{Theoretical Physics Group, Lawrence Berkeley National Laboratory, Berkeley, CA 94720, U.S.A.}

\date{\today}

\begin{abstract}
Axions may be produced in abundance inside stellar cores and then convert into observable $X$-rays in the Galactic magnetic fields.  We focus on the Quintuplet and Westerlund 1 super star clusters, which
host large numbers of hot, young stars including Wolf-Rayet stars; these stars produce axions efficiently through the axion-photon coupling.
We use Galactic magnetic field models to calculate the expected $X$-ray flux locally from axions emitted from these clusters.  We then combine the axion model predictions with archival \new{Nuclear Spectroscopic Telescope Array (NuSTAR)} data from 10 - 80 keV to search for evidence of axions.  We find no significant evidence for axions and 
constrain the axion-photon coupling $g_{a\gamma\gamma} \lesssim 3.6 \times 10^{-12}$ GeV$^{-1}$ for masses $m_a \lesssim 5 \times 10^{-11}$ eV at 95\% confidence.

\end{abstract}
\maketitle

\noindent 

Ultralight \new{axion-like particles} that couple weakly to ordinary matter are natural extensions to the Standard Model. 
For example, string compactifications often predict large numbers of such pseudo-scalar particles that interact with the Standard Model predominantly through dimension-five operators~\cite{Svrcek:2006yi,Arvanitaki:2009fg}.  If an axion couples to quantum chromodynamics (QCD) then it may also solve the strong {\it CP} problem~\cite{Peccei:1977ur,Peccei:1977hh,Weinberg:1977ma,Wilczek:1977pj}; \new{in this work we refer to both the QCD axion and axion-like particles as axions.}

Axions may interact electromagnetically  through the operator \mbox{${\mathcal L} = - g_{a\gamma\gamma} a F_{\mu \nu} \tilde F^{\mu \nu} / 4$}, where $a$ is the axion field, $F$ is the electromagnetic field-strength tensor, with $\tilde F$ its Hodge dual, and $g_{a\gamma\gamma}$ is the dimensionful coupling constant \new{of axions to photons}. This operator allows both the production of axions in stellar plasmas through the Primakoff Process~\cite{Pirmakoff:1951pj,PhysRevD.33.897} and the conversion of axions to photons in the presence of static external magnetic fields.  Strong constraints on $g_{a\gamma\gamma}$ for low-mass axions come from the \new{CERN Axion Solar Telescope (CAST)} experiment~\cite{Anastassopoulos:2017ftl}, which searches for axions produced in the Solar plasma that free stream to Earth and then convert to $X$-rays in the magnetic field of the CAST detector.  CAST has excluded axion couplings $g_{a\gamma\gamma} \gtrsim 6.6 \times 10^{-11}$ GeV$^{-1}$
for axion masses $m_a \lesssim 0.02$ eV at 95\% confidence~\cite{Anastassopoulos:2017ftl}.  Primakoff axion production also opens a new pathway by which stars may cool, and strong limits ($g_{a \gamma \gamma} \lesssim 6.6 \times 10^{-11}$ GeV$^{-1}$ at 95\% confidence for $m_a \lesssim$ keV) are derived from observations of the horizontal branch (HB) star lifetime, which would be modified in the presence of axion cooling~\cite{Ayala:2014pea}. 
\begin{figure}[htb]
\hspace{0pt}
\vspace{-0.2in}
\begin{center}
\includegraphics[width=0.49\textwidth]{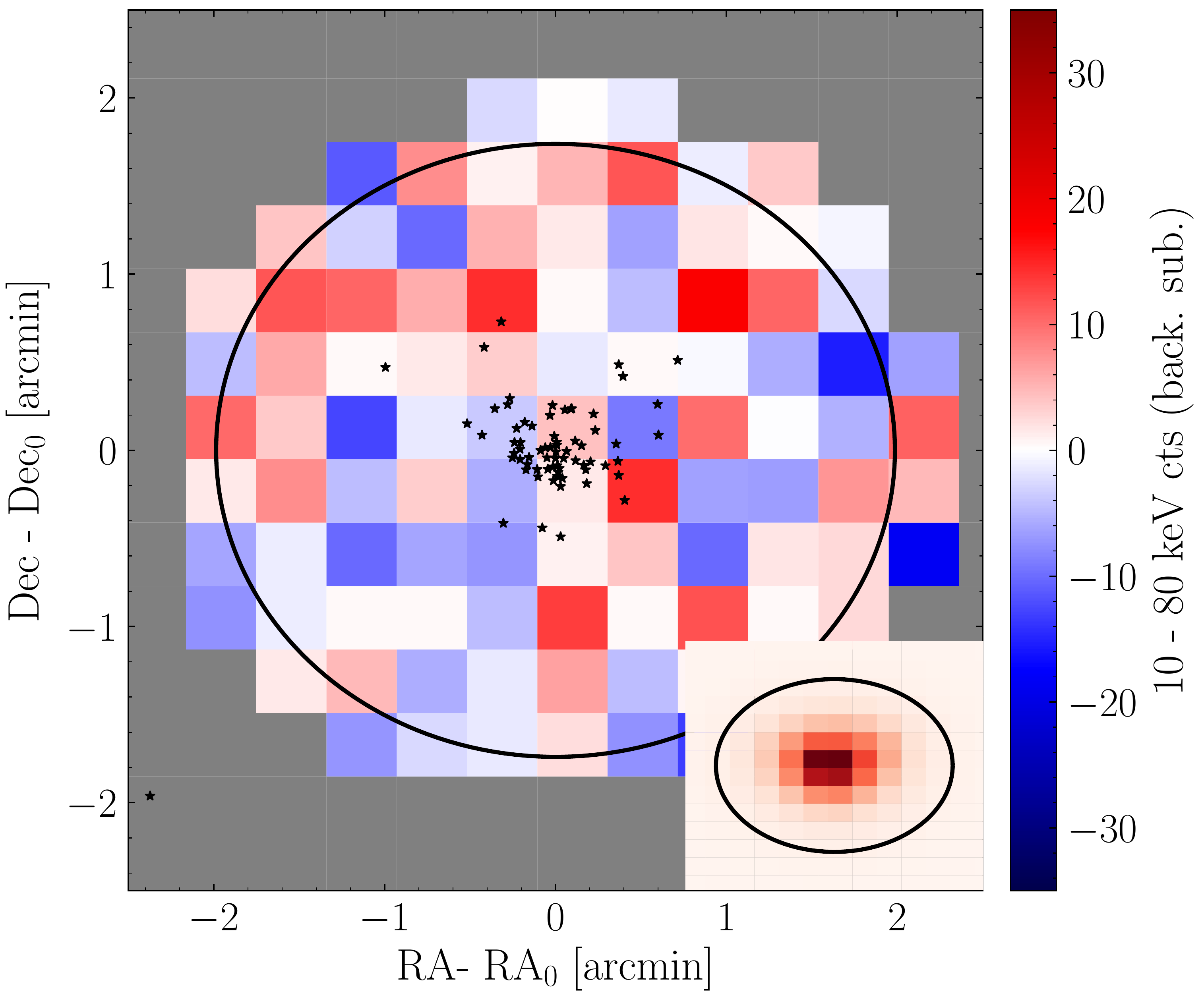}
\caption{The stacked and pixelated background-subtracted count data (10 - 80 keV) from the NuSTAR observations of the Quintuplet SSC.  The locations of the stars are indicated in black, while the 90\% energy containment region for emission associated with the SSC is indicated by the black circle, accounting for the NuSTAR \new{point spread function (PSF)}.  RA$_0$ and DEC$_0$ denote the locations of the cluster center.  We find no evidence for axion-induced emission from this SSC, which would follow the spatial counts template illustrated in the inset panel.  
\label{fig:ill}
}
\end{center}
\end{figure}

In this work, we produce some of the strongest constraints to-date on $g_{a \gamma \gamma}$ for $m_a \lesssim  10^{-9}$ eV through $X$-ray observations with the \new{Nuclear Spectroscopic Telescope Array (NuSTAR) telescope~\cite{Harrison:2013md}} of super star clusters (SSCs).  The SSCs contain large numbers of hot, young, and massive stars, such as Wolf-Rayet (WR) stars.  We show that these stars, and as a result the SSCs, are highly efficient at producing axions with energies $\sim$10--100 keV through the Primakoff process.  These axions may then convert into photons in the Galactic magnetic fields, leading to signatures observable with space-based $X$-ray telescopes such as NuSTAR.  We analyze archival NuSTAR data from the Quintuplet SSC near the Galactic Center (GC) along with the nearby Westerlund 1 (Wd1) cluster and constrain  $g_{a\gamma\gamma} \lesssim 3.6 \times 10^{-12}$ GeV$^{-1}$ at 95\% confidence for $m_a \lesssim 5 \times 10^{-11}$ eV.  In Fig.~\ref{fig:ill} we show the locations of the stars within the Quintuplet cluster that are considered in this work on top of the background-subtracted NuSTAR counts, from 10 - 80 keV, with the point-spread function (PSF) of NuSTAR also indicated. In the Supplementary Material (SM) we show that observations of the Arches SSC yield similar but slightly weaker limits.

Our work builds upon significant previous efforts to use stars as laboratories to search for axions.  Some of the strongest constraints on the axion-matter couplings, for example, come from examining how HB, white dwarf (WD), red giant, and neutron star (NS) cooling would be affected by an axion~\cite{Raffelt:1985nj,Isern:2008nt,Isern:2008fs,Isern:2010wz,Bertolami:2014wua,Ayala:2014pea,Redondo:2013wwa,Viaux:2013lha,Giannotti:2015kwo,Giannotti:2017hny}.
When the stars have large magnetic fields, as is the case for WDs and NSs, the axions can be converted to $X$-rays in the stellar magnetospheres~\cite{Fortin:2018ehg,Dessert:2019sgw,Dessert:2019dos,Buschmann:2019pfp}. 
Intriguingly, in~\cite{Dessert:2019dos,Buschmann:2019pfp} observations of the Magnificent Seven  nearby isolated NSs found evidence for a hard $X$-ray excess consistent with the expected axion spectrum from nucleon bremsstrahlung.  
This work extends these efforts by allowing the axions to convert to $X$-rays not just in the stellar magnetic fields but also in the Galactic magnetic fields~\cite{1995PhLB..344..245C,Giannotti:2017qvz,Meyer:2016wrm}.

\noindent
{\it Axion production in SSCs.---}During helium burning, particularly massive stars may undergo considerable mass loss, especially through either rotation or binary interaction, which can begin to peel away the hydrogen envelope, revealing the hot layers underneath and reversing the cooling trend. Stars undergoing this process are known as WR stars, and these stars are the most important in our analyses. If the star has a small ($<$40\% abundance) remaining hydrogen envelope, it is classified as a WNh star; at $<$5\% 
hydrogen abundance it is classified as a WN star; otherwise, it is classified as WC or WO, which indicates the presence of $>$2\% carbon, and oxygen, respectively, in the atmosphere. 

Axions are produced through the photon coupling $\gagg$ in the high-mass stars in SSCs through the Primakoff process $\gamma + (e^-,Z) \rightarrow a + (e^-,Z)$. This process converts a stellar photon to an axion in the screened electromagnetic field of the nucleons and electrons. The massive stars are high-temperature and low-density and therefore form nonrelativistic nondegenerate plasmas. The Primakoff emission rate was calculated in~\cite{PhysRevD.33.897,PhysRevD.37.1356} as a function of temperature, density, and composition, and is described in detail in the SM.

To compute the axion luminosity in a given star, we use the stellar evolution code Modules for Experiments in Stellar Astrophysics (MESA)~\cite{2011ApJS..192....3P,2013ApJS..208....4P} to find, at any particular time in the stellar evolution, radial profiles of temperature, density, and composition.  The simulation states are specified by an initial metallicity $Z$, an initial stellar mass, an initial rotation velocity, and an age. 
The initial metallicity is taken to be constant for all stars.  In the SM we show that the Quintuplet and Arches clusters, which are both near the GC, are likely to have initial  metallicities in the range $Z \in (0.018, 0.035)$, consistent with the conclusions of previous works \new{which} place the initial metallicities of these clusters near solar (solar metallicity is $Z \approx 0.02$)~\cite{Najarro:2004qm,Najarro:2008ki}.
Note that higher metallicities generally lead to the stars entering the WR classifications sooner, when their cores are cooler.
Rotation may also cause certain massive stars to be classified as WR stars at younger ages.  We model the initial rotation distribution as a Gaussian distribution with mean $\mu_{\rm rot}$ and standard deviation $\sigma_{\rm rot}$ for non-negative rotation speeds~\cite{Hunter:2007be,Brott:2011wc}.  Refs.~\cite{Hunter:2007be,Brott:2011wc} found $\mu_{\rm rot} \approx 100$ km/s and $\sigma_{\rm rot} \approx 140$ km/s, but to assess systematic uncertainties we vary $\mu_{\rm rot}$ between $50$ and $150$ km/s~\cite{Hunter:2007be}.

We draw initial stellar velocities from the velocity distribution described above (from 0 to 500 km/s) and initial stellar masses from the Kroupa initial mass function~\cite{Kroupa:2000iv} (from 15 to 200 $M_\odot$).  We use MESA to evolve the stars from \new{pre-main-sequence (pre-MS)--before core hydrogen ignition--to near-supernova.}  At each time step we assign each stellar model a spectroscopic classification using the definitions in~\cite{Weidner_2010,Hamann:2006tf}.  
We then construct an ensemble of models for each spectroscopic classification by joining together the results of the different simulations that result in the same classification for stellar ages within the age range for star formation in the cluster; for Quintuplet, this age range is between 3.0 and 3.6 Myr~\cite{Clark_2018}.
Note that each simulation generally provides multiple representative models, taken at different time steps. 
In total we compute $10^5$ models per stellar classification.

Quintuplet hosts 71 stars of masses $\gtrsim 50 M_\odot$, with a substantial WR cohort~\cite{Clark_2018}. 
In particular it has 14 WC + WN stars, and we find that these stars dominate the predicted axion flux.  For example, at $g_{a\gamma\gamma} = 10^{-12}$ GeV$^{-1}$ we compute that the total axion luminosity from the SSC (with $Z = 0.035$ and $\mu_{\rm rot} = 150$ km/s) is $2.1_{-0.4}^{+0.7} \times 10^{35}$ erg/s, with WC + WN stars contributing $\sim$70\% of that flux.
Note that the uncertainties arise from performing multiple (500) draws of the stars from our ensembles of representative models.  In the 10 - 80 keV energy range relevant for NuSTAR the total luminosity is $1.7_{-0.3}^{+0.4} \times 10^{35}$ erg/s.  We take $Z = 0.035$ and $\mu_{\rm rot} = 150$ km/s because these choices lead to the most conservative limits.  For example, taking the metallicity at the lower-end of our range ($Z = 0.018$) along with $\mu_{\rm rot} = 100$ km/s the predicted 10 - 80 keV flux increases by $\sim$60\%.  At fixed $Z = 0.035$ changing $\mu_{\rm rot}$ from 150 km/s to 100 km/s increases the total luminosity (over all energies) by $\sim$10\%, though the luminosity in the 10 - 80 keV range is virtually unaffected. 

The Wd1 computations proceed similarly.  Wd1 is measured from parallax to be a distance $d \in (2.2,4.8)$ kpc from the Sun~\cite{2020MNRAS.492.2497A}, accounting for both statistical and systematic uncertainties~\cite{2019MNRAS.486L..10D}.  Wd1 is estimated to have an age between 4.5 and 7.1 Myr from isochrone fitting, which we have broadened appropriately from~\cite{Clark_2019} accounting for expanded distance uncertainties.  In our fiducial analysis we simulate the stars in Wd1 for initial metallicity $Z = 0.035$ and $\mu_{\rm rot} = 150$ km/s as this leads to the most conservative flux predictions, even though it is likely that the metallicity is closer to solar for Wd1~\cite{1998A&AS..127..423P}, in which cases the fluxes are larger by almost a factor of two (see the SM).  We model 153 stars in
Wd1~\cite{Clark_2019}, but the axion flux is predominantly produced by the 8 WC and 14 WN stars.  In total we find that the 10 - 80 keV luminosity, for $g_{a\gamma\gamma} = 10^{-12}$ GeV, is $9.02_{-1.1}^{+1.2} \times 10^{35}$ erg/s, which is $\sim$5 times larger than that from Quintuplet.

\noindent
{\it Axion conversion in Galactic fields.---}The axions produced within the SSCs may convert to $X$-rays in the Galactic magnetic fields.  The axion Lagrangian term ${\mathcal{L}}  = g_{a\gamma\gamma} a {\bf E} \cdot {\bf B}$, written in terms of electric and magnetic fields ${\bf E}$ and ${\bf B}$, causes an incoming axion state to rotate into a polarized electromagnetic wave in the presence of an external magnetic field (see, {\it e.g.},~\cite{Raffelt:1987im}).  
The conversion probability $p_{a\to \gamma}$ depends on the transverse magnetic field, the axion mass $m_a$, and the plasma frequency $\omega_{\rm pl} \approx 3.7 \times 10^{-12} ( n_e / 10^{-2} \, \, {\rm cm}^{-3}  )^{-1/2}$ eV, with $n_e$ the free-electron density (see the SM for an explicit formula). \new{Note that hydrogen absorption towards all of our targets is negligible, being at most $\sim$5\% in the 15-20 keV bin of the Quintuplet analysis~\cite{2006MNRAS.371...38W}.   }

To compute the energy-dependent conversion probabilities $p_{a \to \gamma}$ for our targets we need to know the magnetic field profiles and electron density distributions along the lines of sight. For our fiducial analysis we use the regular components of the JF12 Galactic magnetic field model~\cite{2012ApJ...757...14J, Jansson_2012} and the YMW16 electron density model~\cite{Yao_2017} (though in the SM we show that the \texttt{ne2001}~\cite{Cordes:2002wz} model gives similar results), though the JF12 model does not cover the inner kpc of the Galaxy.
 Outside of the inner kpc the conversion probability for Quintuplet is dominated by the out-of-plane (X-field) component in the JF12 model.  We conservatively assume that the magnitude of the vertical magnetic field within the inner kpc is the same as the value at 1 kpc ($|B_{z}| \approx 3$ $\mu$G), \new{as illustrated in \mbox{Supp. Fig. S6}}.  In our fiducial magnetic field model the conversion probability is $p_{a \to \gamma} \approx 2.4 \times 10^{-4}$ ($7 \times 10^{-5}$) for $g_{a\gamma\gamma} = 10^{-12}$ GeV$^{-1}$ for axions produced in the Quintuplet SSC with $m_a \ll 10^{-11}$ eV and $E = 80$ keV ($E= 10$ keV).  Completely masking the inner kpc reduces these conversion probabilities to $p_{a \to \gamma} \approx 1.0 \times 10^{-4}$ ($p_{a \to \gamma} \approx 3.2 \times 10^{-5}$), for $E = 80$ keV ($E= 10$ keV).  On the other hand, changing global magnetic field model to that presented in~\cite{2011ApJ...738..192P} (PTKN11), which has a larger in-plane component than the JF12 model but no out-of-plane component, leads to conversion probabilities at $E = 80$ and $10$ keV of $p_{a \to \gamma} \approx 4.9 \times 10^{-4}$ and $p_{a \to \gamma} \approx 4.2 \times 10^{-5}$, respectively, with the inner kpc masked.

The magnetic field is likely larger than the assumed 3 $\mu$G within the inner kpc.
Note that the local interstellar magnetic field, as measured directly by the {\it Voyager} missions~\cite{2020A&A...633L..12I}, indirectly by the Interstellar Boundary Explorer~\cite{2016ApJ...818L..18Z}, inferred from polarization measurements of nearby stars~\cite{2010ApJ...724.1473F}, and inferred from pulsar dispersion measure and the rotation measure data~\cite{2010A&A...513A..28S},  has magnitude $B \sim 3$ $\mu$G, and all evidence points to the field rising significantly in the inner kpc~\cite{Ferriere:2009dh}.  For example, Ref.~\cite{Crocker_2010} bounded the magnetic field within the inner 400 pc to be at least 50 $\mu$G, and more likely 100 $\mu$G (but less than $\sim$400 $\mu$G~\cite{Crocker_2011}), by studying non-thermal radio emission in the inner Galaxy. 
Localized features in the magnetic field in the inner kpc may also further enhance the conversion probability beyond what is accounted for here.  For example, the line-of-sight to the Quintuplet cluster overlap with the GC radio arc non-thermal filament, which has a $\sim$3 mG vertical field over a narrow filament of cross-section $\sim$$(10\, \, {\rm pc})^2$ (see, {\it e.g.},~\cite{Guenduez:2019cwe}).  Accounting for the magnetic fields structures described above in the inner few hundred pc may enhance the conversion probabilities by over an order of magnitude relative to our fiducial scenario (see the SM).  

When computing the conversion probabilities for Wd1 we need to account for the uncertain distance $d$ to the SSC (with currently-allowable range given above).
In the JF12 model we find the minimum $p_{a \to \gamma} / d^2$ (for $m_a \ll 10^{-11}$ eV) is obtained for $d \approx 2.6$ kpc, which is thus the value we take for our fiducial distance in order to be conservative.  At this distance the conversion probability is $p_{a\to\gamma} \approx 2.4 \times 10^{-6}$ ($p_{a\to\gamma} \approx 1.5 \times 10^{-6}$) for $E = 10$ keV ($E = 80$ keV), assuming $g_{a\gamma\gamma} = 10^{-12}$ GeV$^{-1}$ and $m_a \ll 10^{-11}$ eV. We note that the conversion probabilities are over 10 times larger in the PTKN11 model (see the SM), since there is destructive interference (for $d  \approx 2.6$ kpc) in the JF12 model towards Wd1. 
We do not account for turbulent fields in this analysis; inclusion of these fields may further increase the conversion probabilities for Wd1, although we leave this modeling for future work.

\noindent
{\it Data analysis.---}We reduce and analyze 39 ks 
of archival NuSTAR data from Quintuplet with observation ID \texttt{40010005001}.  This observation was performed as part of the NuSTAR Hard X-ray Survey of the GC Region~\cite{Mori:2015vba,Hong:2016qjq}. 
 The NuSTAR data reduction was performed with the HEASoft software version 6.24~\cite{1999ascl.soft12002B}. This process leads to a set of counts, exposure, and background maps for every energy bin and for each exposure (we use data from both Focal Plane Modules A and B).  The astrometry of each exposure is calibrated independently using the precise location of the source \mbox{1E 1743.1-2843}~\cite{Porquet:2003jv}, which is within the field of view.   
 The background maps account for the cosmic $X$-ray background, reflected solar X-rays, and instrumental backgrounds such as Compton-scattered gamma rays and detector and fluorescence emission lines~\cite{Wik:2014boa}.  We then stack and re-bin the data sets to construct pixelated images in each of the energy bins.  We use 14 5-keV-wide energy bins between 10 and 80 keV.  We label those images $d_i = \{ c_i^p \}$, where $c_i^p$ stands for the observed counts in energy bin $i$ and pixel $p$.  The pixelation used in our analysis is illustrated in Fig.~\ref{fig:ill}.
 
 For the Wd1 analysis we reduced Focal Plane Module A and B data \new{totaling 138 ks} from observation IDs \texttt{80201050008}, \texttt{80201050006}, and \texttt{80201050002}.  This set of observations was performed to observe outburst activity of the Wd1 magnetar CXOU J164710.2--45521~\cite{Borghese:2019con}, which we mask at $0.5'$ in our analysis. (The magnetar is around 1.5' away from the cluster center.)  Note that in~\cite{Borghese:2019con} hard $X$-ray emission was only detected with the NuSTAR data from 3 - 8 keV from CXOU J164710.2--45521 -- consistent with this, removing the magnetar mask does not affect our extracted spectrum for the SSC above \new{10} keV.  We use the magnetar in order to perform astrometric calibration of each exposure independently.
The  Wd1 exposures suffer from ghost-ray contamination~\cite{2017arXiv171102719M} from a nearby point source that is outside of the NuSTAR field of view at low energies (below $\sim$15 keV)~\cite{Borghese:2019con}. \new{(Ghost-ray contamination refer to those photons that reflect only a single time in the mirrors.)} The ghost-ray contamination affects our ability to model the background below 15 keV and so we remove the 10 - 15 keV energy bin from our analysis.
 
 In each energy bin we perform a Poissonian template fit over the pixelated data to constrain the number of counts that may arise from the template associated with axion emission from the SSC.  To construct the signal template we use a spherically-symmetric approximation to the NuSTAR PSF~\cite{An:2014hua} and we account for each of the stars in the SSC individually in terms of spatial location and expected flux, which generates a non-spherical and extended template.  We label the set of signal templates by $S_i^p$. We search for emission associated with the signal templates by profiling over background emission.  We use the set of background templates described above and constructed when reducing the data, which we label $B_i^p$.
  \begin{figure}[htb]
\hspace{0pt}
\vspace{-0.2in}
\begin{center}
\includegraphics[width=0.49\textwidth]{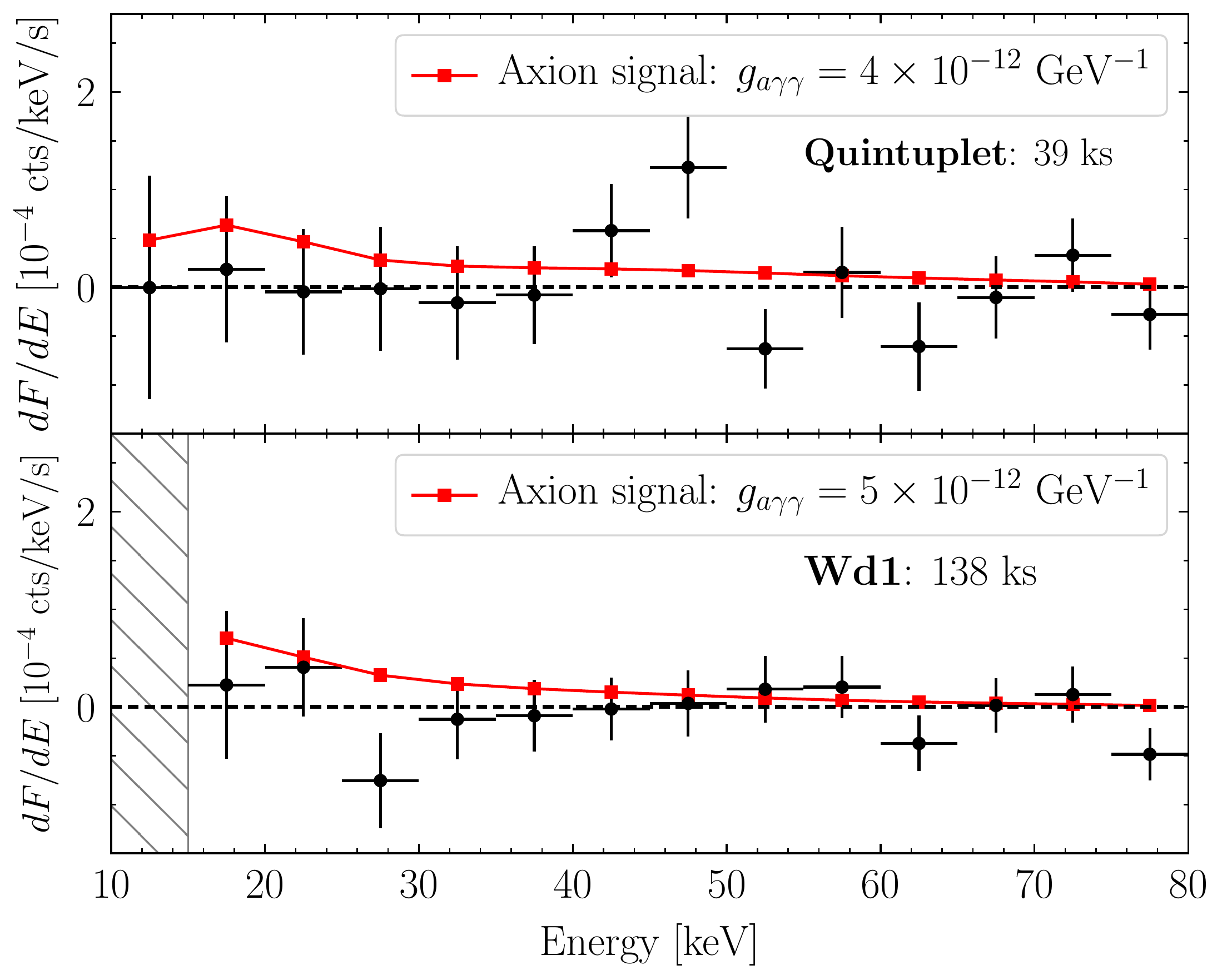}
\caption{The spectra associated with the axion-induced templates from the Quintuplet and Wd1 SSCs constructed from the NuSTAR data analyses, with best-fit points and 1$\sigma$ uncertainties indicated.  In red we show the predicted spectra from an axion with $m_a \ll 10^{-11}$ eV and indicated $g_{a\gamma\gamma}$. Note that for Wd1 we do not analyze the 10 - 15 keV energy bin because of ghost-ray contamination.
\label{fig:flux_spectra}
}
\end{center}
\end{figure}

 Given the set of signal and background templates we construct a Poissonian likelihood in each energy bin:
 \es{eq:LL}{
 p_i(d_i | \{S_i,A_B\} ) = \sum_p { {(\mu_i^p)}^{c_i^p} e^{-\mu_i^p} \over c_i^p!} \,,
 }
 with $\mu_i^p = S_i S_i^p + A_B B_i^p$.  We then construct the profile likelihood $p_i(d_i | \{S_i \})$ by maximizing the log likelihood at each fixed $S_i$ over the nuisance parameter $A_B$. Note that when constructing the profile likelihood we use the region of interest (ROI) where we mask pixels further than 2.0' from the SSC center.  The 90\% containment radius of NuSTAR is $\sim$1.74', independent of energy, as indicated in Fig.~\ref{fig:ill}.  We use a localized region around our source to minimize possible systematic biases from background mismodeling.  However, as we show in the SM our final results are not strongly dependent on the choice of ROI.  
 We also show in the SM that if we inject a synthetic axion signal into the real data and analyze the hybrid data, we correctly recover the simulated axion parameters. 
 
 The best-fit flux values and 1$\sigma$ uncertainties extracted from the profile likelihood procedure are illustrated in Fig.~\ref{fig:flux_spectra} for the Quintuplet and Wd1  data sets.
 We compare the spectral points to the axion model prediction to constrain the axion model.  More precisely, we combine the profile likelihoods together from the individual energy bins to construct a joint likelihood that may be used to search for the presence of an axion signal: $p(d | \{m_a , g_{a \gamma \gamma}\} ) = \prod_i p_i\big[d_i | R_i(m_a, g_{a \gamma \gamma} )\big]$, where $R_i(m_a, g_{a \gamma \gamma} )$ denotes the predicted number of counts in the $i^{\rm th}$ energy bin given an axion-induced $X$-ray spectrum with axion model parameters  $\{m_a, g_{a \gamma \gamma}\}$.  The values $R_i(m_a, g_{a \gamma \gamma} )$ are computed using the forward-modeling matrices constructed during the data reduction process.  
 
\begin{figure}[htb]  
\hspace{0pt}
\vspace{-0.2in}
\begin{center}
\includegraphics[width=0.49\textwidth]{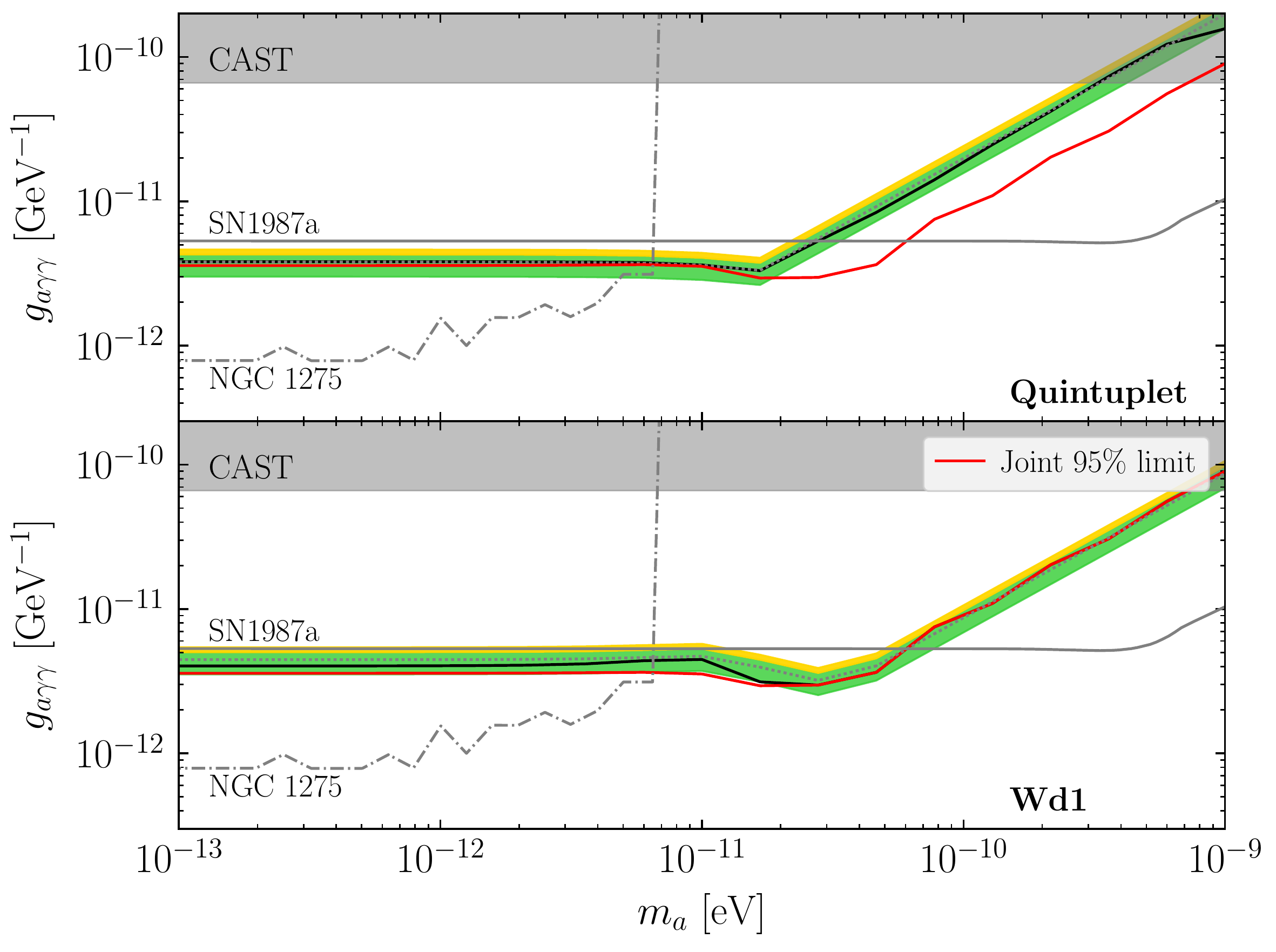}
\caption{ The 95\% upper limits (black) on $g_{a\gamma\gamma}$ as a function of the axion mass from the Quintuplet and Wd1 data analyses.  We compare the limits to the 1$\sigma$ (green band) and 2$\sigma$ (yellow band) expectations under the null hypothesis, along with the median expectations (dotted).  The joint 95\% upper limit, combining Quintuplet and Wd1, is also indicated (expected joint limit not shown).  At low masses our limits may be surpassed by those from searches for $X$-ray spectral modulations from NGC 1275~\cite{Reynolds:2019uqt},  \new{though we caution that those limits have been called into question recently, as discussed further in the text}~\cite{Libanov:2019fzq}. 
\label{fig:limits}
}
\end{center}
\end{figure}

 In Fig.~\ref{fig:limits} we illustrate the 95\% power-constrained~\cite{Cowan:2011an} upper limits on $g_{a\gamma\gamma}$ as a function of the axion mass $m_a$ found from our analyses.  The joint limit (red in Fig.~\ref{fig:limits}), combining the Quintuplet and Wd1 profile likelihoods, becomes $g_{a\gamma\gamma} \lesssim 3.6 \times 10^{-12}$ GeV$^{-1}$ at low axion masses. 
 At fixed $m_a$ the upper limits are constructed by analyzing the test statistic $q(g_{a\gamma\gamma} | m_a) \equiv 2 \ln p(d | \{m_a , g_{a \gamma \gamma}\} ) -2 \ln p(d | \{m_a , \bar g_{a \gamma \gamma}\} )$, where $\bar g_{a \gamma \gamma}$ is the signal strength that maximizes the likelihood, allowing for the possibility of negative signal strengths as well.  The 95\% upper limit is given by the value $g_{a \gamma \gamma} > \bar g_{a \gamma \gamma}$ such that $q(g_{a\gamma\gamma} | m_a) \approx 2.71$ (see, {\it e.g.},~\cite{Cowan:2010js}).  The 1$\sigma$ and 2$\sigma$ expectations for the 95\% upper limits under the null hypothesis, constructed from the Asimov procedure~\cite{Cowan:2010js}, are also shown in  Fig.~\ref{fig:limits}. 
 The evidence in favor of the axion model is $\sim$$0.3\sigma$ (0$\sigma$) local significance at low masses for Quintuplet (Wd1). 

 We compare our upper limits with those found from the CAST experiment~\cite{Anastassopoulos:2017ftl}, the non-observation of gamma-rays from SN1987a~\cite{Payez:2014xsa} (see also~\cite{Raffelt:1990yz,Raffelt:2006cw,Chang:2016ntp} along with~\cite{Bar:2019ifz}, who recently questioned the validity of these limits), and the NGC 1275 $X$-ray spectral modulation search~\cite{Reynolds:2019uqt}.  
It was recently pointed out, however, that the limits in~\cite{Reynolds:2019uqt} are highly dependent on the intracluster magnetic field models and could be orders of magnitude weaker, when accounting for both regular and turbulent fields~\cite{Libanov:2019fzq}.  \new{The CAST limits are stronger than ours for $m_a \gtrsim 10^{-9}$ eV and rely on less modeling assumptions, since CAST searches for axions produced in the Sun, though we have made conservative choices in our stellar modeling.}

\noindent
{\it Discussion.---}We present limits on the axion-photon coupling $g_{a\gamma\gamma}$ from a search with NuSTAR hard $X$-ray data for axions emitted from the hot, young stars within SSCs and converting to $X$-rays in the Galactic magnetic fields.  We find the strongest limits from analyses of data towards the Quintuplet and Wd1 clusters.
Our limits represent some of the strongest and most robust limits to-date on $g_{a\gamma\gamma}$ for low-mass axions. 
We find no evidence for axions.   
Promising targets for future analyses could be nearby supergiant stars, such as Betelgeuse~\cite{1995PhLB..344..245C,Giannotti:2020abc}, or young NSs such as Cas A. 

\noindent
{\it Acknowledgments.---We  thank Fred Adams, Malte Buschmann, Roland Crocker, Ralph Eatough, Glennys Farrar, Katia Ferri\`ere, Andrew Long, Kerstin Perez, and Nick Rodd for helpful comments and discussions.  This  work  was  supported  in  part  by  the  DOE Early  Career  Grant  DESC0019225  and  through  computational resources and services provided by Advanced Research  Computing  at  the  University  of  Michigan,  Ann Arbor.  Figures and Supplementary data are provided in~\cite{SM}. }

\bibliography{axion}

\clearpage

\onecolumngrid
\begin{center}
  \textbf{\large Supplementary Material: $X$-ray Searches for Axions from Super Star Clusters}\\[.2cm]
Christopher Dessert, Joshua W. Foster, Benjamin R. Safdi
\end{center}

\onecolumngrid
\setcounter{equation}{0}
\setcounter{figure}{0}
\setcounter{table}{0}
\setcounter{section}{0}
\setcounter{page}{1}
\makeatletter
\renewcommand{\theequation}{S\arabic{equation}}
\renewcommand{\thefigure}{S\arabic{figure}}

This Supplementary Material contains additional results and explanations of our methods that clarify and support the results presented in the main Letter.  First, we present additional details regarding the data analyses, simulations, and calculations performed in this work.  We then show additional results beyond those presented in the main Letter.  In the last section we provide results of an auxiliary analysis used to derive the metallicity range considered in this work.   

\section{Methods: Data Reduction, Analysis, Simulations, and Calculations}

In this section we first provide additional details needed to reproduce our NuSTAR data reduction, before giving extended discussions of our MESA simulations, axion luminosity calculations, and conversion probability calculations.

\subsection{Data Reduction and analysis}
To perform the NuSTAR data reduction, we use the NuSTARDAS software included with HEASoft 6.24~\cite{1999ascl.soft12002B}. We first reprocess the data with the NuSTARDAS task \texttt{nupipeline}, which outputs calibrated and screened events files. We use the strict filtering for the South Atlantic Anomaly. We then create counts maps for both focal plane modules (FPMs) of the full NuSTAR FOV with \texttt{nuproducts} in energy bins of width $5$ keV from $5-80$ keV.\footnote{\new{We use 5 keV-wide energy bins as a compromise between having narrow energy bins that allow us to resolve the spectral features in our putative signal (see Fig.~\ref{fig:flux_spectra}) and having wide-enough bins that allow to accurately determine the background template normalizations in our profile likelihood analysis procedure.  However, small-to-moderate changes to the bins sizes ({\it e.g.}, increasing them by a factor of 2) lead to virtually identical results.}} We additionally generate the ancillary response files (ARFs) and the redistribution matrix files (RMFs) for each FPM. We generate the corresponding exposure maps with \texttt{nuexpomap}, which produces exposure maps with units [s]. To obtain maps in exposure units [cm$^2$ s keV] that we can use to convert from counts to flux, we multiply in the mean effective areas in each bin with no PSF or vignetting correction.

Once the data is reduced, we apply the analysis procedure described in the main text to measure the spectrum associated with the signal template in each energy bin.  However, to compare the signal-template spectrum to the axion model prediction, we need to know how to forward-model the predicted axion-induced flux, which is described in more detail later in the SM, through the instrument response.  In particular, we 
pass the signal flux prediction through the detector response to obtain the expected signal counts that we can compare to the data:
\es{eq:for_mod}{
	\mu_{S,i}^e({\bm \theta}_{\rm S}) =t^e \int dE^\prime{\rm RMF}_i^e(E^\prime) {\rm ARF}^e(E^\prime) S(E^\prime|{\bm \theta}_{\rm S}) \,.
}
Here, $t^e$ is the exposure time corresponding to the exposure $e$ in [s], while the signal is the expected intensity spectrum in [erg/cm$^2$/s/keV]. We have now obtained the expected signal counts $\mu_{S,i}^e({\bm \theta}_{\rm S})$ that may be integrated into the likelihood given in~\eqref{eq:LL}.

\subsection{MESA Simulations}
\label{sec:MESA}

MESA is a one-dimensional stellar evolution code which solves the equations of stellar structure to simulate the stellar interior at any point in the evolution. In our fiducial analysis, we construct models at a metallicity Z = 0.035, initial stellar masses from 15 to 200 $M_\odot$, and initial surface rotations from 0 km/s to 500 km/s as indicated in the main text.  We use the default inlist for high-mass stars provided with MESA. This inlist sets a number of parameters required for high-mass evolution, namely the use of Type 2 opacities. We additionally use the Dutch wind scheme~\cite{Glebbeek_2009} as in the high rotation module.  

On this grid, we simulate each star from the pre-MS phase until the onset of neon burning around $1.2 \times 10^9$ K. At that point, the star only has a few years before undergoing supernova. Given that no supernova has been observed in the SSCs since the observations in 2012-2015, this end-point represents the most evolved possible state of stars in the SSCs at time of observation.  The output is a set of radial profiles at many time steps along the stellar evolution. The profiles describe, for example, the temperature, density, and composition of the star. These profiles allow us to compute the axion spectrum at each time step by integrating the axion volume emissivity over the interior. 

Here we show detailed results for a representative star of mass 85 $M_\odot$ with initial surface rotation of 300 km/s. This star is a template star for the WC phase (and other WR phases) in the Quintuplet Cluster, which dominates the Quintuplet axion spectrum in the energies of interest.
\begin{figure}[htb]  
\hspace{0pt}
\vspace{-0.2in}
\begin{center}
\includegraphics[width=0.49\textwidth]{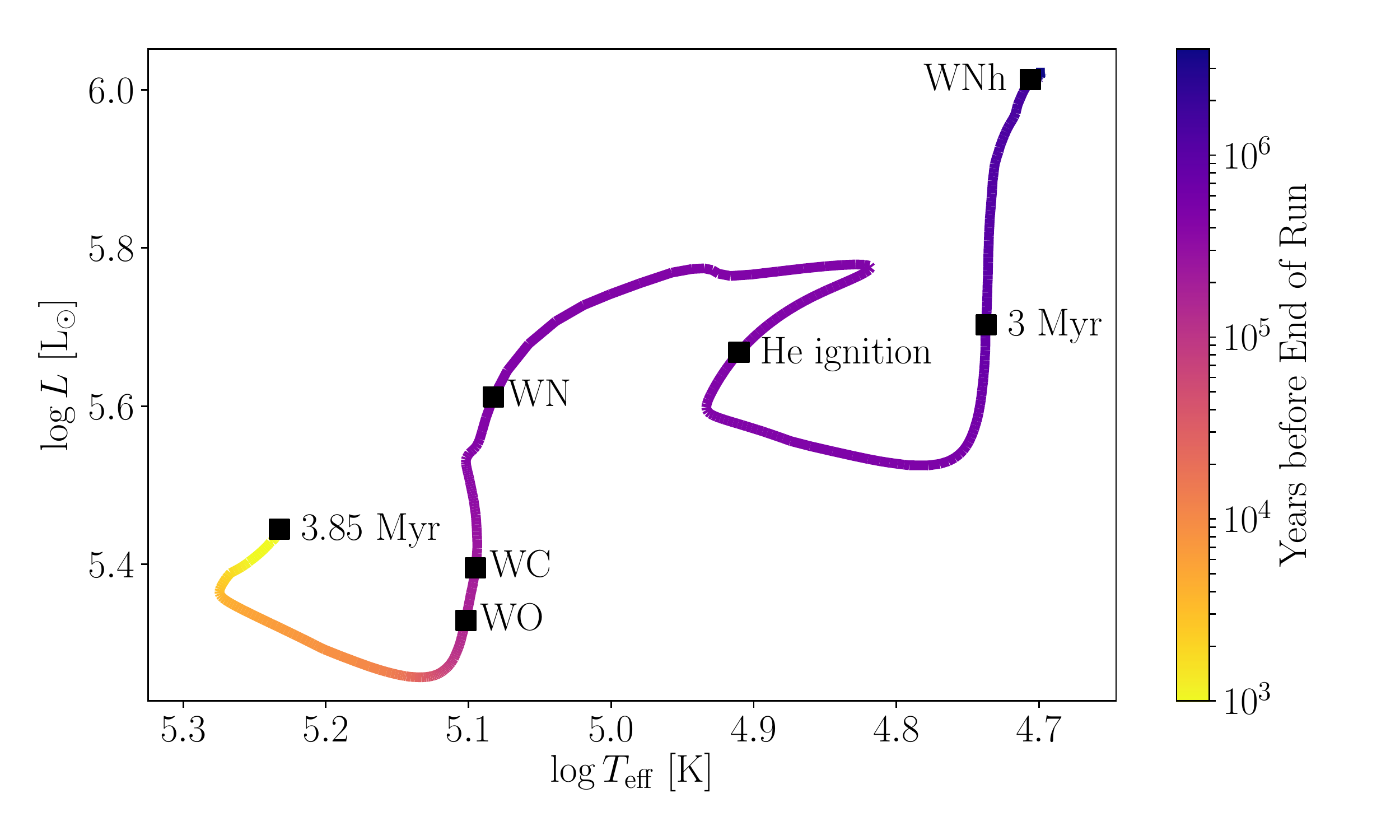}
\includegraphics[width=0.49\textwidth]{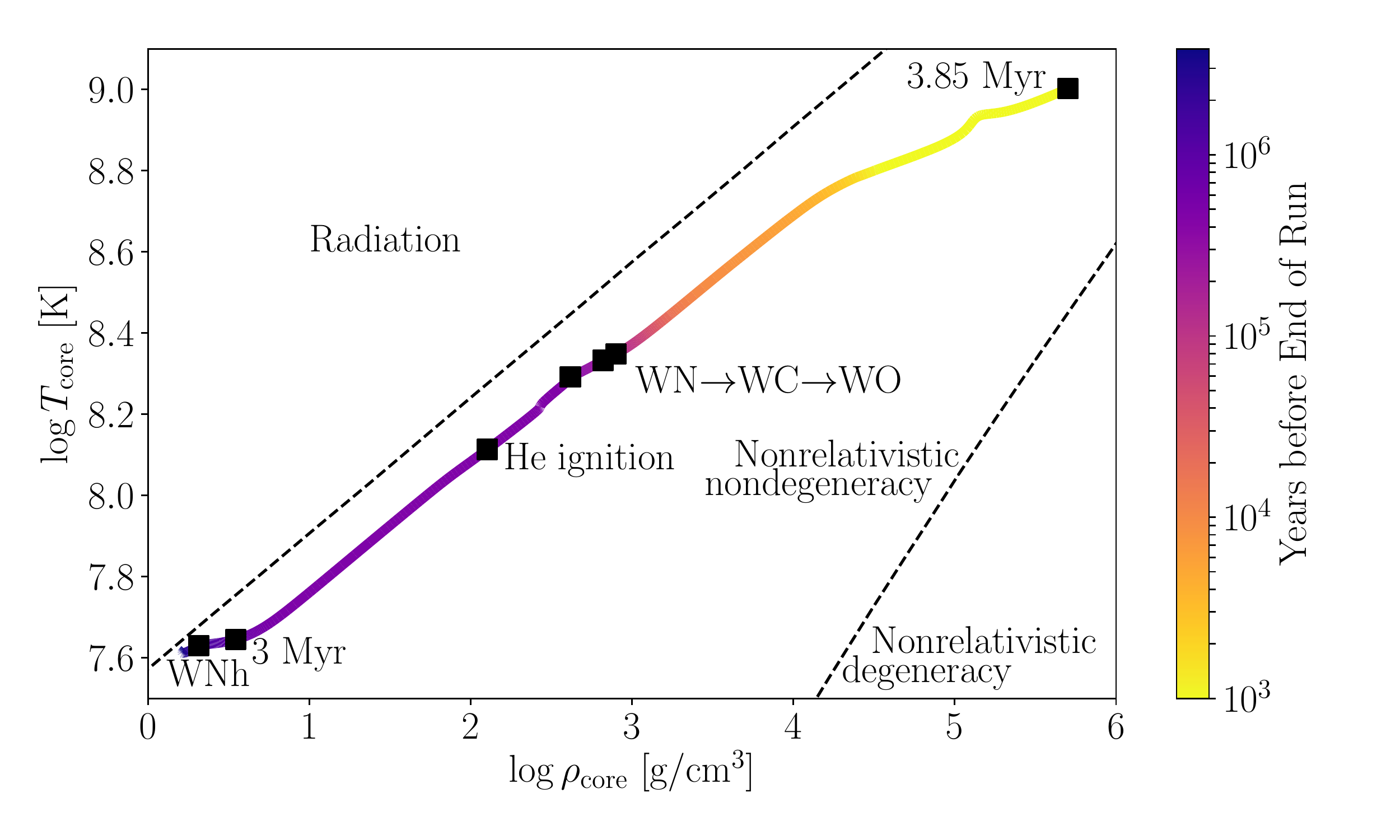}
\caption{(Left) The HR diagram for the Quintuplet template star of mass 85 $M_\odot$ and initial surface rotation of 300 km/s. The coloring indicates the year before the run was stopped, approximately a few years from supernova. We mark with black squares, in order of occurrence, when the star enters the WNh phase, when it is 3 Myr old, when its core undergoes helium ignition, when it enters the WN, WC, and WO phases, and finally when the run ends at 3.85 Myr. (Right) A $\log$T-$\log\rho$ diagram for the template star with the same points of interest marked. We also show the relevant degeneracy zones, showing that the star is entirely in the nonrelativistic nondegenerate regime.
\label{fig:MESA_HR_Trho}
}
\end{center}
\end{figure}
In the left panel of Fig.~\ref{fig:MESA_HR_Trho}, we show the Hertzsprung–Russell (HR) diagram for our template star. The star's life begins on the MS, where it initiates core hydrogen burning. Eventually, the core runs out of hydrogen fuel and is forced to ignite helium to prevent core collapse (see Fig.~\ref{fig:MESA_abund} left). Because helium burns at higher temperatures, the star contracts the core to obtain the thermal energy required to ignite helium (see Fig.~\ref{fig:MESA_core}). At the same time, the radiation pressure in stellar winds cause heavy mass loss in the outer layers, which peels off the hydrogen envelope (see Fig.~\ref{fig:MESA_MR}). When the surface is 40\% hydrogen, the star enters the WNh phase; when it is 5\% hydrogen, the star enters the WN phase. Further mass loss begins to peel off even the helium layers, and the star enters the WC and WO phases when its surface is 2\% carbon and oxygen by abundance~\cite{Hamann:2006tf}, respectively (see Fig.~\ref{fig:MESA_abund} right).

\begin{figure}[htb]  
\hspace{0pt}
\vspace{-0.2in}
\begin{center}
\includegraphics[width=0.49\textwidth]{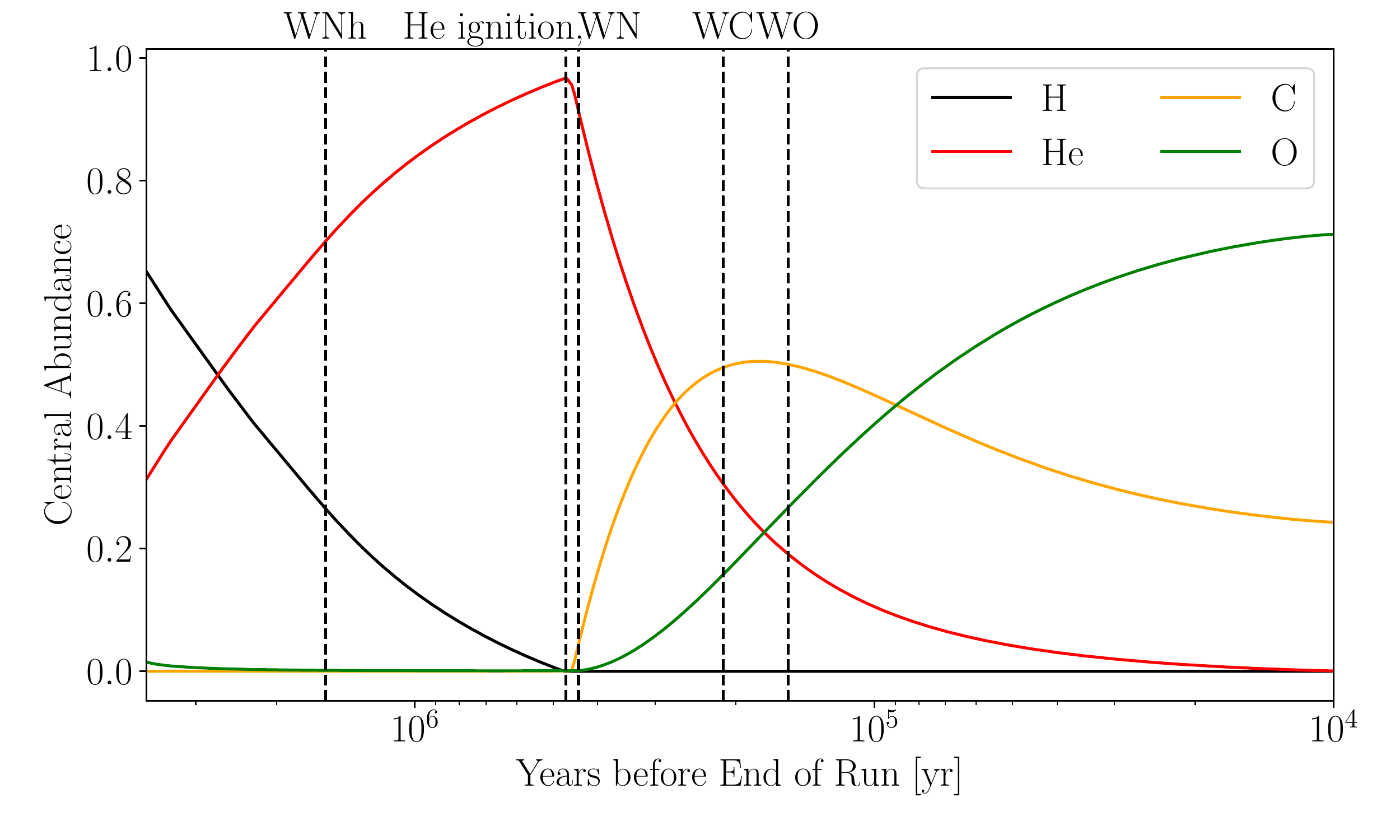}
\includegraphics[width=0.49\textwidth]{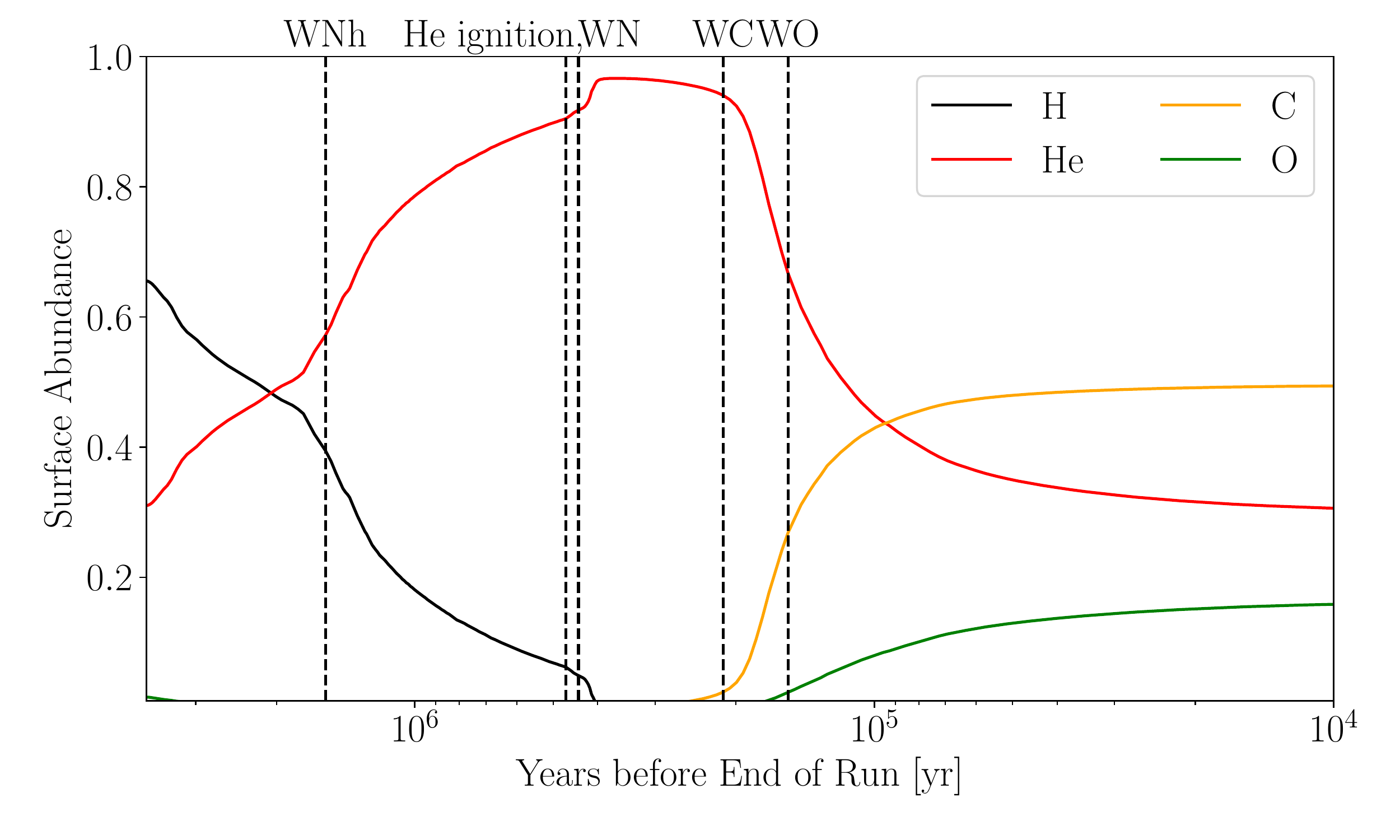}
\caption{(Left) The abundances of hydrogen (black), helium (red), carbon (yellow), and oxygen (green) in the center of the star as a function of time, for the simulation described in Fig.~\ref{fig:MESA_HR_Trho}. With dashed-black vertical lines, we mark several points of interest: ``WNh'' indicates the time the star enters the WNh phase, ``He ignition'' when its core undergoes helium ignition, and ``WN'',``WC'', and ``WO'' indicate the beginning of the WN, WC, and WO phases, respectively. (Right) The same as in the left panel, but for surface abundances. 
\label{fig:MESA_abund}
}
\end{center}
\end{figure}

\begin{figure}[htb]  
\hspace{0pt}
\vspace{-0.2in}
\begin{center}
\includegraphics[width=0.49\textwidth]{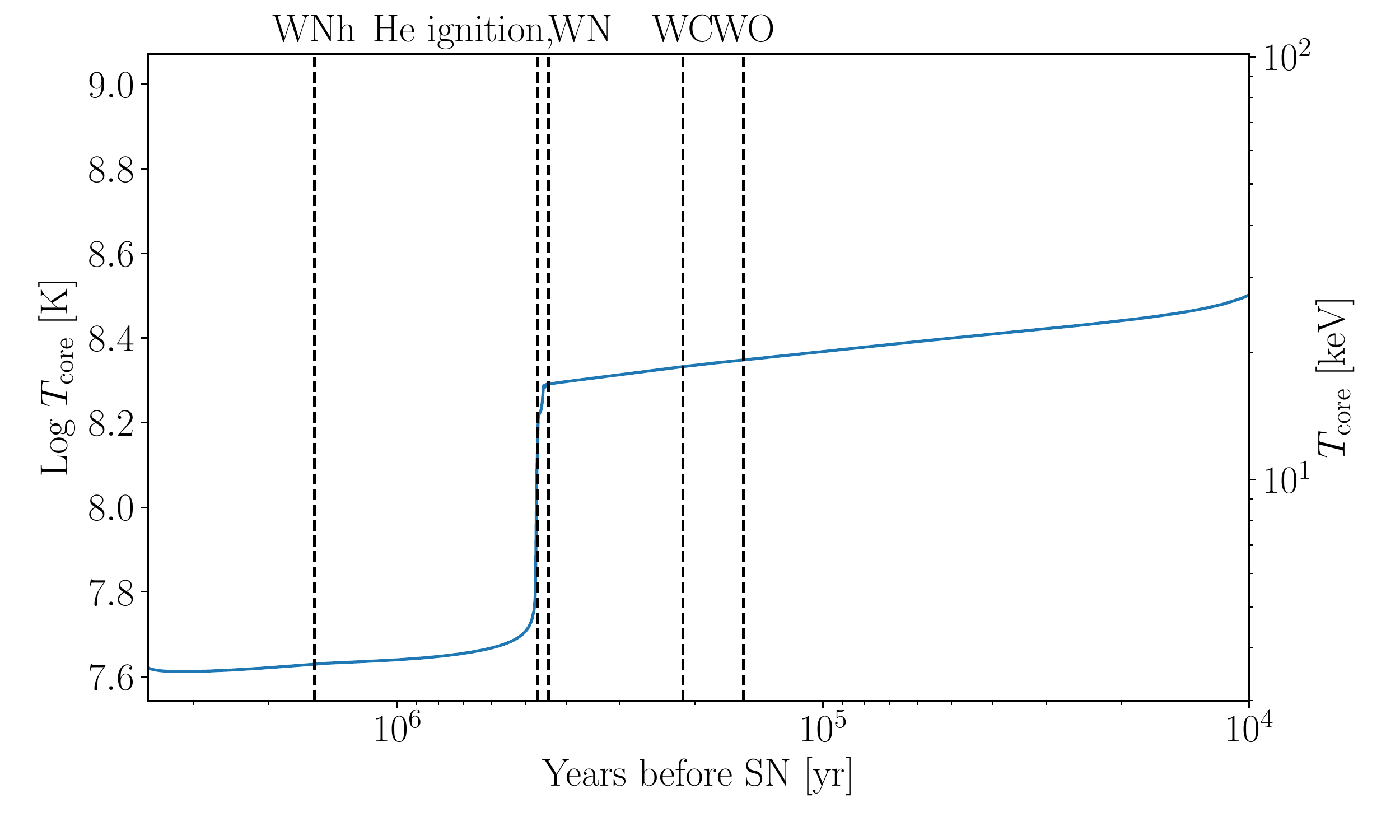}
\includegraphics[width=0.49\textwidth]{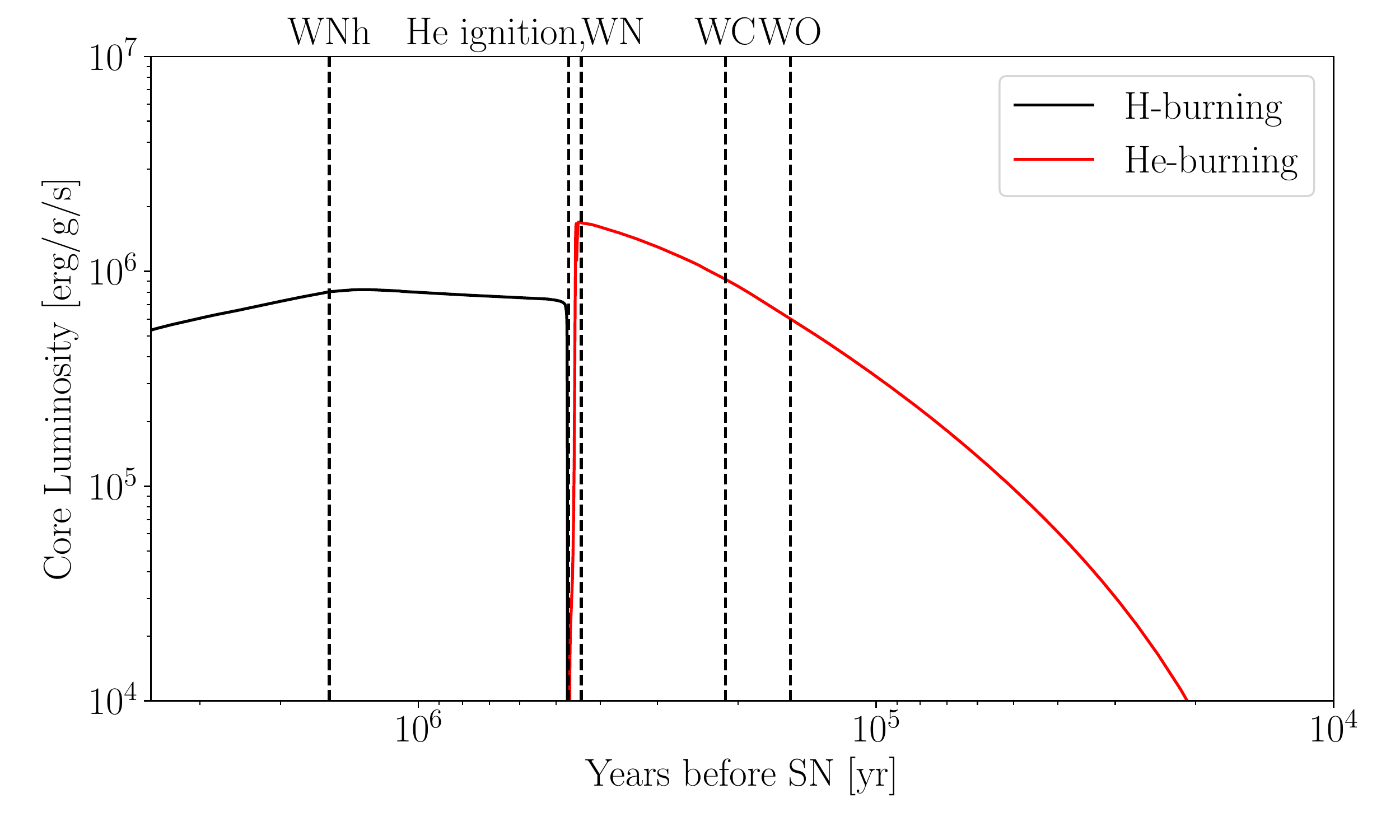}
\caption{(Left) The stellar core temperature as a function of time for the simulation described in Fig.~\ref{fig:MESA_HR_Trho}. (Right) The hydrogen and helium luminosities in the core through the CNO cycle and the triple-alpha process, respectively. The dashed-black vertical lines retain their meanings from Fig.~\ref{fig:MESA_abund}. 
\label{fig:MESA_core}
}
\end{center}
\end{figure}

\begin{figure}[htb]  
\hspace{0pt}
\vspace{-0.2in}
\begin{center}
\includegraphics[width=0.49\textwidth]{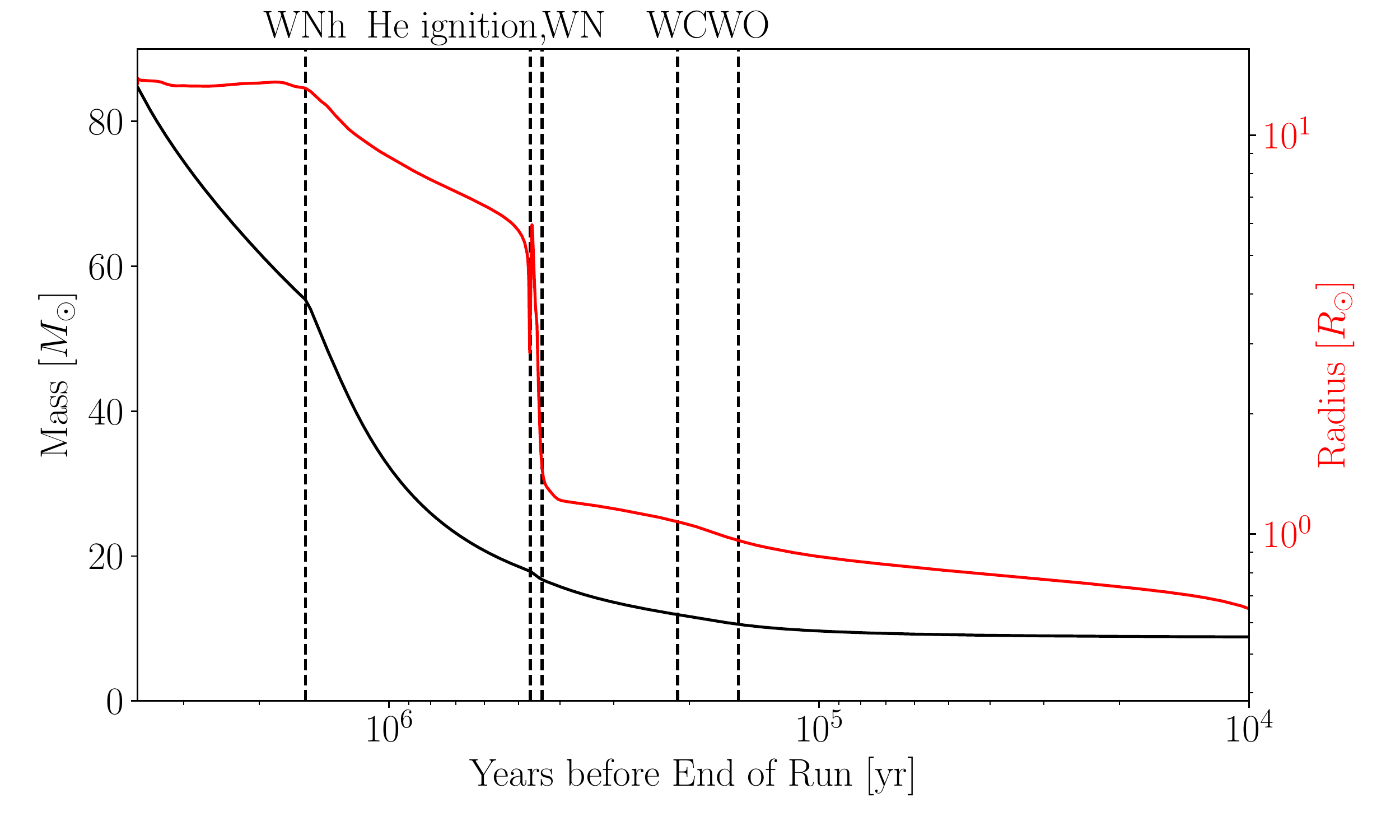}
\caption{The stellar mass (black) and radius (red) as a function of time from the simulation described in Fig.~\ref{fig:MESA_HR_Trho}. The dashed-black vertical lines retain their meanings from Fig.~\ref{fig:MESA_abund}. 
\label{fig:MESA_MR}
}
\end{center}
\end{figure}

\subsection{Axion Production in SSCs}
\label{sec:prod}

In this section we overview how we use the output of the MESA simulations to compute axion luminosities and spectra.

\subsubsection{The Axion Energy Spectrum}
 
Here we focus on the calculation of the axion energy spectrum [erg/cm$^2$/s/keV]. The axion production rate is \cite{1990PhR...198....1R}
\begin{equation}
    \Gamma_p(E) = \dfrac{\gagg^2 T \kappa^2}{32\pi}\left[\left(1+\dfrac{\kappa^2}{4E^2}\right)\ln{\left(1+\dfrac{4E^2}{\kappa^2}\right)}-1\right] \,,
\end{equation}
where $\kappa^2 = \frac{4\pi\alpha}{T}\sum_i{Z_i^2n_i}$ gives the Debye screening scale, which is the finite reach of the Coulomb field in a plasma and cuts off the amplitude. To obtain the axion energy spectrum, this is to be convolved with the photon density, such that
\es{}{
\dfrac{dL_p}{dE}(E) &= \dfrac{1}{\pi^2}\dfrac{E^3}{e^{E/T}-1}\Gamma_p(E)\\
    &= \dfrac{\gagg^2}{8\pi^3}\dfrac{\xi^2T^3E}{e^{E/T}-1}\left[\left(E^2+\xi^2T^2\right)\ln{\left(1+\dfrac{E^2}{\xi^2T^2}\right)}-E^2\right] \,,
    }
where we have defined the dimensionless parameter $\xi^2 = \dfrac{\kappa^2}{4T^2}$. To obtain the axion emissivity for a whole star, we integrate over the profiles produced with MESA, and we show results for this calculation in the next section. Finally, the axion-induced photon spectrum at Earth is given by
\begin{equation}
    \dfrac{dF}{dE}(E) = P_{a\rightarrow\gamma}(E)\dfrac{1}{4\pi d^2}\dfrac{dL_a}{dE}(E) \,,
\end{equation}{}
with the conversion probability $P_{a\rightarrow\gamma}$ computed later.

\subsubsection{Results for Template Star}
\label{sec:MESA_axion}

In this section, we show our expectation for the axion luminosity from our template star. 

\begin{figure}[htb]
\hspace{0pt}
\vspace{-0.2in}
\begin{center}
\includegraphics[width=0.49\textwidth]{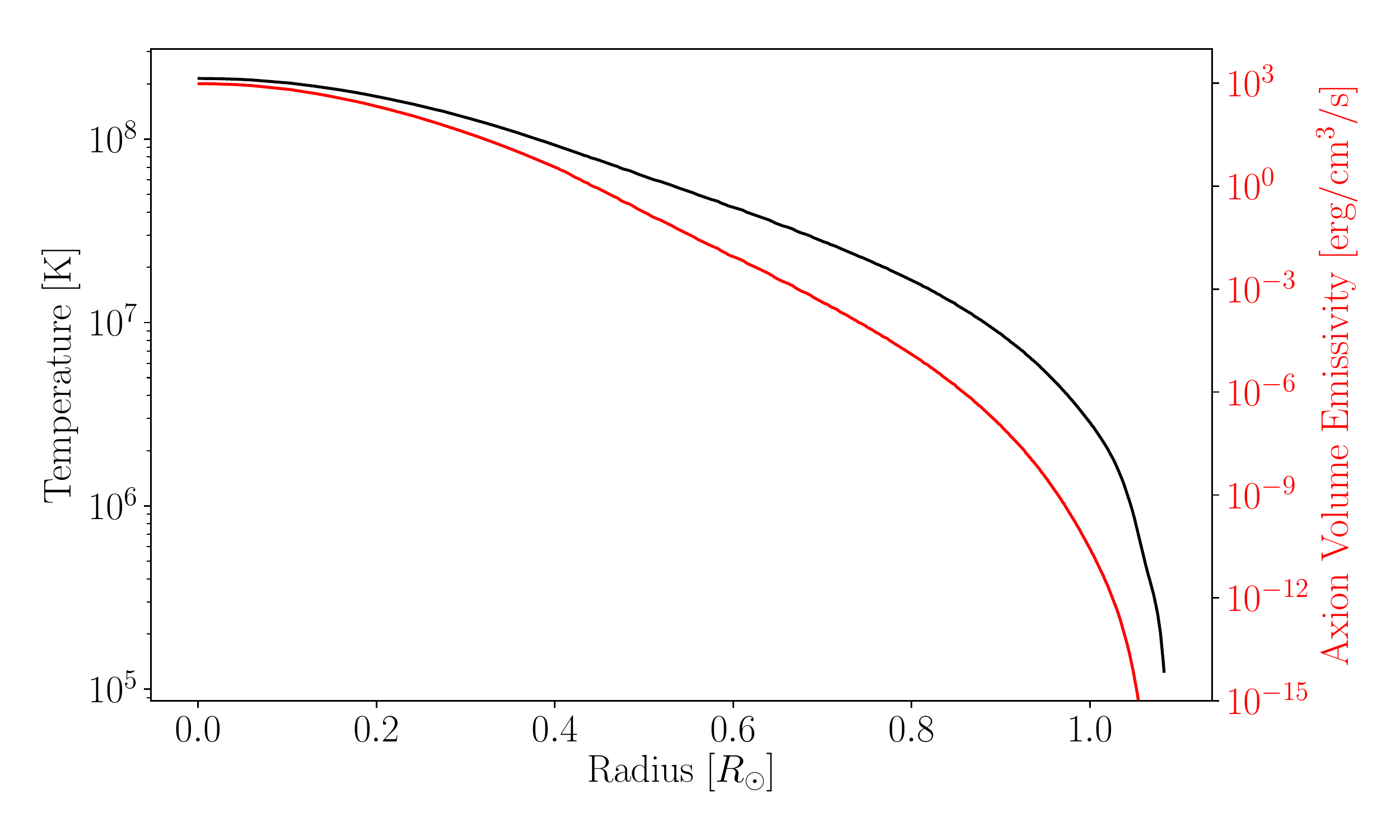}
\includegraphics[width=0.49\textwidth]{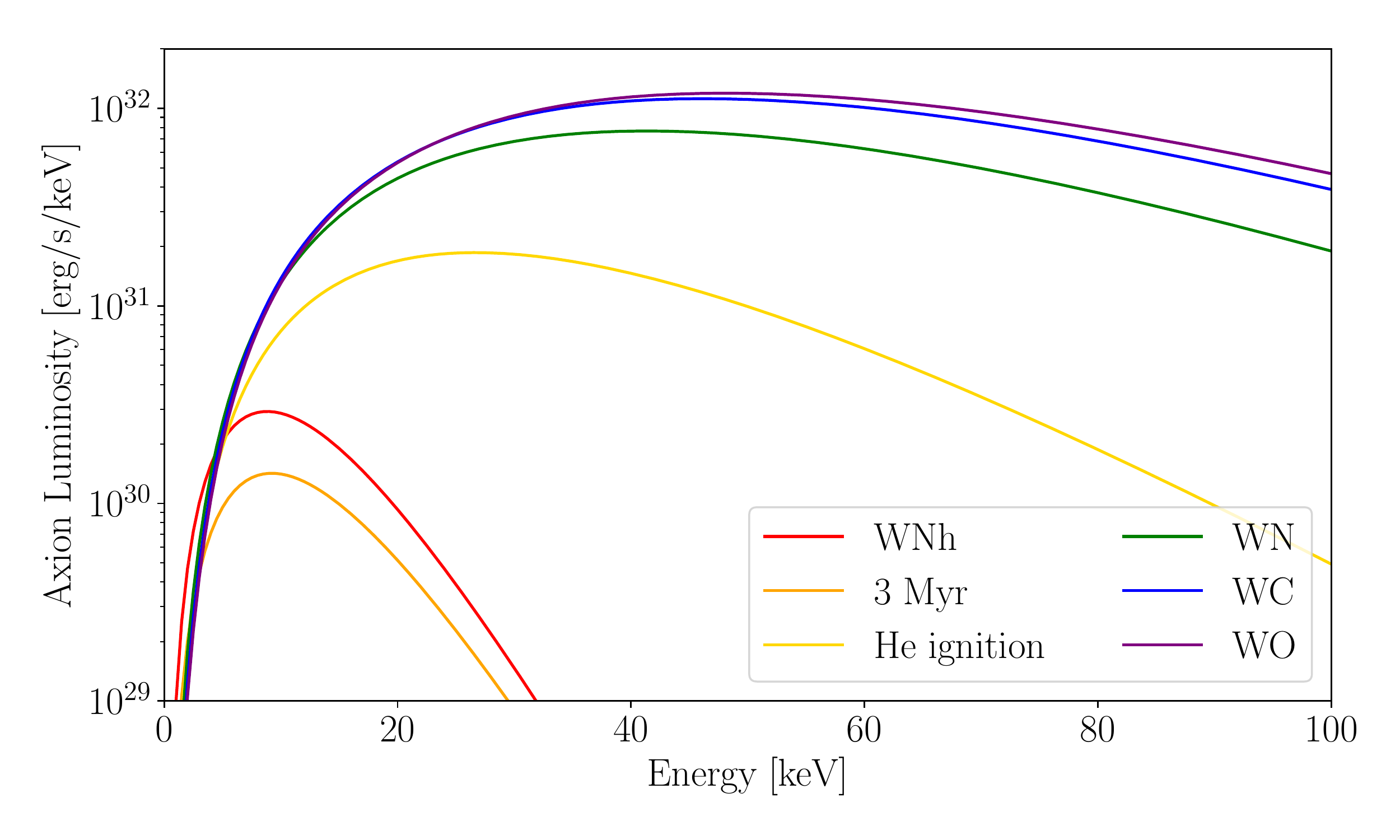}
\caption{(Left) Axion volume emissivity over the interior of the star. In this figure we have taken the stellar model to be the one at the start of the WC stage and fixed $\gagg = 10^{-12}$ GeV$^{-1}$. For comparison purposes, we also show the temperature profile. (Right) Axion luminosity spectrum for those same stages marked in Fig.~\ref{fig:MESA_abund}.
\label{fig:axion_profile}
}
\end{center}
\end{figure}

In the left panel of Fig.~\ref{fig:axion_profile}, we show the axion emissivity from the radial slices of the MESA profile, using the model at the start of the WC evolutionary stage. As expected, the stellar core is by far the most emissive due to its high temperature and density. We also show the temperature profile in the star. Note that the axion volume emissivity does not have the same profile shape as the temperature because the emissivity also depends on the density and composition which are highly nonuniform over the interior. 

In the right panel of Fig.~\ref{fig:axion_profile}, we show how the axion luminosity changes over the stellar lifetime. We see that before helium ignition, the axion luminosity is rather low, and the axion spectrum reaches its maximum around 10 keV, owing to the low core temperature\textemdash the star is still hydrogen burning at core temperatures well below 10 keV. During helium ignition, the luminosity increases quickly due to the sudden increase in temperature. During helium burning, the core temperature continues to increase; for this reason, more evolved stars will be more luminous in axions.

\subsection{Magnetic field model and conversion probability}
\label{sec:prob}

When the axion-to-photon conversion probability $p_{a \to \gamma}$ is sufficiently less than unity, it may be approximated by~\cite{Raffelt:1987im}:
\es{integral}{
p_{a \to \gamma} ={g_{a\gamma\gamma}^2 \over 4} \sum_{i=1,2} \left| \int_0^d dr' B_i(r') e^{i \Delta_a r' - i \int_0^{r'} dr'' \Delta_{||} (r'')} \right|^2 \,,
}
where $B_i$, for $i=1,2$, denote the two orthogonal projections of the magnetic field onto axes perpendicular to the direction of propagation.  The integrals are over the line of sight, with the source located a distance $d$ from Earth, and $r = 0$ denoting the location of the source.  We have also defined $\Delta_a \equiv - m_a^2 / (2 E)$ and $\Delta_{||}(r) \equiv - \omega_{\rm pl}(r)^2 / (2 E)$, with $E$ the axion energy and $\omega_{\rm pl}(r)$ the location-dependent plasma mass.  The plasma mass may be related to the number density of free electrons $n_e$ by $\omega_{\rm pl} \approx 3.7 \times 10^{-12} ( n_e / 10^{-2} \, \, {\rm cm}^{-3}  )^{-1/2}$ eV.
To perform the integral we need to know (i) the free electron density along the line of sight to the target, and (ii) the orthogonal projections of the magnetic field along the line-of-sight.  In this section we give further details behind the electron-density and magnetic-field profiles used in this Letter.

The Quintuplet and Arches SSCs are both $\sim$30 pc away from the GC and thus are expected to have approximately the same conversion probabilities for conversion on the ambient Galactic magnetic fields.  It is possible, however, that local field configurations near the GC could enhanced the conversion probabilities for one or both of these sources.
For example, the axions are expected to travel through or close to the GC radio arc, which has a strong magnetic field $\sim$mG over a cross-section $\sim$$(10 \, \, {\rm pc})^2$~\cite{Guenduez:2019cwe}.  Magnetic fields within the clusters themselves may also be important.    

\begin{figure}[htb]
\hspace{0pt}
\vspace{-0.2in}
\begin{center}
\includegraphics[width=0.49\textwidth]{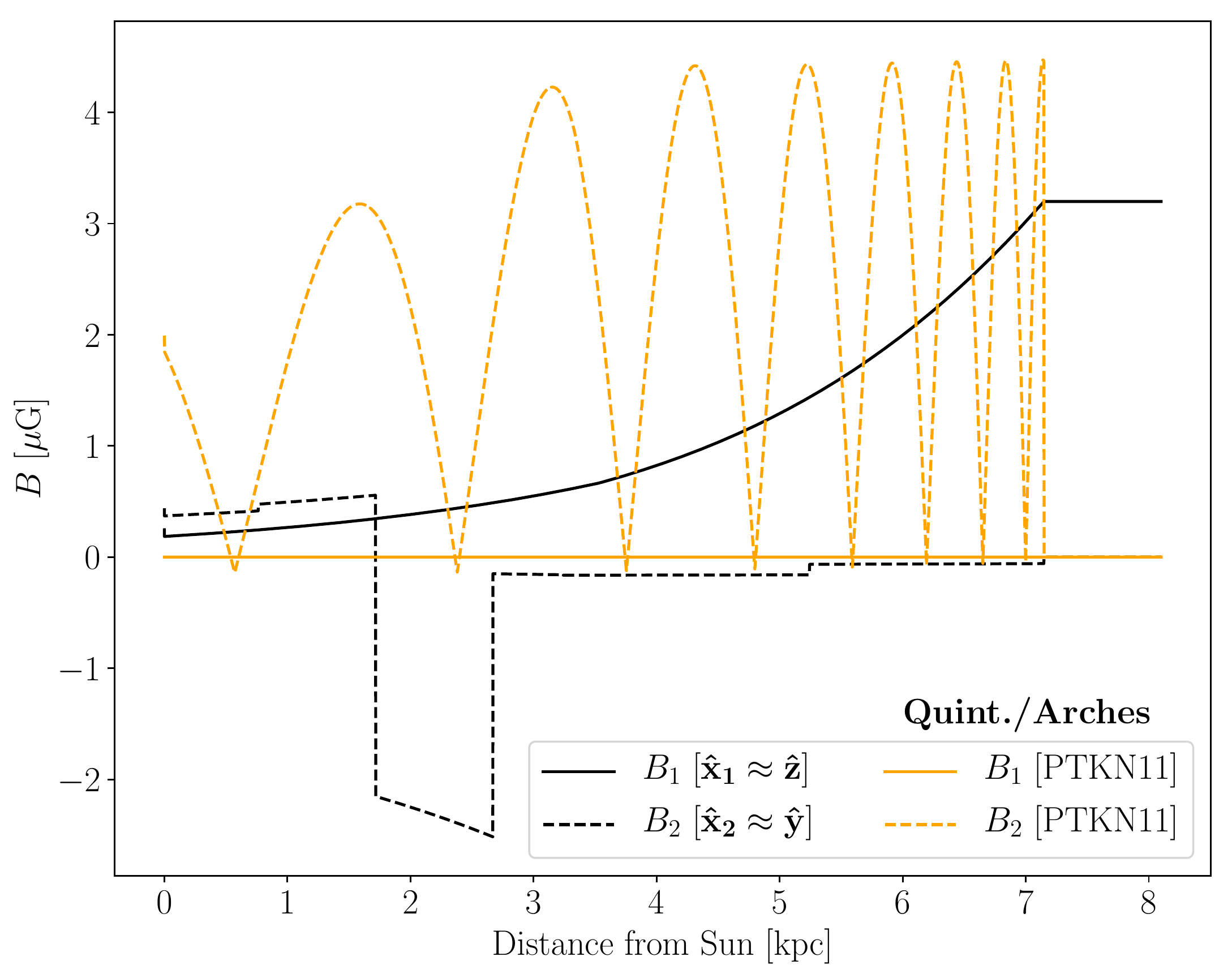}
\includegraphics[width=0.49\textwidth]{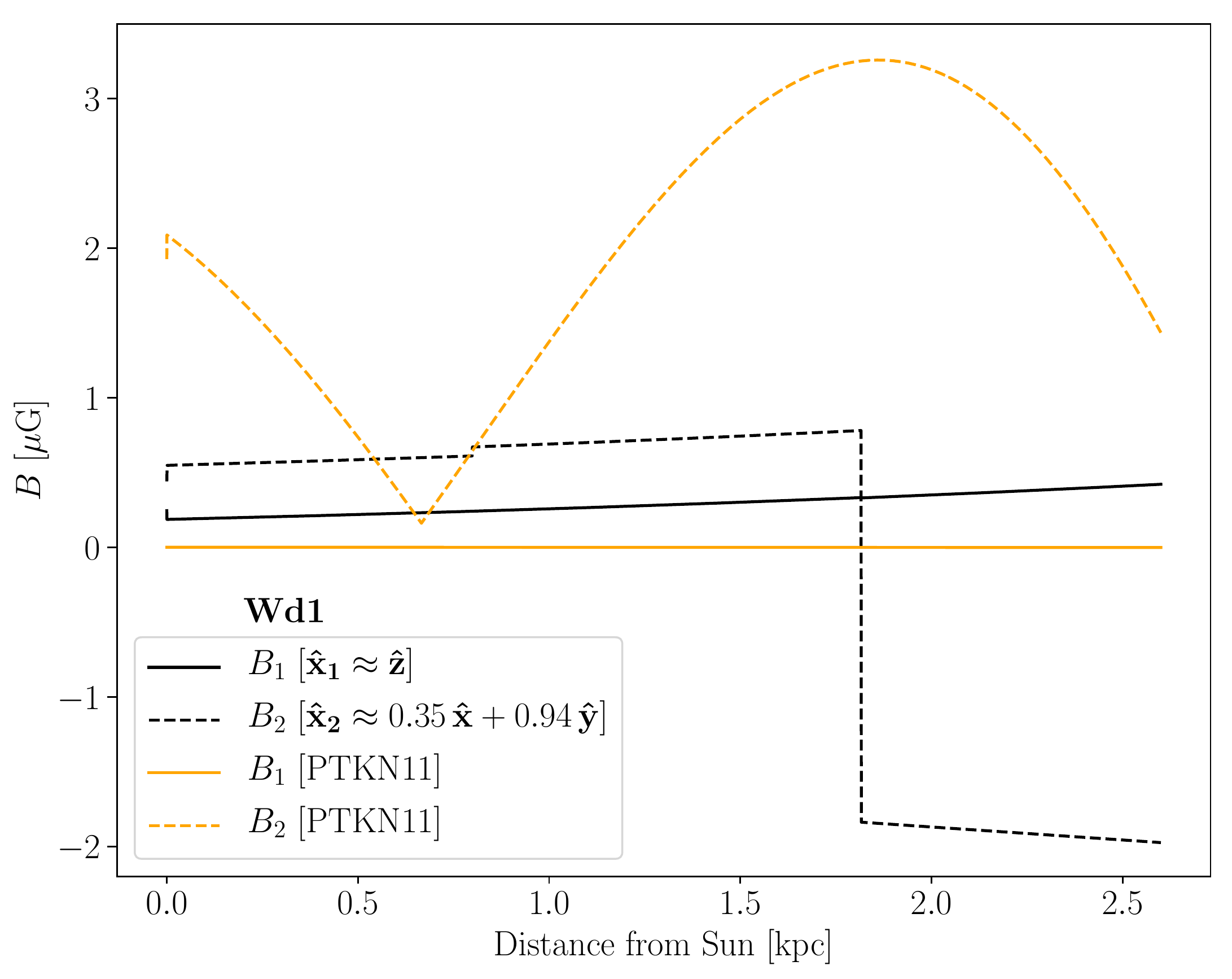}
\caption{We denote the projections of the Galactic magnetic field onto the plane normal to the propagation direction by $B_1$, $B_2$.  (Left) The transverse magnetic field components in our fiducial model (the JF12 model, black) and alternate model (PTKN11, orange) towards the Quintuplet and Arches clusters.  Note that in our fiducial $B$-field model we extend the JF12 model to distances less than 1 kpc from the GC using the field values at 1 kpc.  The true magnetic field values in the inner kpc almost certainly surpass those from this conservative model (see text for details).  (Right) The two field components towards the Wd1 cluster, which is taken to be at a distance of 2.6 kpc from the Sun.  The conversion probabilities towards Wd1 are much larger in the alternate model (PTKN11) than in our fiducial model (JF12), though we stress that random fields are not included and could play an important role in the conversion probabilities towards Wd1.
\label{fig:fid-field-GC}
}
\end{center}
\end{figure}

Our fiducial magnetic field model for Quintuplet and Arches is illustrated in the left panel of Fig.~\ref{fig:fid-field-GC}.  
In the right panel we show the magnetic field profiles relevant for the Wd1 observations.  The components of the $B$-field along the two transverse directions are denoted by $B_1$ and $B_2$.  For the Quintuplet and Arches analyses, the propagation direction is very nearly aligned with $-{\bf \hat x}$ (in Galactic coordinates), so we may take $B_1$ to point in the ${\bf \hat z}$ direction, towards the north Galactic pole, and $B_2$ to point in the direction ${\bf \hat y}$ (the approximate direction of the local rotation).  Note that the targets are slightly offset from the origin of the Galactic coordinate system, so the actual basis vectors have small components in the other directions.  As Wd1 is essentially within the plane of the disk, one of the transverse components points approximately in the ${\bf \hat z}$ direction ($B_1$).

The dominant magnetic field towards the GC within our fiducial $B$-field model is the vertical direction ($B_1$), which is due to the out-of-plane $X$-shaped halo component in the JF12 model~\cite{2012ApJ...757...14J, Jansson_2012}.  However, in the JF12 model that component is cut off within 1 kpc of the GC, due to the fact that in becomes difficult to model the $B$-field near the GC.  The $B$-field is expected to continue rising near the GC -- for example, in~\cite{Crocker_2010} it was claimed that the $B$-field should be at least 50 $\mu$G (and likely 100 $\mu$G) within the inner 400 pc.  However, to be conservative in our fiducial $B$-field model we simply extend the $B$-field to the GC by assuming it takes the value at 1 kpc (about 3 $\mu$G) at all distances less than 1 kpc from the GC.  We stress that this field value is likely orders of magnitude less than the actual field strength, but this assumption serves to make our results more robust.  The extended field model is illustrated in Fig.~\ref{fig:fid-field-GC}.

To understand the level of systematic uncertainty arising from the $B$-field models we also show in Fig.~\ref{fig:fid-field-GC} the magnetic field profiles for the alternative ordered $B$-field model PTKN11~\cite{2011ApJ...738..192P}.  This model has no out-of-plane component, but the regular $B$-field within the disk is stronger than in the JF12 model.  In the case of Quintuplet and Arches we find, as discussed below, that the PTKN11 model leads to similar but slightly enhanced conversion probabilities relative to the JF12 model.  On the other hand, the conversion probabilities in the PTKN11 model towards Wd1 are significantly larger than in the JF12 model.

There is a clear discrepancy in Fig.~\ref{fig:fid-field-GC} between the magnetic field values observed at the solar location, in both the JF12 model and the PTKN11 model, and the local magnetic field strength, which is $\sim$3 $\mu$G~\cite{2016ApJ...818L..18Z}.  The reason is that the magnetic field profiles shown in Fig.~\ref{fig:fid-field-GC} are only the regular components; additional random field components are expected.  For example, in the JF12 model the average root-mean-square random field value at the solar location is $\sim$6.6 $\mu$G~\cite{2012ApJ...757...14J, Jansson_2012}.  The random field components could play an important role in the axion-to-photon conversion probabilities, especially for the nearby source Wd1, but to accurately account for the random field components one needs to know the domains over which the random fields are coherent.  It is expected that these domains are $\sim$100 pc~\cite{Jansson_2012}, in which case the random fields may dominate the conversion probabilities, but since the result depends sensitively on the domain sizes, which remain uncertain, we conservatively neglect the random-field components from the analyses in this work (though this would be an interesting subject for future work).

\begin{figure}[htb]
\hspace{0pt}
\vspace{-0.2in}
\begin{center}
\includegraphics[width=0.49\textwidth]{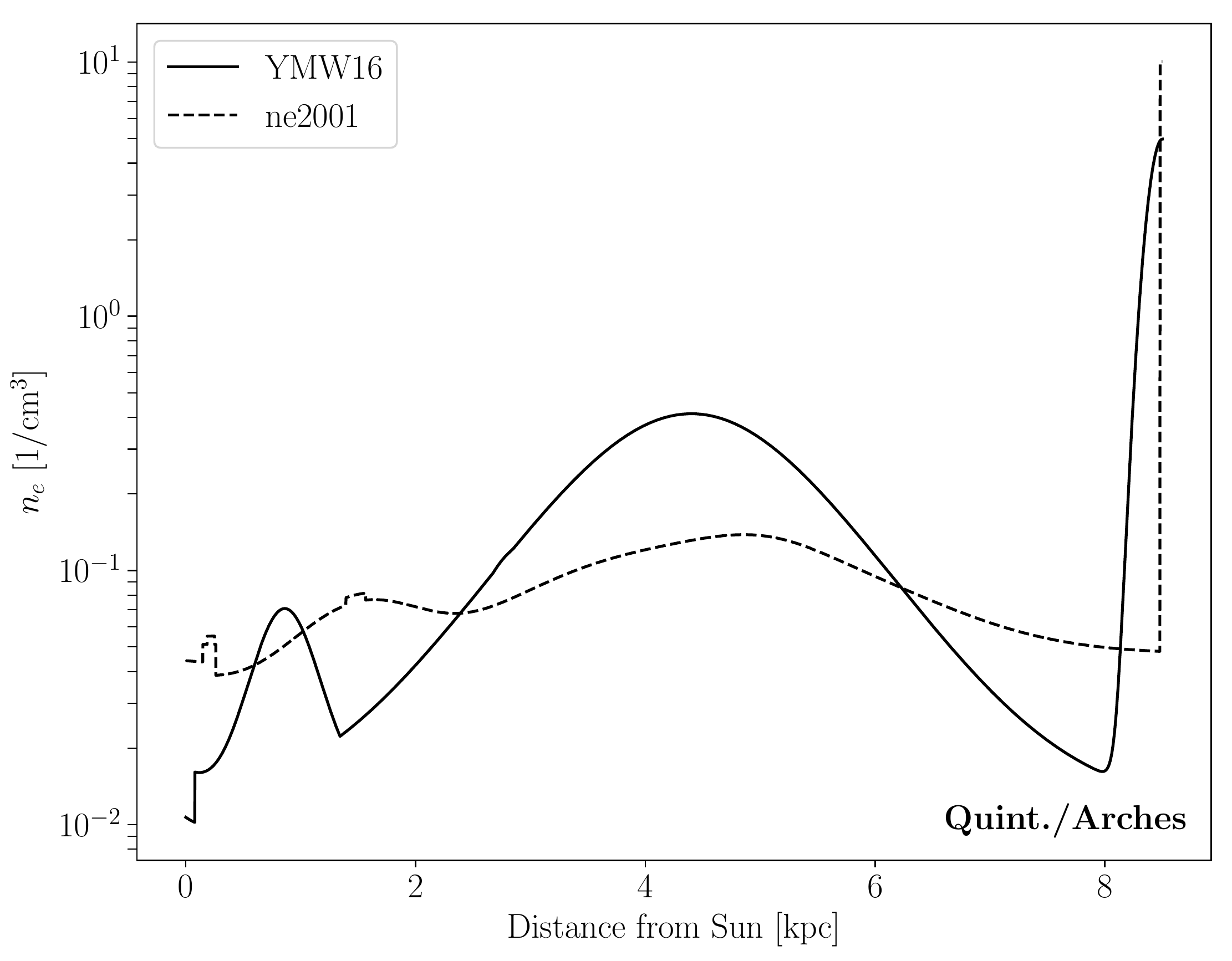}
\includegraphics[width=0.49\textwidth]{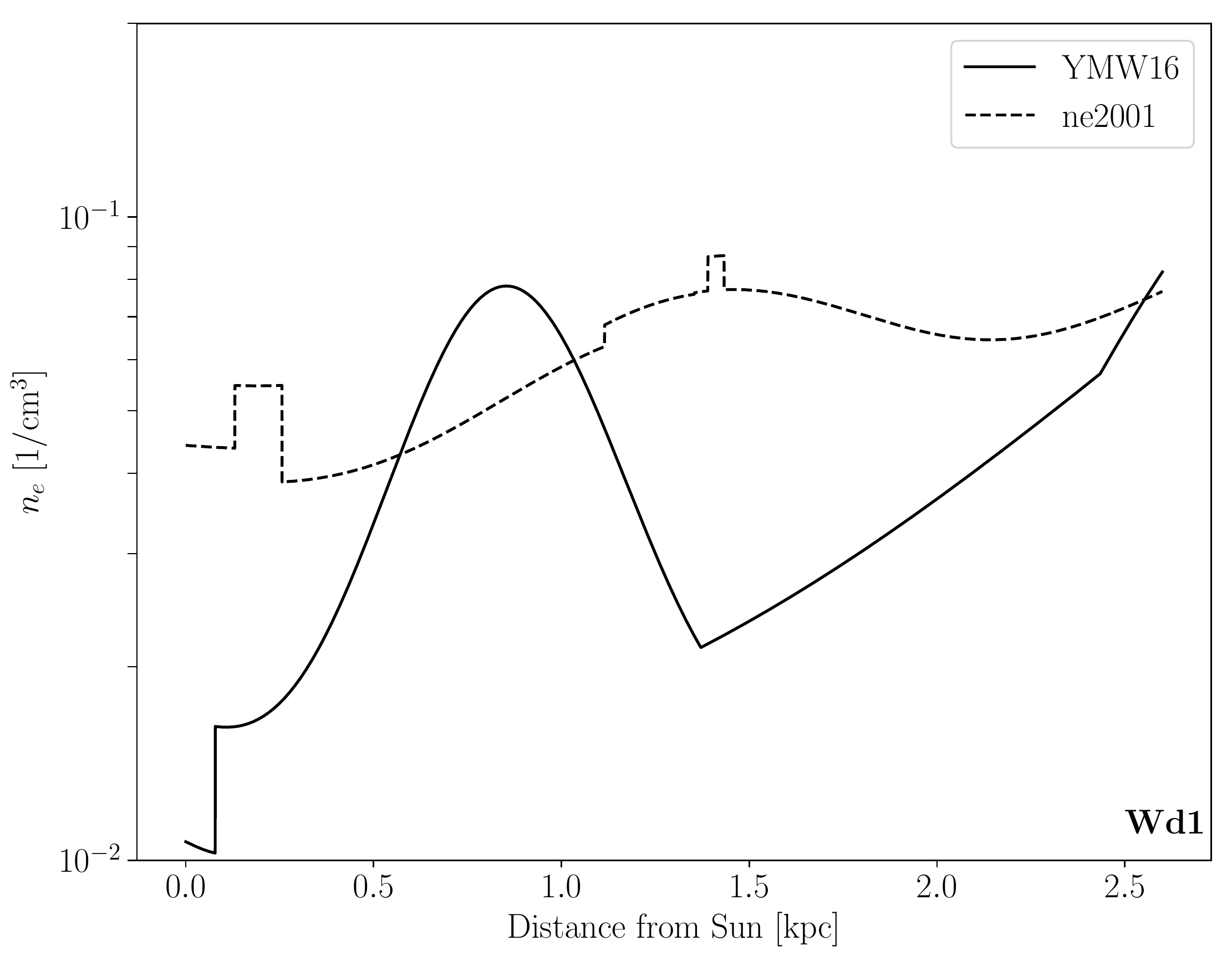}
\caption{(Left) The free electron density $n_e$ towards the GC in our fiducial model (YMW16) and the alternate model (\texttt{ne2001}).  (Right) As in the left panel but towards the Wd1 cluster. The free-electron density gives the photon an effective mass and thus affects the axion-photon conversion probability.
\label{fig:ne_mods}
}
\end{center}
\end{figure}

To compute the conversion probabilities we also need the free-electron densities.  We use the YMW16 model~\cite{Yao_2017} as our fiducial model, but we also compare our results to those obtained with the older \texttt{ne2001} model~\cite{Cordes:2002wz} to assess the possible effects of mismodeling the free-electron density.  In the left panel of Fig.~\ref{fig:ne_mods} we compare the free electron densities between the two models as a function of distance away from the Sun towards the GC, while in the right panel we show the free electron densities towards Wd1.  The differences between these models result in modest differences between the computed conversion probabilities, as discussed below.

\begin{figure}[htb]
\hspace{0pt}
\vspace{-0.2in}
\begin{center}
\includegraphics[width=0.49\textwidth]{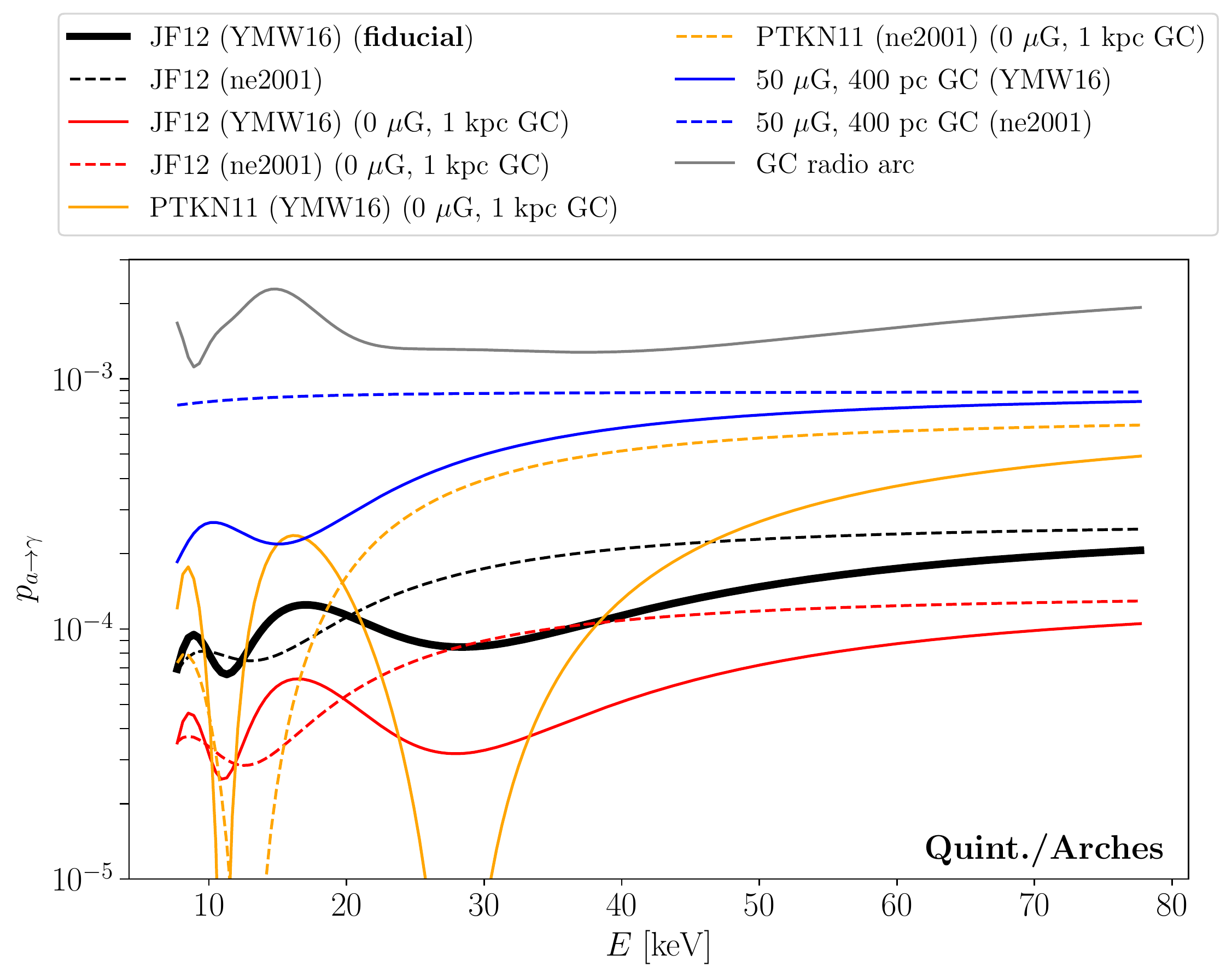}
\includegraphics[width=0.49\textwidth]{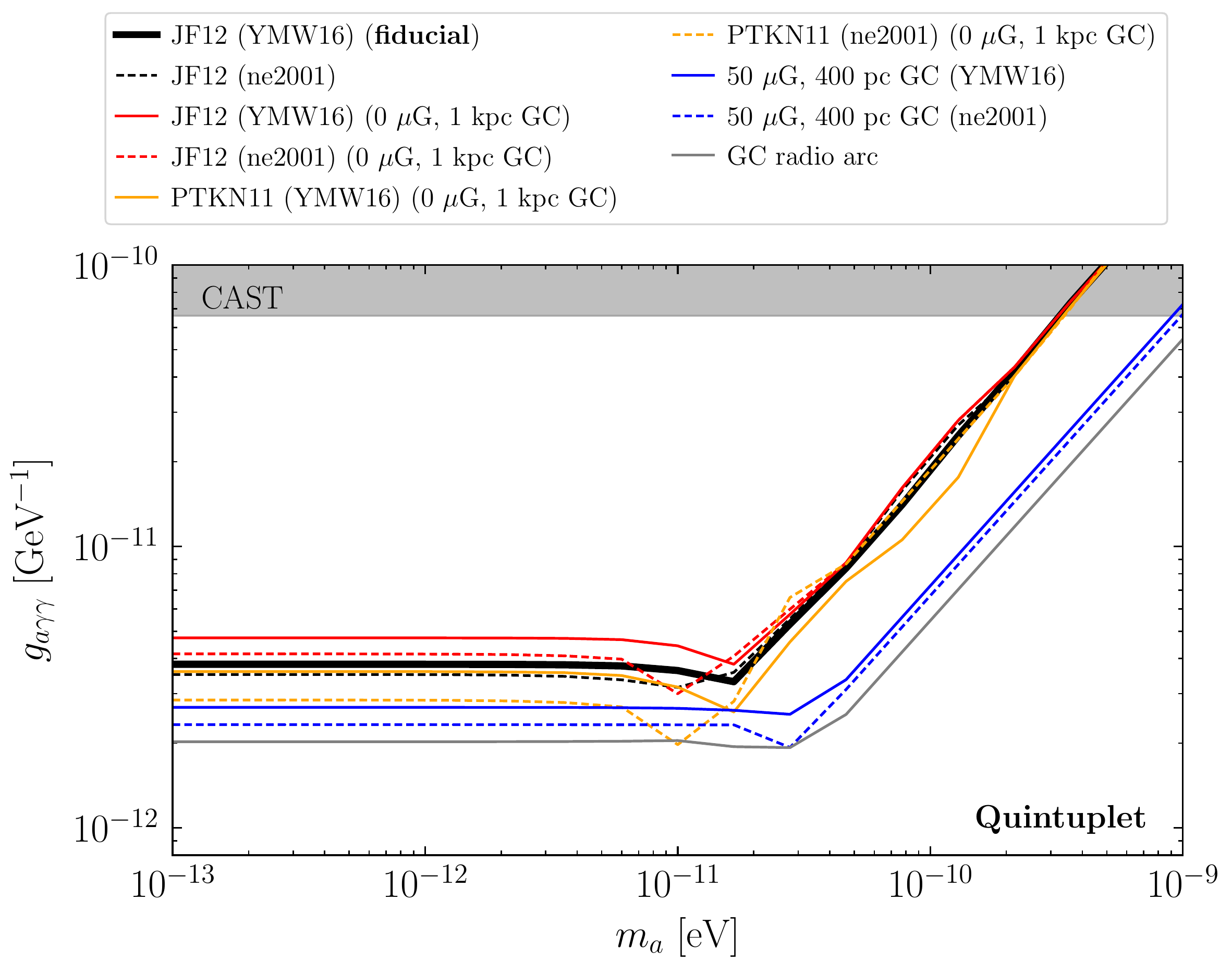}
\includegraphics[width=0.475\textwidth]{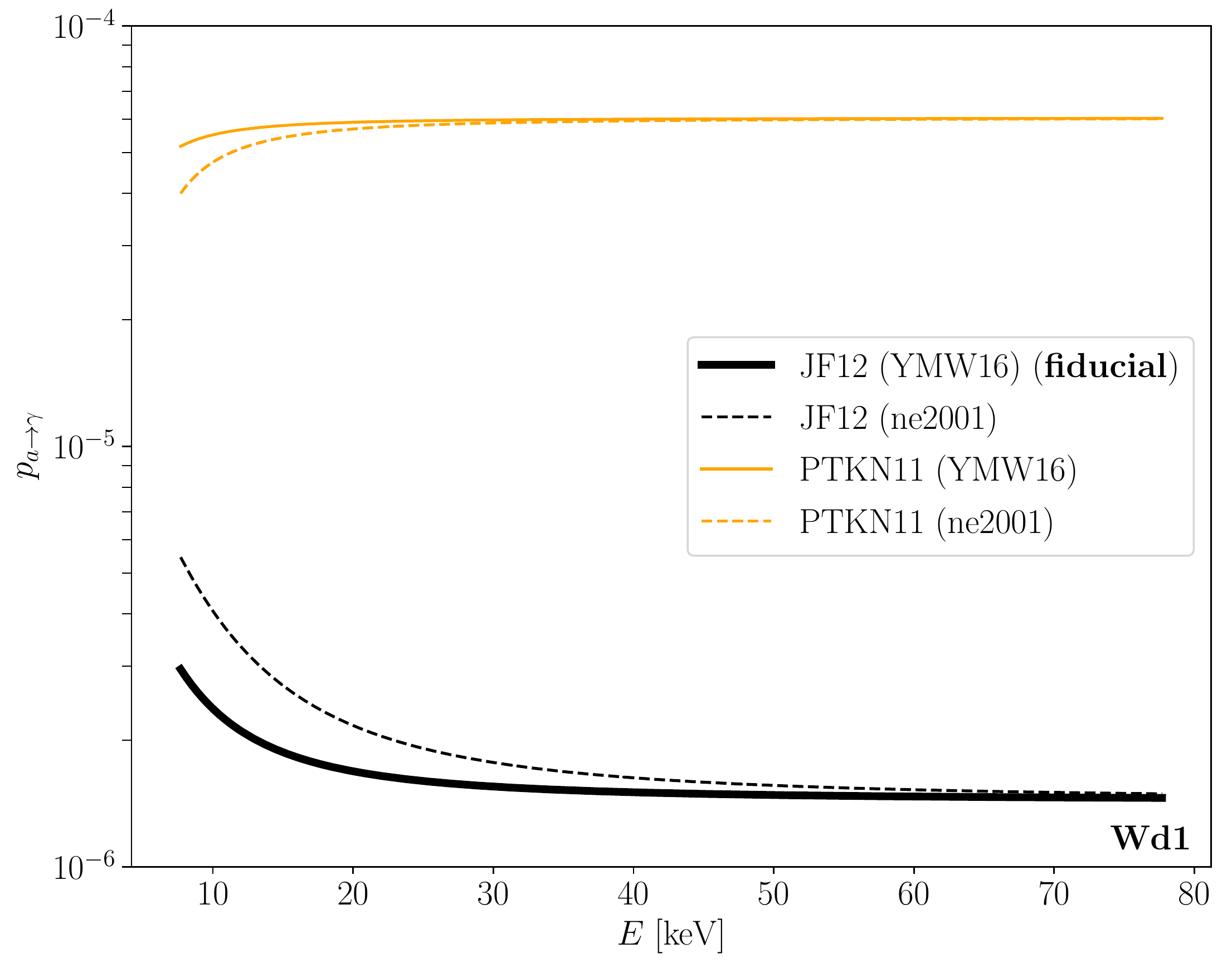}
\includegraphics[width=0.49\textwidth]{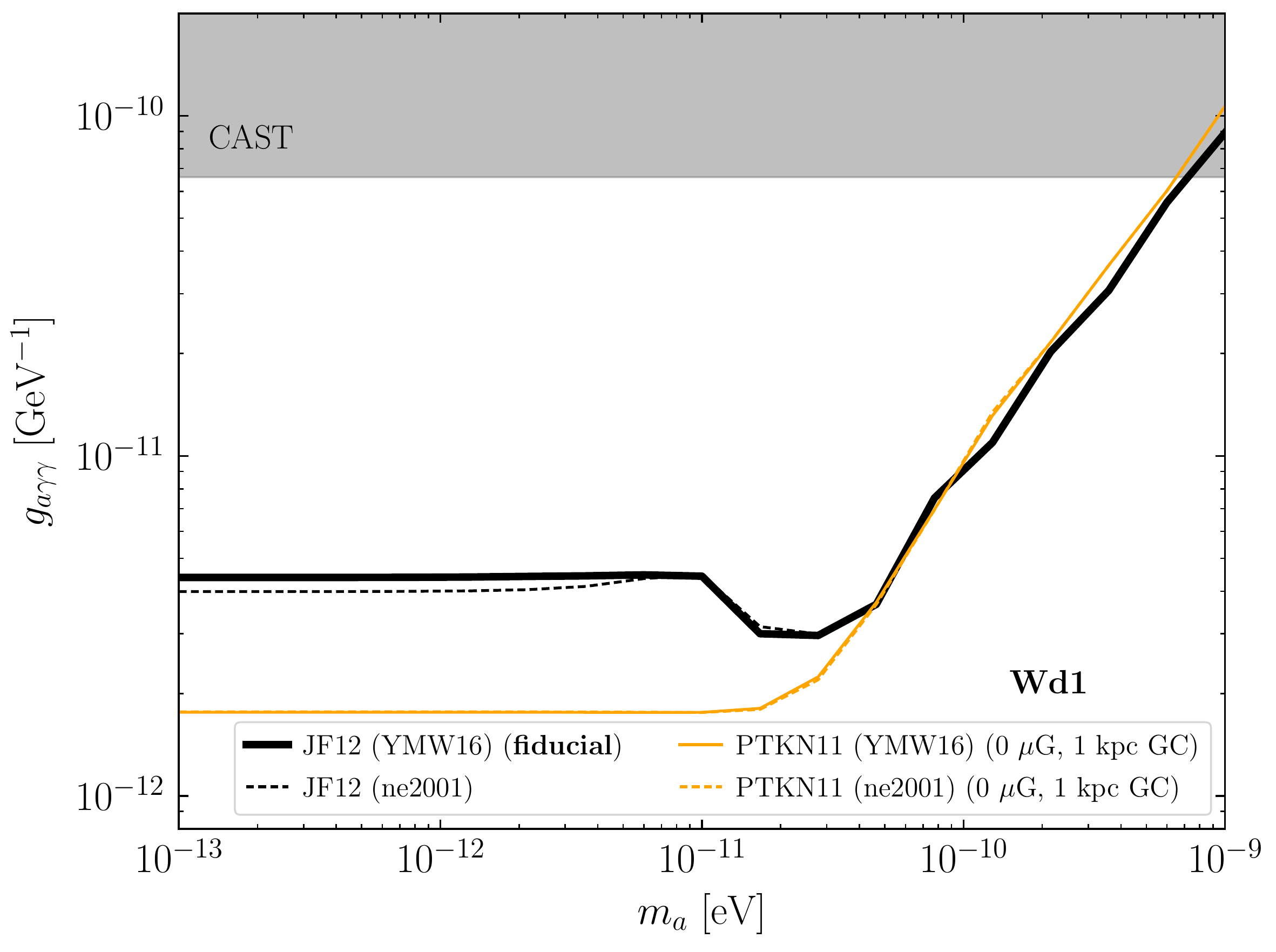}
\caption{
(Left Column) The axion-photon conversion probabilities $p_{a \to \gamma}$, assuming $g_{a\gamma\gamma} = 10^{-12}$ GeV$^{-1}$, computed as a function of the axion energy $E$ (and assuming $m_a \ll 10^{-10}$ eV) using the formula given in~\eqref{integral}.  (Top Left) The conversion probabilities for axions produced in the Quintuplet or Arches clusters for different modeling assumptions for the Galactic magnetic field and free-electron density.  Our fiducial result is shown in solid black.  Note that the plasma mass, induced by the free-electron density, becomes more important at lower axion energies and induces the lower-energy features.  The dashed black curve shows the effect of changing from the YMW16 free-electron model to the \texttt{ne2001} model.  Removing the $B$-field within the inner kpc leads to the results in red, while only modeling a 50 $\mu$G field in the inner 400 pc leads to the results in blue.  Changing to the PTKN11 model (and masking the inner kpc) gives the results in orange.  We estimate that if the axions traverse the GC radio arc, located near the Quintuplet and Arches clusters, the conversion probabilities could be enhanced to the values shown in grey. (Bottom Left) As in the top left panel but for axions emitted from the Wd1 cluster. (Right Column) The effects of the different conversion probability models on the 95\% upper limits on $g_{a\gamma\gamma}$ for Quintuplet (top right) and Wd1 (bottom right).  Note that Arches is similar to Quintuplet, since they are both assumed to have the same conversion-probability models.
\label{fig:conv_probs}
}
\end{center}
\end{figure}

Combining the magnetic field models in Fig.~\ref{fig:fid-field-GC} and the free-electron models in Fig.~\ref{fig:ne_mods} we may compute the axion-photon conversion probabilities, for a given axion energy $E$.  These conversion probabilities are presented in the left panels of Fig.~\ref{fig:conv_probs} (assuming $g_{a\gamma\gamma} = 10^{-12}$ GeV$^{-1}$ and $m_a \ll 10^{-11}$ eV).  In the top left panel we show the results for Quintuplet and Arches, while the bottom left panel gives the conversion probabilities for Wd1, computed under both free-electron models and various magnetic field configurations. 

In the top left panel our fiducial conversion probability model is shown in solid black. Changing to the \texttt{ne2001} model would in fact slightly enhance the conversion probabilities at most energies, as shown in the dotted black, though the change is modest.  Completely removing the $B$-field within 1 kpc of the GC leads only to a small reduction to the conversion probabilities, as indicated in red.  Changing magnetic field models to that of~\cite{2011ApJ...738..192P} (PTKN11), while also removing the $B$-field within the inner kpc, leads to slightly enhanced conversion probabilities, as shown in orange (for both the YMW16 and \texttt{ne2001} $n_e$ models).  Note that the conversion probabilities exhibit clear constructive and destructive interference behavior in this case at low energies, related to the periodic nature of the disk-field component, though including the random field component it is expected that this behavior would be largely smoothed out.   

As discussed previously the magnetic field is expected to be significantly larger closer in towards the GC than in our fiducial $B$-field model.  As an illustration in blue we show the conversion probabilities computed, from the two different free-electron models, when we only include a $B$-field component of magnitude $50$ $\mu$G pointing in the ${\bf \hat z}$ direction within the inner 400 kpc (explicitly, in this case we do not include any other $B$-field model outside of the inner 400 kpc).  The conversion probabilities are enhanced in this case by about an order of magnitude across most energies relative to in our fiducial model.  The inner Galaxy also likely contains localized regions of even strong field strengths, such as non-thermal filaments with $\sim$mG ordered fields.  As an illustration of the possible effects of such fields on the conversion probabilities, in Fig.~\ref{fig:conv_probs} we show in grey the result we obtain for the conversion probability when we assume that the axions traverse the GC radio arc, which we model as a 10 kpc wide region with a vertical field strength of 3 mG and a free-electron density $n_e = 10$ cm$^{-3}$~\cite{YusefZadeh:2003qx,Guenduez:2019cwe}.  Due to modeling uncertainties in the non-thermal filaments and the ambient halo field in the inner hundreds of pc, we do not include such magnetic-field components in our fiducial conversion probability model.  However, we stress that in the future, with a better understanding of the Galactic field structure in the inner Galaxy, our results could be reinterpreted to give stronger constraints.  

The Wd1 conversion probabilities change by over an order of magnitude going between the JF12 and PTKN11 models, as seen in the bottom left panel of Fig.~\ref{fig:conv_probs}, though it is possible that this difference would be smaller when random fields are properly included on top of the JF12 model (though again, we chose not to do this because of sensitivity to the random-field domain sizes).  

The effects of the different conversion probabilities on the $g_{a\gamma\gamma}$ limits may be seen in the top right panel for Quintuplet (Arches gives similar results, since the conversion probabilities are the same) and Wd1 in the bottom right panel of Fig.~\ref{fig:conv_probs}.  Note that the observed fluxes scale linearly with $p_{a \to \gamma}$ but scale like $g_{a\gamma\gamma}^4$, so differences between conversion probability models result in modest differences to the $g_{a\gamma\gamma}$ limits.  Still, it is interesting to note that the Wd1 limits with the PTKN11 model are stronger than the fiducial Quintuplet limits, which emphasizes the importance of better understanding the $B$-field profile towards Wd1.  For Quintuplet (and also Arches) we see that depending on the field structure in the inner $\sim$kpc, the limits may be slightly stronger and extend to slightly larger masses (because of field structure on smaller spatial scales) than in our fiducial $B$-field mode.

\section{Extended Data Analysis Results}

In this section we present additional results from the data analyses summarized in the main Letter.

\subsection{Quintuplet}

In this subsection we give extended results for the Quintuplet data analysis.  Our main focus is to establish the robustness of the flux spectra from the NuSTAR data analysis (shown in Fig.~\ref{fig:flux_spectra}) that go into producing the limits on $g_{a\gamma\gamma}$ shown in Fig.~\ref{fig:limits}.  

\subsubsection{Data and templates}

\begin{figure}[htb]  
\hspace{0pt}
\vspace{-0.2in}
\begin{center}
\includegraphics[width=0.325\textwidth]{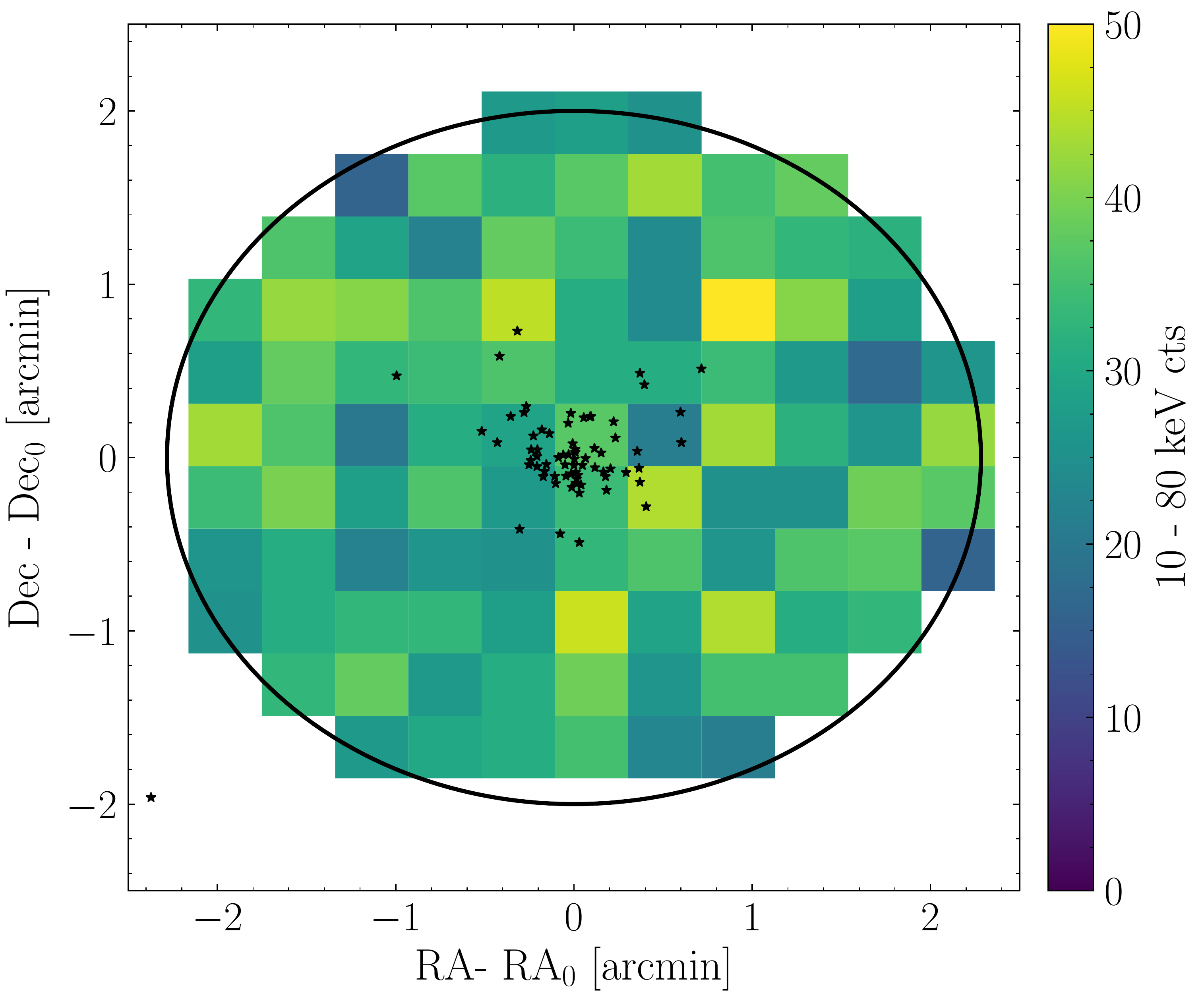}
\includegraphics[width=0.325\textwidth]{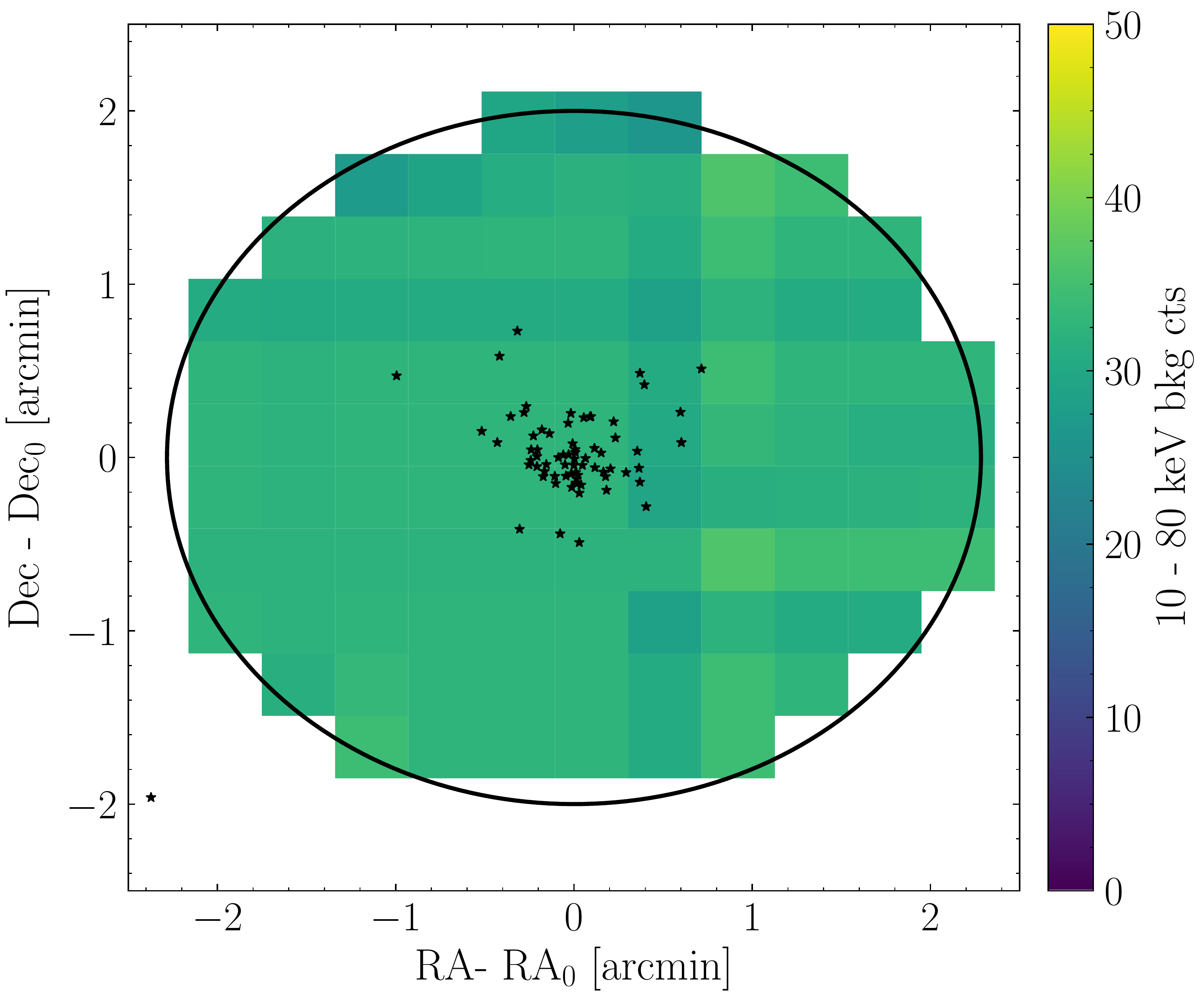}
\includegraphics[width=0.325\textwidth]{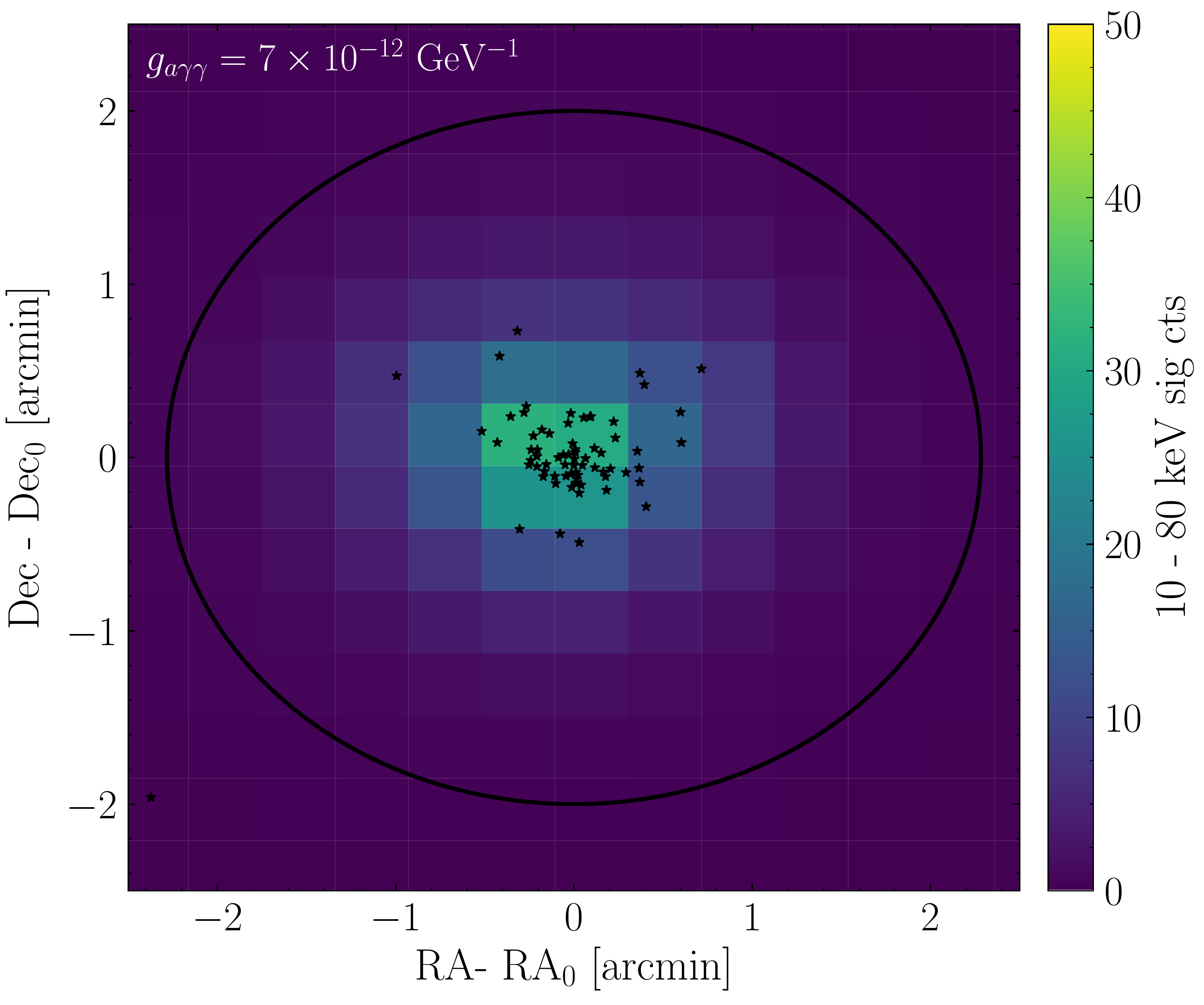}
\caption{(Left) As in Fig.~\ref{fig:ill}, but for the total observed counts between 10 - 80 keV instead of the background-subtracted counts. (Right) The best-fit background model, summed from 10 - 80 keV, for the Quintuplet data set shown in the left panel. (Right) The predicted axion-induced signal template from Quintuplet, in counts, normalized for an axion with $g_{a\gamma\gamma} = 7 \times 10^{-12}$ GeV$^{-1}$ and $m_a \ll 10^{-11}$ eV. 
}
\label{fig:all-maps-quint}
\end{center}
\end{figure}

First we take a closer look at the stacked data and models that go into the Quintuplet data analysis.  The stacked counts data in the vicinity of Quintuplet are shown in the left panel of Fig.~\ref{fig:all-maps-quint}.  We show the counts summed from 10 - 80 keV.  Note that the circle in that figure indicates $2'$, which the radius of our fiducial analysis ROI.\footnote{Note that ROIs for all of our analyses are centered upon the center of axion fluxes in RA and DEC, though the distinction between the center of fluxes and the SSC center is minimal for all of our targets.}  As in Fig.~\ref{fig:ill} we also indicate the locations of the individuals stars in Quintuplet that may contribute axion-induced $X$-ray flux.  The middle panel shows the expected background flux from our background template.  The template is generally uniform over the ROI, with small variations.  On the other hand, the right panel shows the axion-induced signal counts template, normalized for $g_{a\gamma\gamma} = 7 \times 10^{-12}$ GeV$^{-1}$, which is localized about the center of the SSC.  Note that the signal template is generated by accounting for the PSF of NuSTAR in addition to the locations and predicted fluxes of the individual stars.  

\subsubsection{Axion Luminosity}

\begin{figure}[htb]  
\hspace{0pt}
\vspace{-0.2in}
\begin{center}
\includegraphics[width=0.495\textwidth]{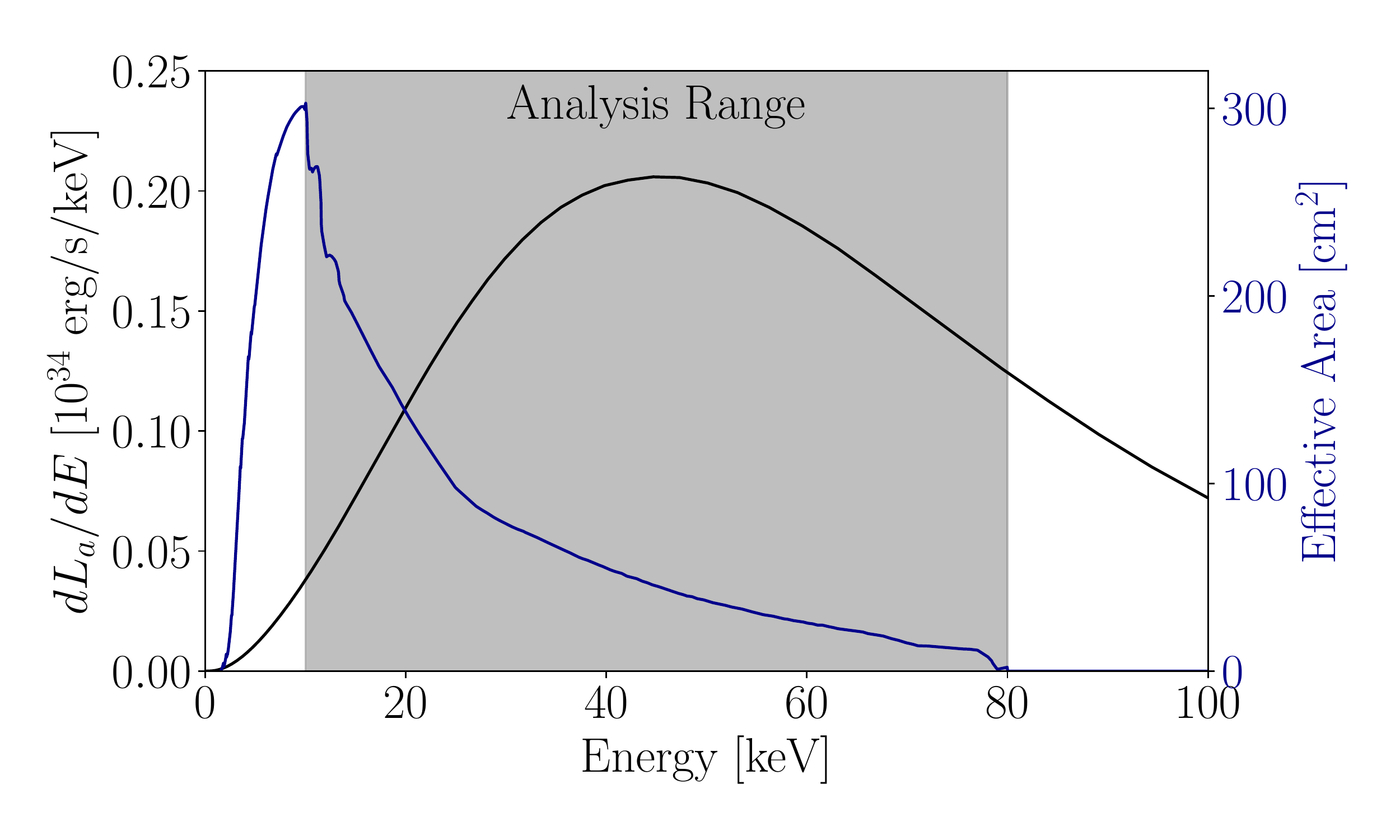}
\includegraphics[width=0.495\textwidth]{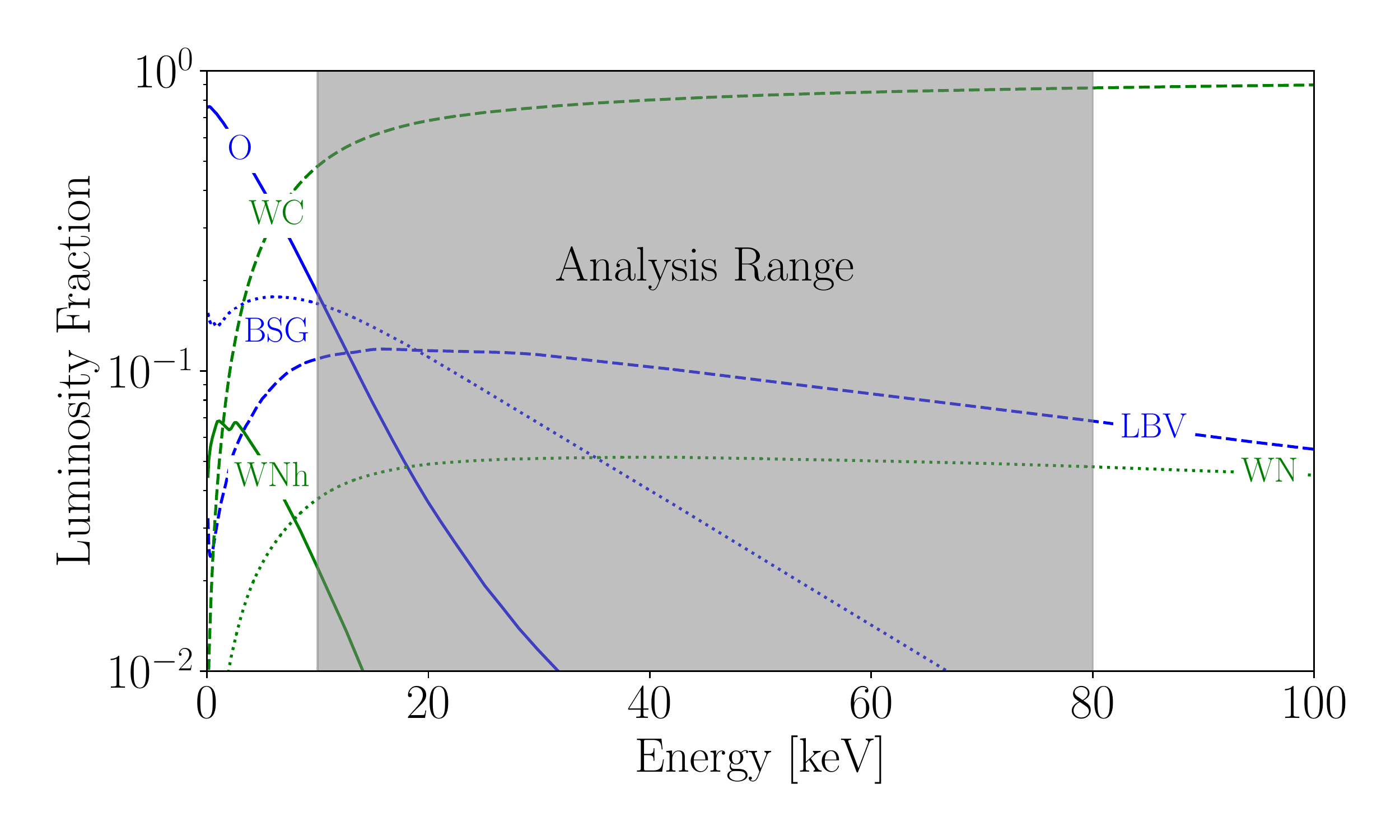}\\
\includegraphics[width=0.695\textwidth]{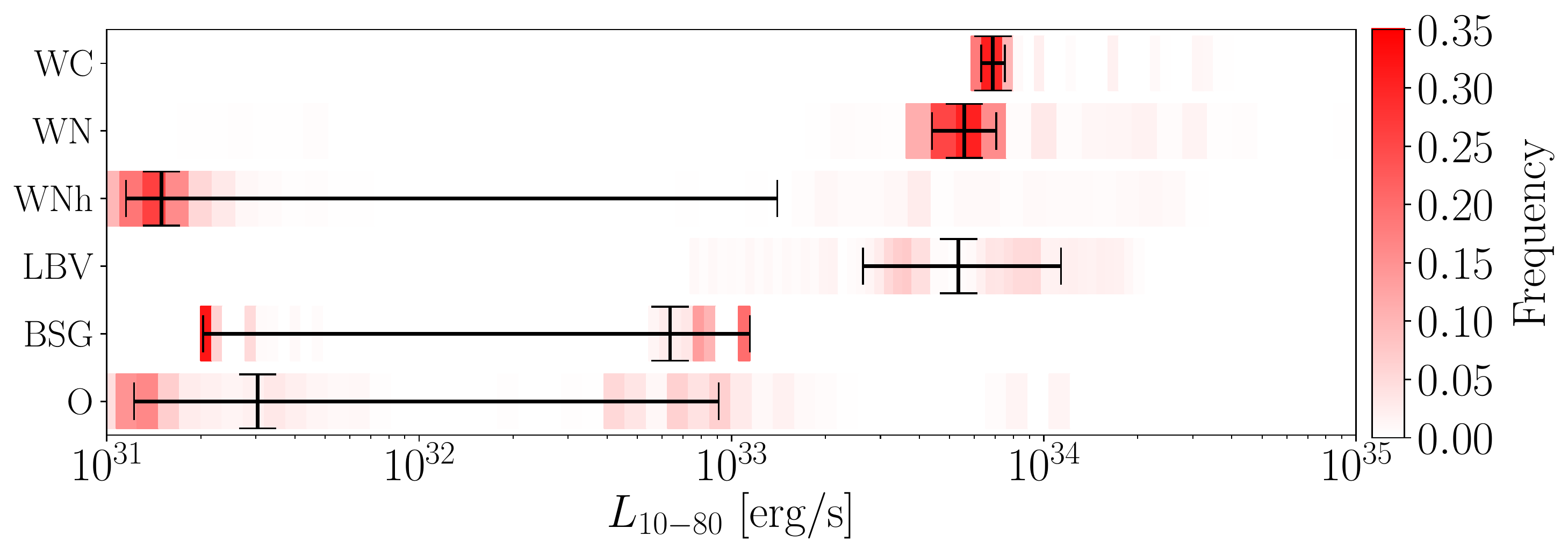}
\caption{(Upper Left) The Quintuplet axion spectrum assuming $g_{a\gamma\gamma} = 10^{-12}$ GeV$^{-1}$ (black) plotted against the NuSTAR effective area (blue). The analysis range, from 10 - 80 keV, is shaded in red. (Upper Right) The individual contributions of each stellar classification to the Quintuplet axion spectrum. The analysis range is again shaded.  (Bottom) The 10-80 keV luminosity distribution assigned to each stellar classification (per star) in Quintuplet. In red we show the frequency with which each luminosity occurs, while the black error bars show the mean and 1$\sigma$ band.
\label{fig:Quint_subtypes}
}
\end{center}
\end{figure}

We now show the axion luminosity and spectra that go into the right panel of Fig.~\ref{fig:all-maps-quint}. For each star in the cluster, we assign it a set of possible MESA models based on its spectral classification as described in the main text. In the upper left panel of Fig.~\ref{fig:Quint_subtypes}, we show the mean expected axion luminosity, as a function of energy, of the Quintuplet cluster, assuming $g_{a\gamma\gamma} = 10^{-12}$ GeV$^{-1}$. The luminosity peaks around 40 keV, but the effective area of NuSTAR, also shown, rapidly drops above 10 keV. Due to the much higher effective area at low energies, most of the sensitivity is at lower energies. There is also considerable flux above 80 keV, although NuSTAR does not have sensitivity at these energies. In the upper right panel, we show the median contribution of each spectral classification in Quintuplet to this luminosity, summed over all stars with the given classification. For all energies of interest, the WC stars dominate the cluster luminosity. This is because WR stars have the hottest cores and there are 13 WC stars in Quintuplet (there is 1 WN star). In the bottom panel, we show the 10 - 80 keV luminosity distribution for each spectral classification, along with the 1$\sigma$ containment bands and the mean expectation. The distribution depends principally on whether or not core helium is ignited while the star is assigned a given classification. The O, BSG, and WNh stars all can be either hydrogen or helium burning, in which case they have 10 - 80 keV luminosities of $\sim 10^{31}$ or $\sim 10^{33}$ erg/s, respectively\textemdash recall that the jump in temperature during helium ignition is a factor $\sim$ 3. The LBV phase is always core helium burning, and the star may go supernova in this phase. The same is true of the WR phases WN and WC, although the stars undergoing supernova in this phase are typically more massive.

\begin{table*}[]
\centering
\begin{tabular}{|c||c|c|c|c|c|c|c|}
\hline
           & O & BSG   & LBV     & WNh   & WC $+$ WN & tot & tot (10-80 keV)    \\ \hline
$N_{\rm star}$	
& $\begin{array}{c} 37 \end{array}  $
&$ \begin{array}{c} 7 \end{array}   $  
&$  \begin{array}{c} 2 \end{array}    $ 
&$  \begin{array}{c} 5 \end{array}    $ 
&$  \begin{array}{c} 14 \end{array} $ 
& 65 
& 65
\\ \hline
$\begin{array}{c}
z = 0.018 \\ \mu_{\rm rot} = 100 \, \,  {\rm km/s  }	
\end{array}$
& $\begin{array}{c}  3.0_{-1.5}^{+1.7} \times 10^{33} \end{array}$ 
& $1.3_{-0.9}^{+0.9} \times 10^{33}$        
& $1.9_{-1.7}^{+2.1} \times 10^{34}$      
& $5.9_{-5.8}^{+5.8} \times 10^{34}$        
& $2.8_{-0.8}^{+2.6} \times 10^{35}$   
& $3.8_{-1.0}^{+2.6} \times 10^{35}$
& $2.8_{-0.6}^{+1.6} \times 10^{35}$
\\ \hline
$\begin{array}{c}
z = 0.035 \\ \mu_{\rm rot} = 100 \, \,  {\rm km/s } 	
\end{array}$
& $\begin{array}{c}  1.9_{-0.9}^{+2.9} \times 10^{34} \end{array}$ 
& $3.5_{-1.2}^{+1.2} \times 10^{33}$        
& $1.4_{-0.7}^{+1.1} \times 10^{34}$      
& $7.4_{-7.3}^{+30} \times 10^{33}$        
& $1.7_{-0.4}^{+0.9} \times 10^{35}$   
& $2.3_{-0.5}^{+0.9} \times 10^{35}$
& $1.7_{-0.3}^{+0.5} \times 10^{35}$ \\
\hline 
$\begin{array}{c}
z = 0.035 \\ \mu_{\rm rot} = 150 \, \,  {\rm km/s  }	
\end{array}$
& $\begin{array}{c}  3.4_{-2.3}^{+2.4} \times 10^{34} \end{array}$ 
& $3.6_{-1.3}^{+1.2} \times 10^{33}$        
& $1.4_{-0.8}^{+1.2} \times 10^{34}$      
& $4.3_{-4.2}^{+22} \times 10^{33}$        
& $1.5_{-0.3}^{+0.7} \times 10^{35}$   
& $2.1_{-0.4}^{+0.7} \times 10^{35}$
& $1.7_{-0.3}^{+0.4} \times 10^{35}$
\\ \hline
\end{tabular}
\caption{The number of stars $N_{\rm star}$ for each stellar class in the Quintuplet cluster, along with the predicted axion luminosities (all in erg/s).  
Note that Quintuplet is $\sim$30 pc away from the GC.  Except in the last column, the axion luminosities are summed over all energies.  All entries assume $g_{a \gamma\gamma} = 10^{-12}$ GeV$^{-1}$ and are summed over all stars for the given stellar class. 
}
\label{tab:SSC-Quint}
\end{table*}

The luminosities in Fig.~\ref{fig:Quint_subtypes} are computed for our fiducial choices of $Z = 0.035$ and $\mu_{\rm rot} = 150$ km/s.  To better understand the importance of these choices we show in Tab.~\ref{tab:SSC-Quint} how the luminosities depend on the initial metallicity $Z$ and mean rotation speed $\mu_{\rm rot}$.  Note that each entry in that table shows the luminosity summed over the stellar sub-types (with the number of stars indicated), and except in the two last columns the luminosities are summed over all stars.  The uncertainties in the entries in Tab.~\ref{tab:SSC-Quint} come from performing 500 draws from the representative models and account for the variance expected from star-to-star within a given classification.  As discussed in the main text, the 10 - 80 keV luminosity could be $\sim$70\% larger than in our fiducial model, depending on the initial $Z$ and $\mu_{\rm rot}$. 

\subsubsection{Injecting an axion signal}

\begin{figure}[htb]  
\hspace{0pt}
\vspace{-0.2in}
\begin{center}
\includegraphics[width=0.325\textwidth]{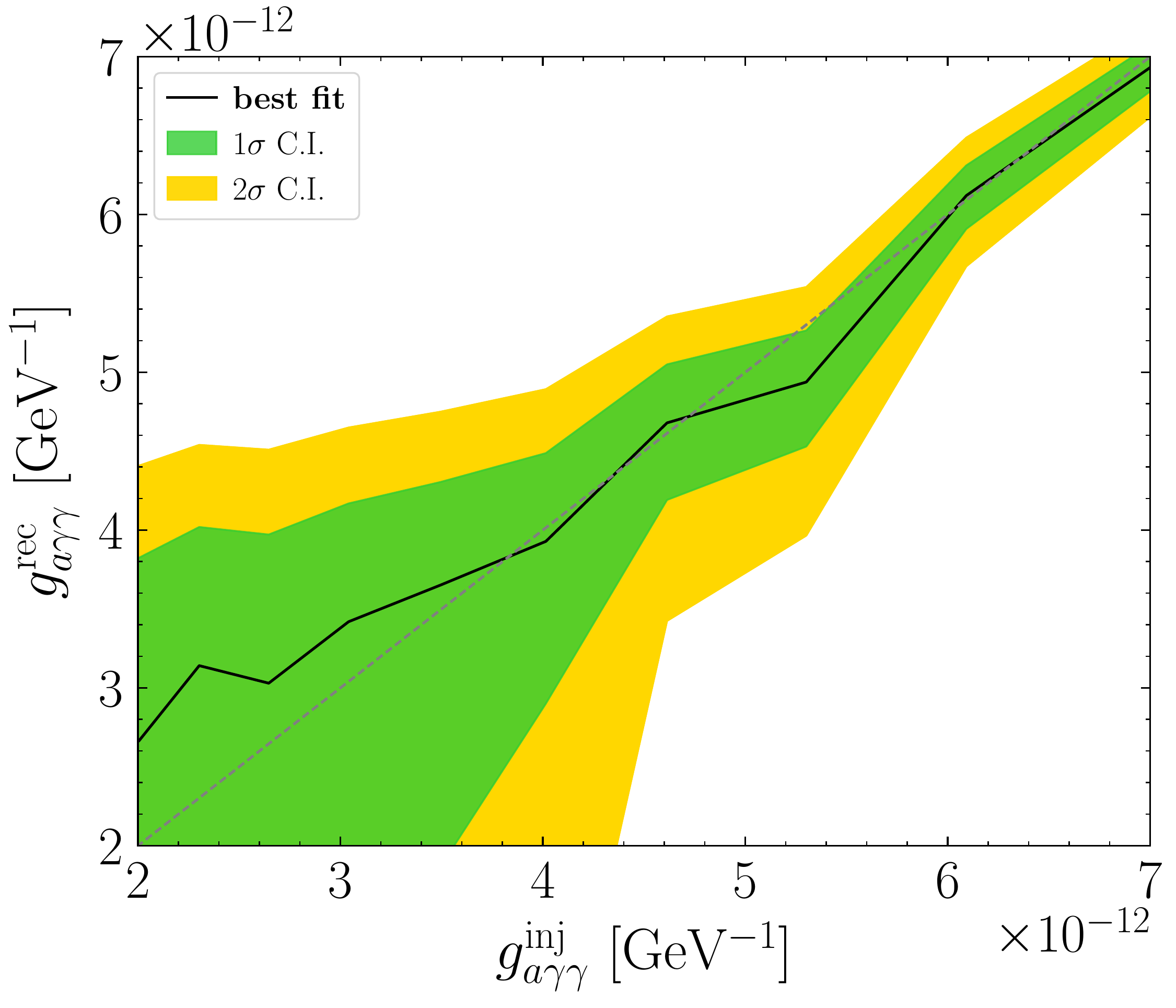}
\includegraphics[width=0.325\textwidth]{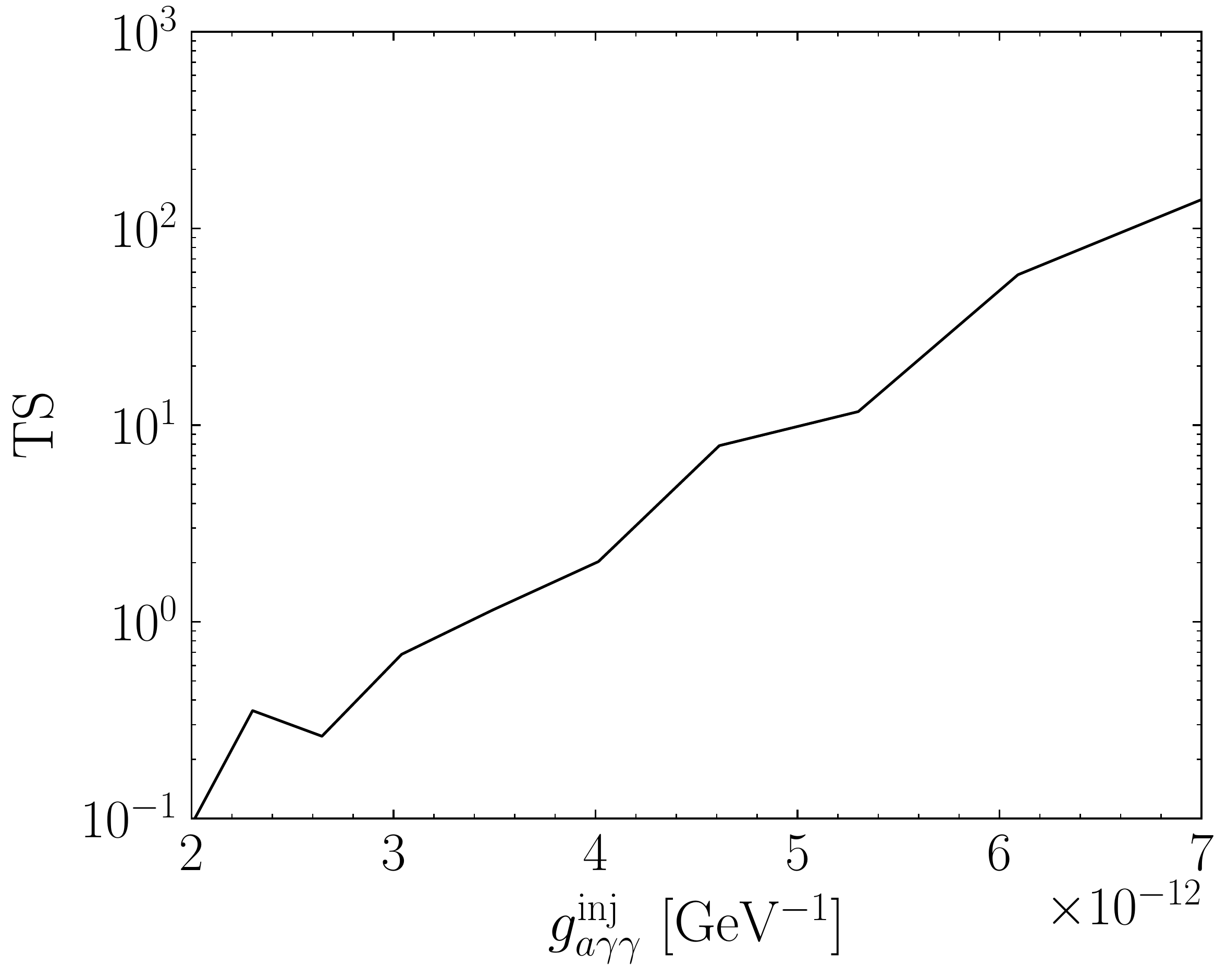}
\includegraphics[width=0.325\textwidth]{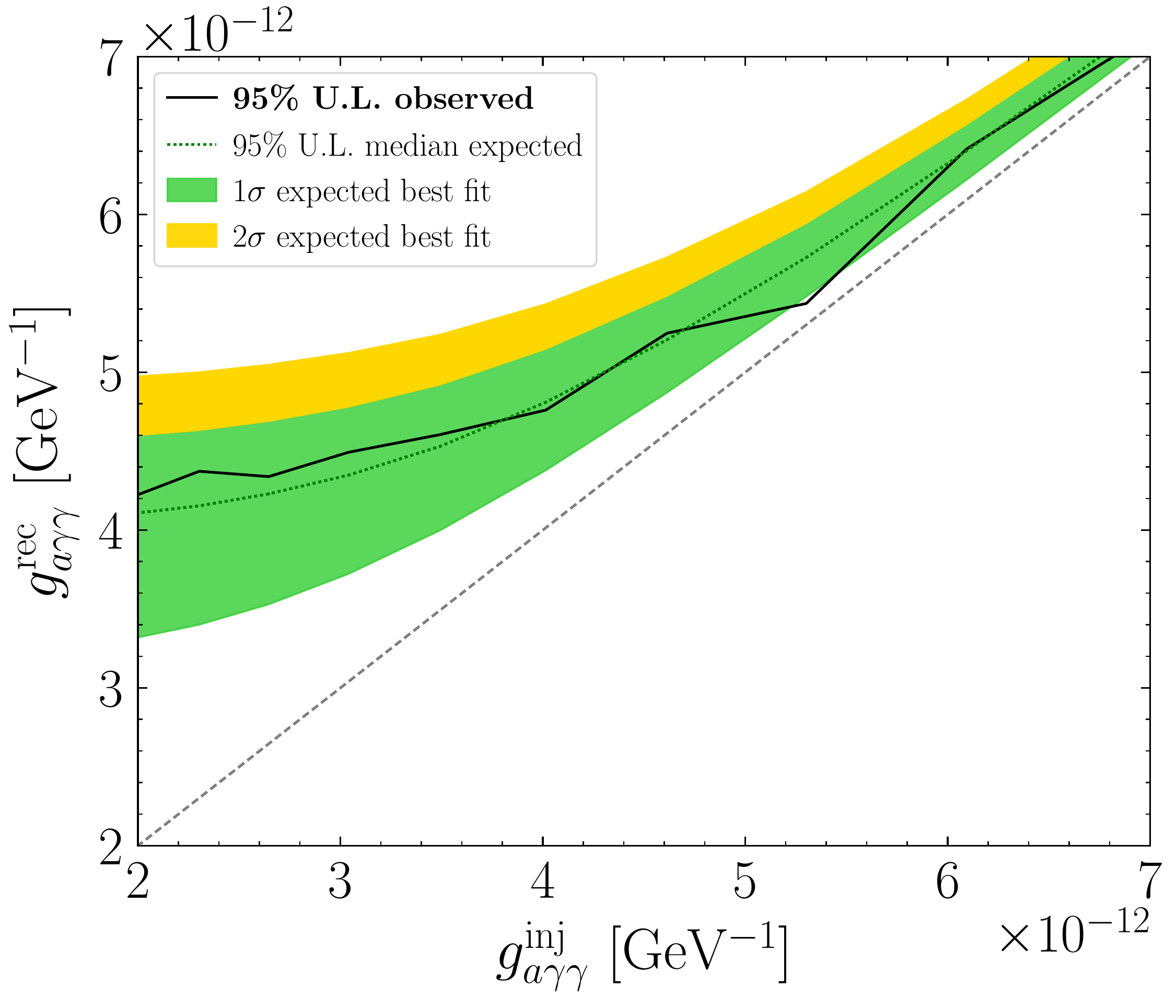}
\caption{ (Left) We inject a synthetic axion signal into the Quintuplet NuSTAR data with axion coupling $g_{a \gamma\gamma}^{\rm inj}$.  We then pass the hybrid synthetic plus real data through our analysis pipeline and show the best-fit coupling $g_{a\gamma\gamma}^{\rm rec}$, along with the recovered 1$\sigma$ and 2$\sigma$ uncertainties. (Middle) The discovery TS for the axion signal for the test illustrated in the left panel.  The square root of the TS is approximately the discovery significance.  (Right) The 95\% upper limit recovered for the injected signal test.  Importantly, the 95\% upper limit is above the injected signal value, for all injected signal strengths, and the upper limit is consistent with the 68\% and 95\% expectations for the upper limit under the null hypothesis, which are indicated in green and gold, respectively.
\label{fig:inj-quint}
}
\end{center}
\end{figure}

As a first test of the robustness of the Quintuplet analysis we inject a synthetic axion signal into the real stacked data and then pass the hybrid real plus synthetic data through our analysis pipeline.  Our goal from this test is to ensure that if a real axion signal were in the data with sufficiently high coupling to photons then we would be able to detect it.  The results from this test are shown in Fig.~\ref{fig:inj-quint}.

The left panel of Fig.~\ref{fig:inj-quint} shows the best-fit $g_{a\gamma\gamma}^{\rm rec.}$ as a function of the simulated $g_{a\gamma\gamma}^{\rm inj.}$ used to produce the axion-induced counts that are added to the real NuSTAR stacked data.  Importantly, as we increase the injected signal strength the recovered signal parameter converges towards the injected value, which is indicated by the dashed curve.  Note that the band shows the 68\% containment region for the recovered signal parameter from the analysis.  As the injected signal strength increases, so to does the  significance of the axion detection.  This is illustrated in the middle panel, which shows the discovery TS as a function of the injected signal strength.  Recall that the significance is approximately $\sqrt{{\rm TS}}$.  Perhaps most importantly, we also verify that the 95\% upper limit does not exclude the injected signal strength.  In the right panel of Fig.~\ref{fig:inj-quint} we show the 95\% upper limit found from the analyses of the hybrid data sets at different $g_{a\gamma\gamma}^{\rm inj}$.  Recall that all couplings above the $g_{a\gamma\gamma}^{\rm rec}$ curve are excluded, implying that indeed we do not exclude the injected signal strength.  Moreover, the 95\% upper limit is consistent with the expectation for the limit under the signal hypothesis, as indicated by the shaded regions at 1$\sigma$ (green) and 2$\sigma$ (yellow) containment.  Note that we do not show the lower 2$\sigma$ containment region, since we power-constrain the limits.  These regions were computed following the Asimov procedure~\cite{Cowan:2011an}. 

\subsubsection{Changing region size}
\begin{figure}[htb]
\hspace{0pt}
\vspace{-0.2in}
\begin{center}
\includegraphics[width=0.55\textwidth]{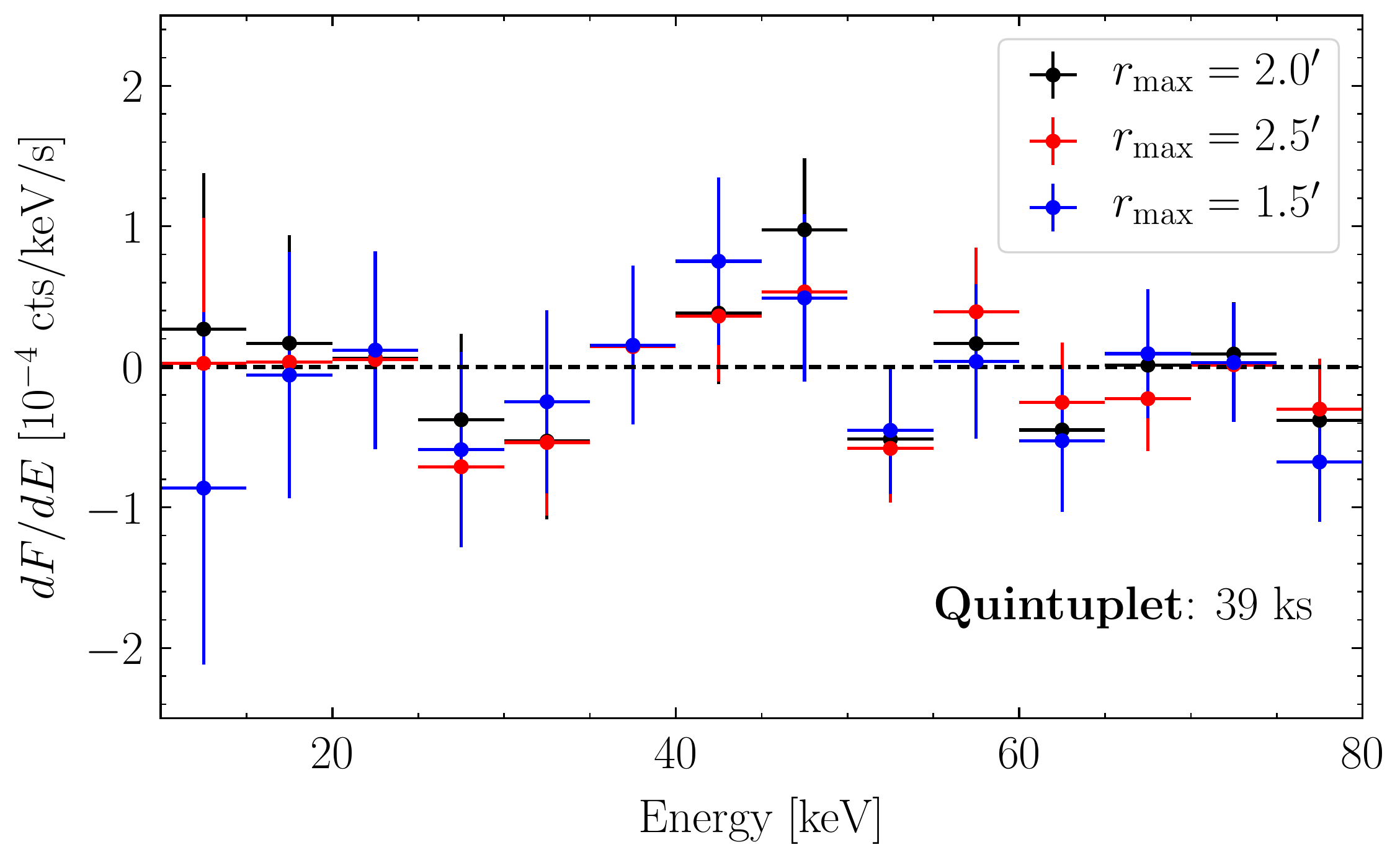}
\caption{As in Fig.~\ref{fig:flux_spectra}, except for different ROI sizes, as indicated. 
\label{fig:flux_spectra-syst}
}
\end{center}
\end{figure}

As a systematic test of the data analysis we consider the sensitivity of the inferred spectrum associated with the axion model template to the ROI size.  In our fiducial analysis, with spectrum shown in Fig.~\ref{fig:flux_spectra}, we use an ROI size of $r_{\rm max} = 2'$.  Here we consider changing the ROI size to $r_{\rm max} = 1.5'$ and $2.5'$.  The resulting spectra are shown in Fig.~\ref{fig:flux_spectra-syst}.  The spectrum does not appear to vary significantly when extracted using these alternate ROIs, indicating that significant sources of systematic uncertainty related to background mismodeling are likely not at play.

\subsection{Westerlund 1}

In this subsection we provide additional details and cross-checks of the Wd1 analysis.

\subsubsection{Data and templates}

\begin{figure}[htb]  
\hspace{0pt}
\vspace{-0.2in}
\begin{center}
\includegraphics[width=0.49\textwidth]{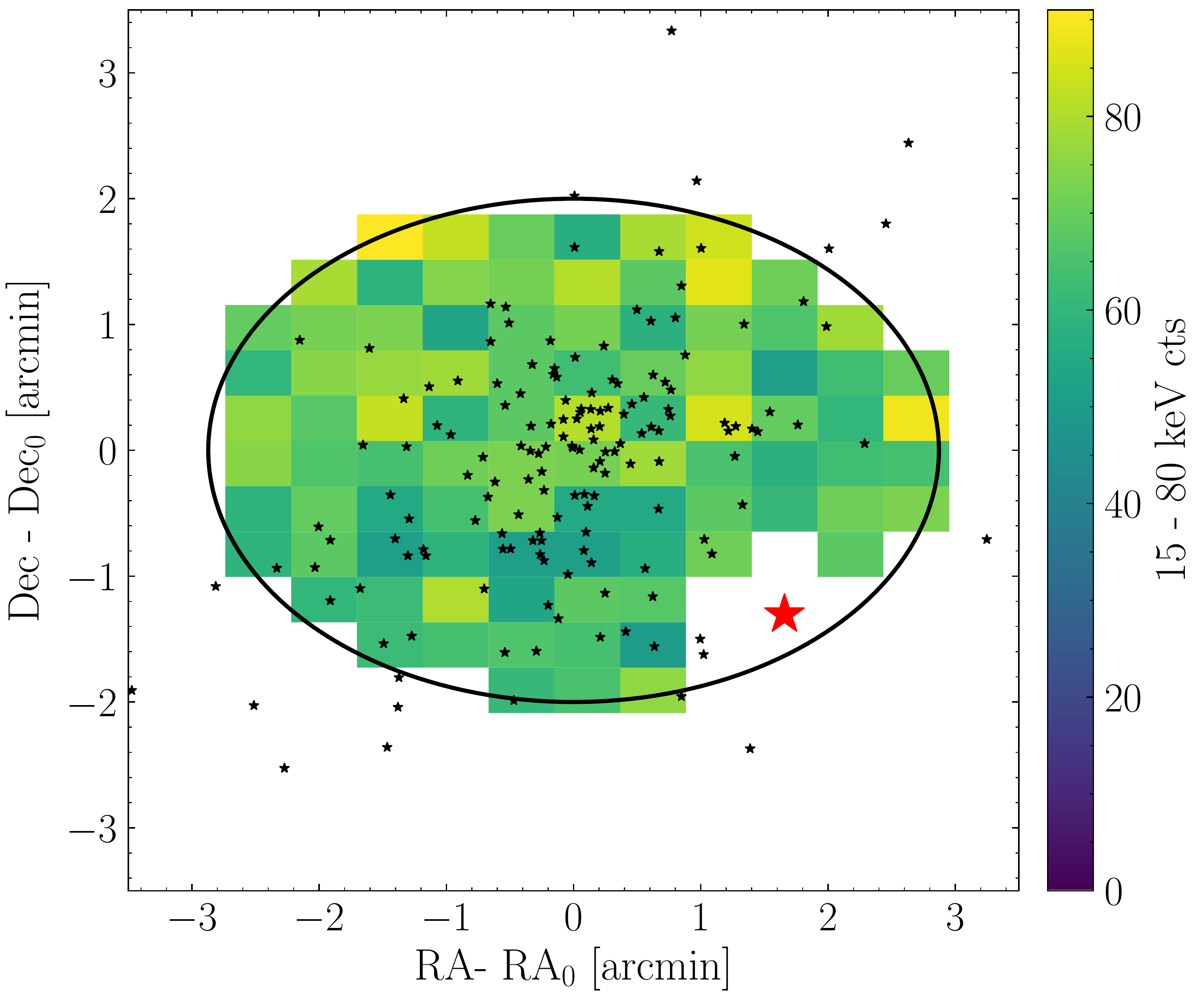}
\includegraphics[width=0.49\textwidth]{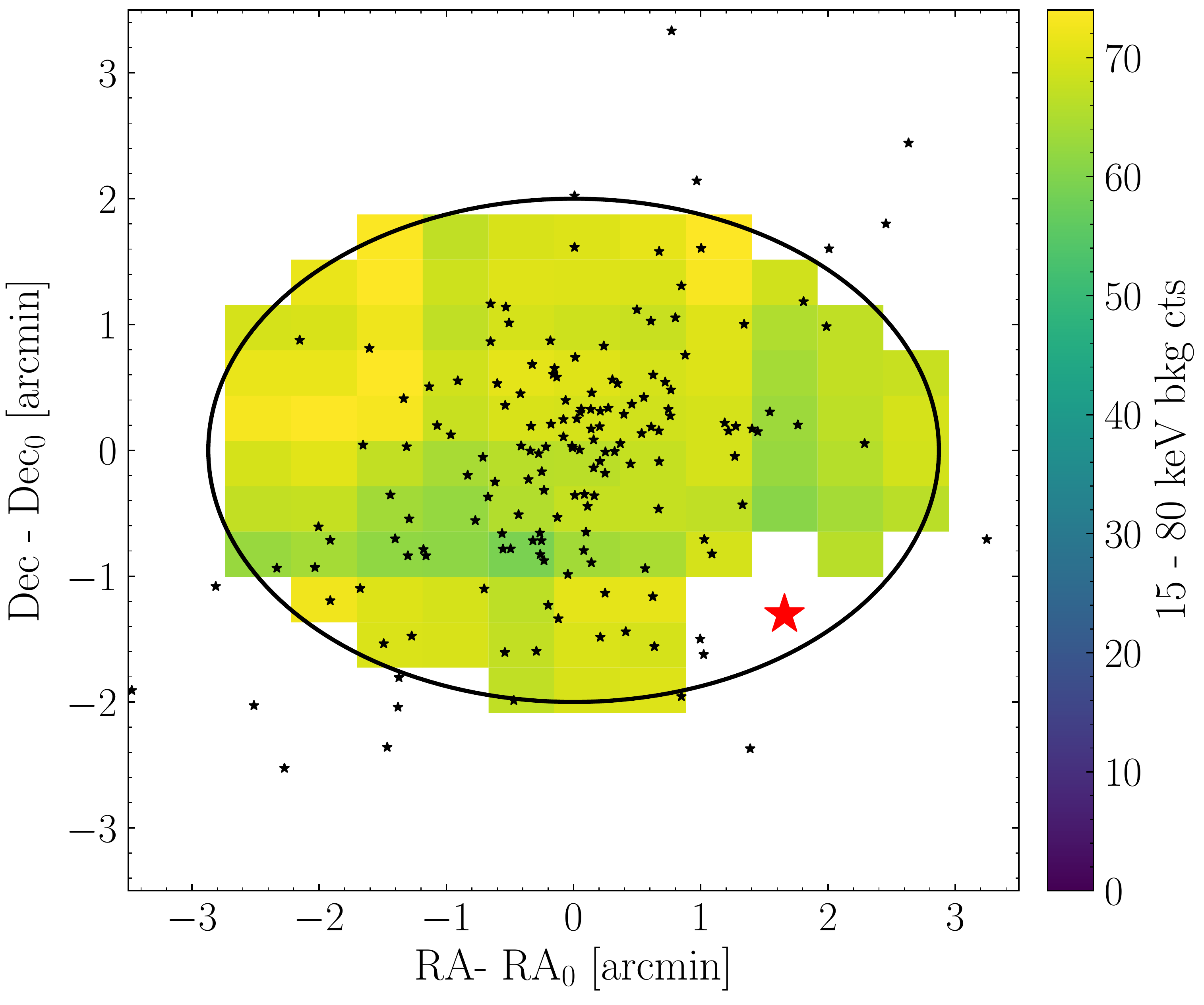}
\includegraphics[width=0.49\textwidth]{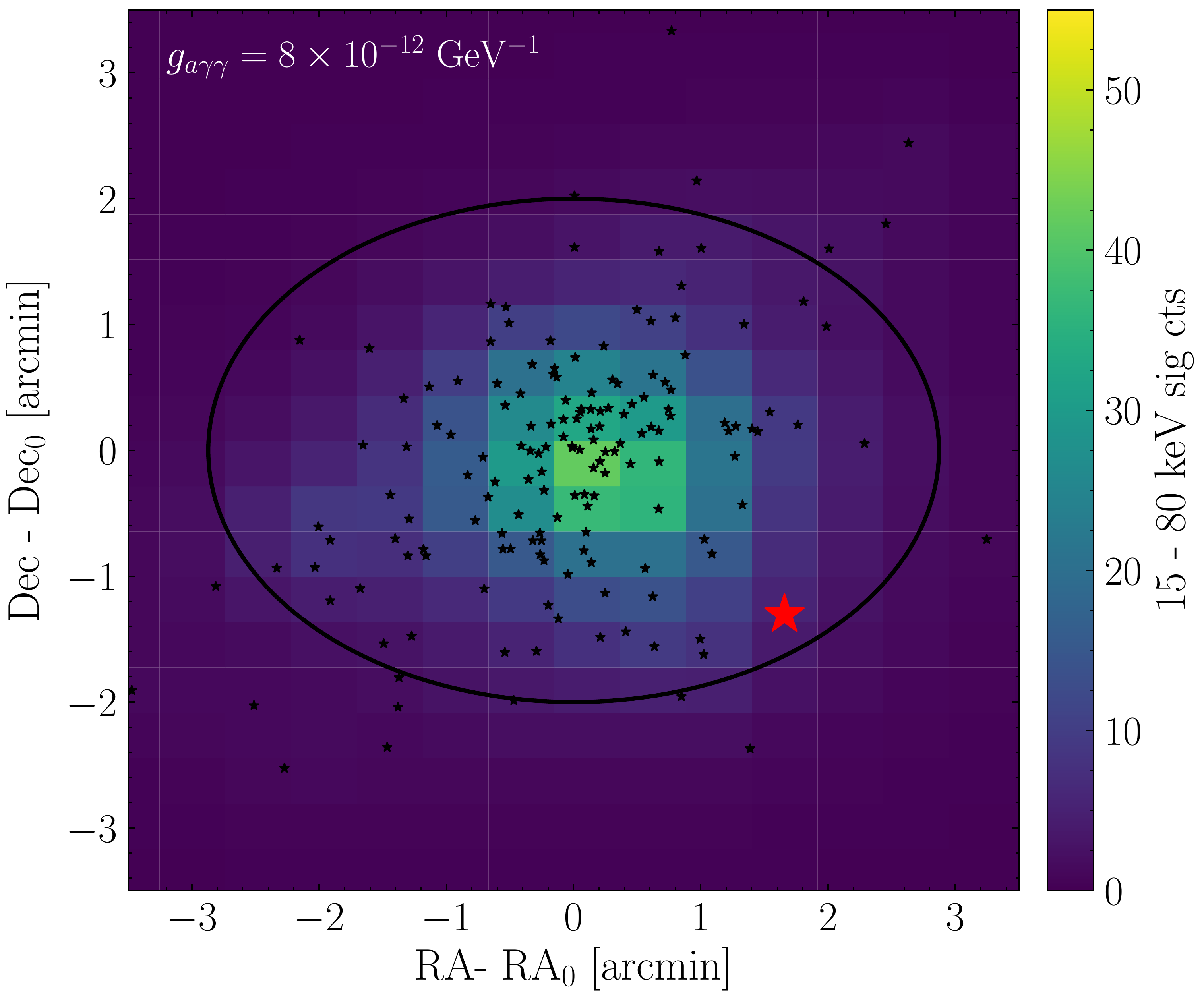}
\includegraphics[width=0.49\textwidth]{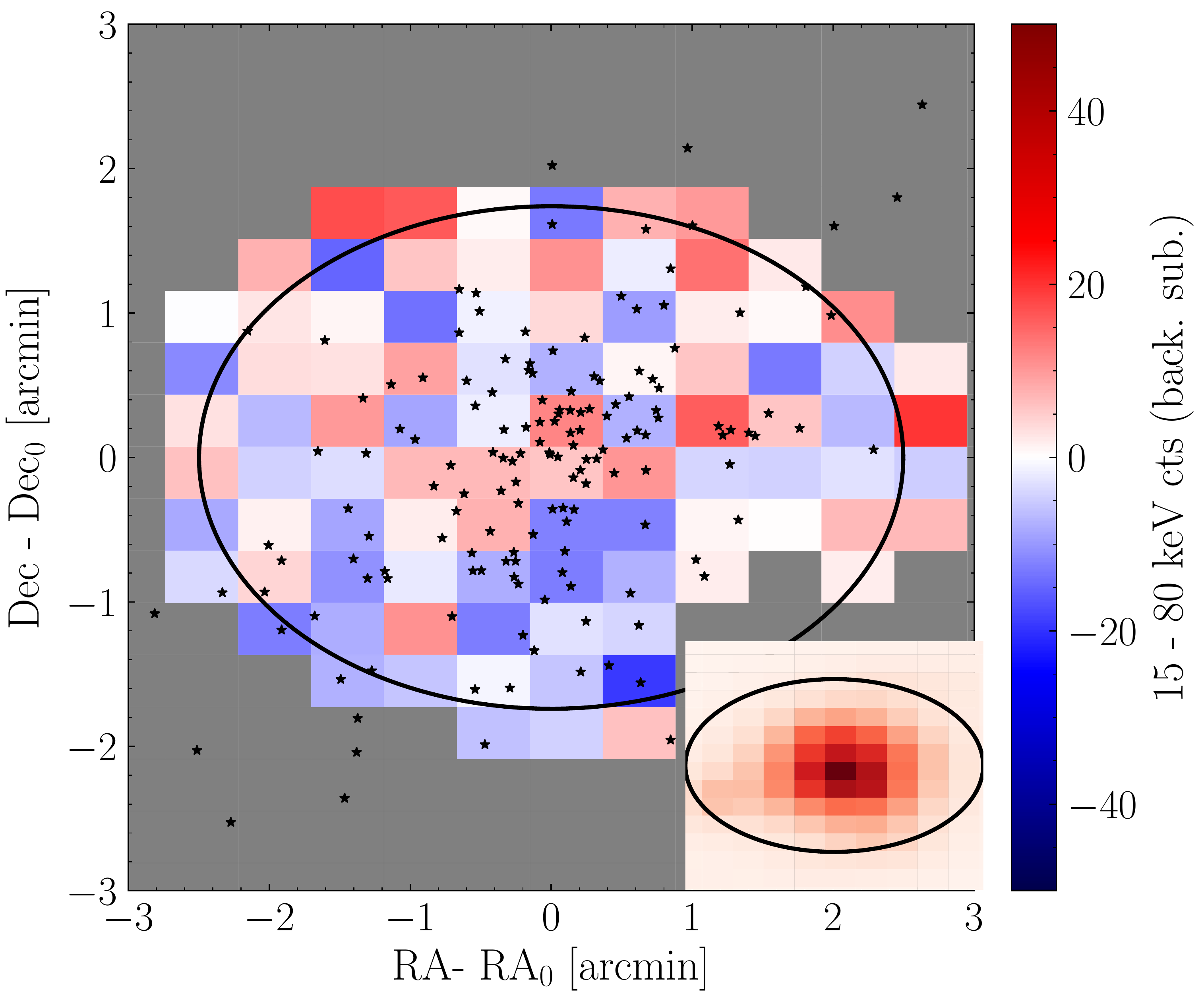}
\caption{As in Fig.~\ref{fig:all-maps-quint}, but for the Wd1 cluster NuSTAR analysis.  The red star indicates the location of the magnetar CXOU J164710.2--45521, which is masked at 0.5'.  Also shown is the background-subtracted count data, as in Fig.~\ref{fig:ill}.
\label{fig:all-maps-Wd1}
}
\end{center}
\end{figure}

In Fig.~\ref{fig:all-maps-Wd1} we show, in analogy with Fig.~\ref{fig:all-maps-quint}, the data, background, and signal maps summed from 15 - 80 keV.  We note that the background templates are summed using their best-fit normalizations from the fits to the null hypothesis of background-only emission.  The signal template is noticeably extended in this case beyond a point-source template and is shown for $g_{a\gamma\gamma} = 8 \times 10^{-12}$ GeV$^{-1}$ and $m_a \ll 10^{-11}$ eV.  The location of the magnetar CXOU J164710.2--45521 is indicated by the red star.    

\subsubsection{Axion Luminosity}

\begin{figure}[htb]  
\hspace{0pt}
\vspace{-0.2in}
\begin{center}
\includegraphics[width=0.495\textwidth]{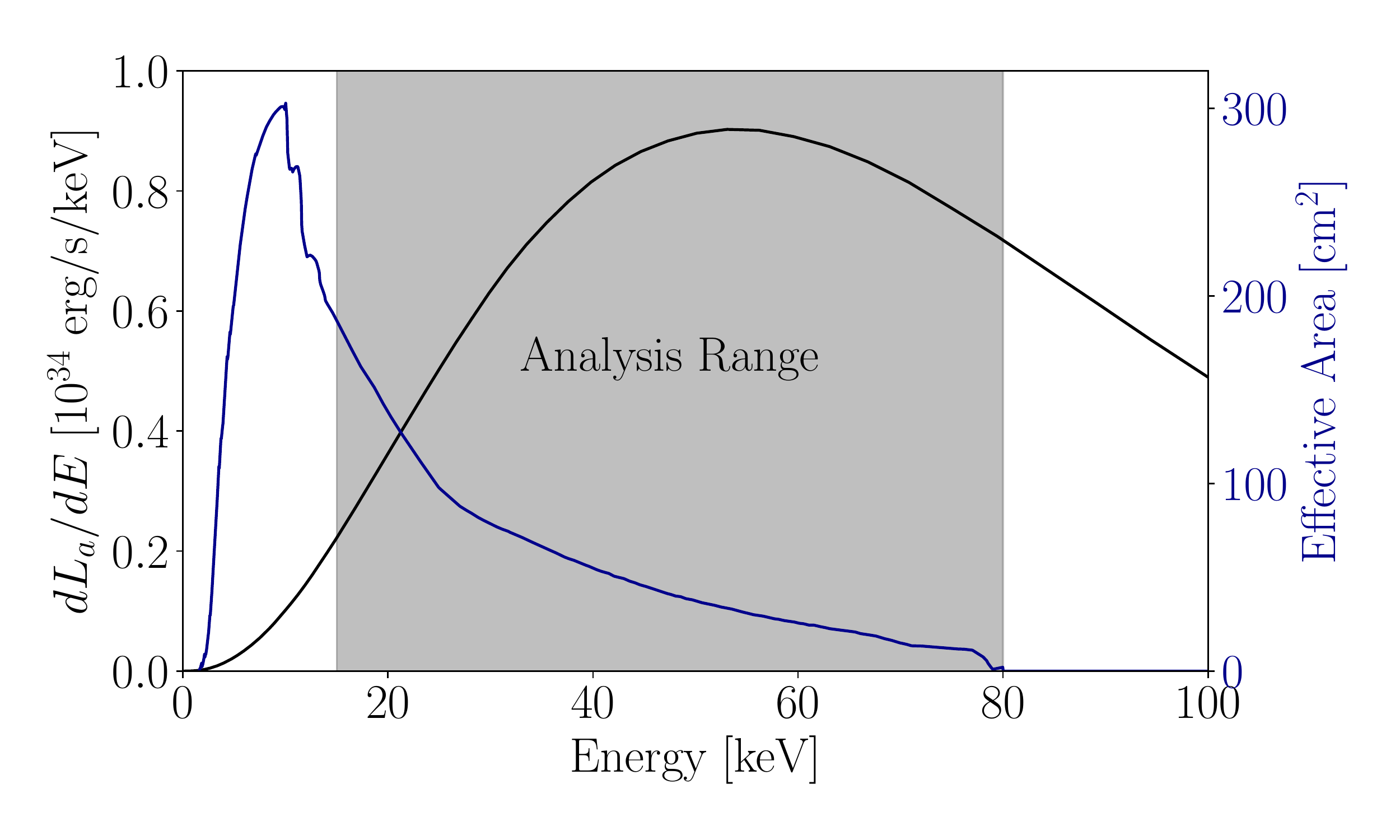}
\includegraphics[width=0.495\textwidth]{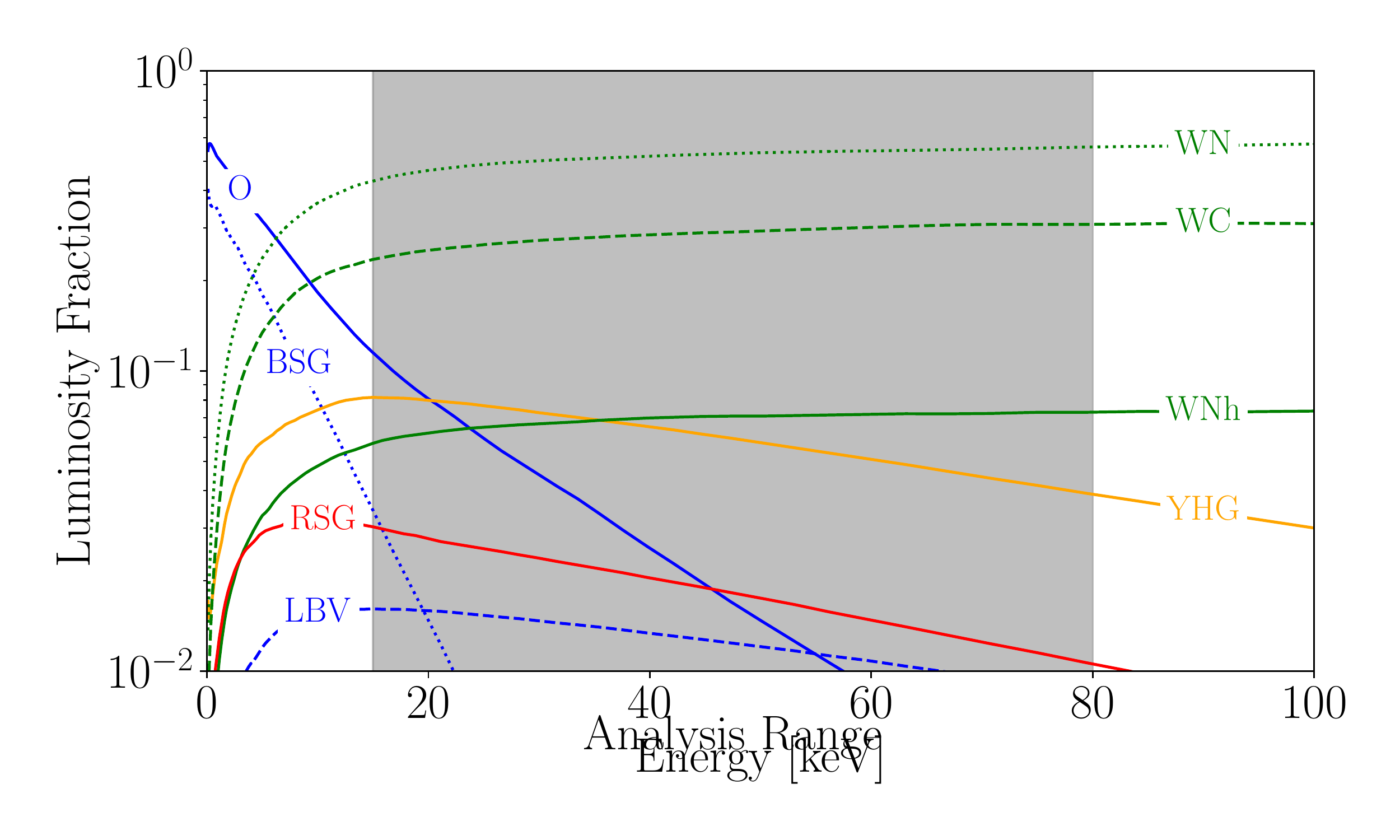}\\
\includegraphics[width=0.695\textwidth]{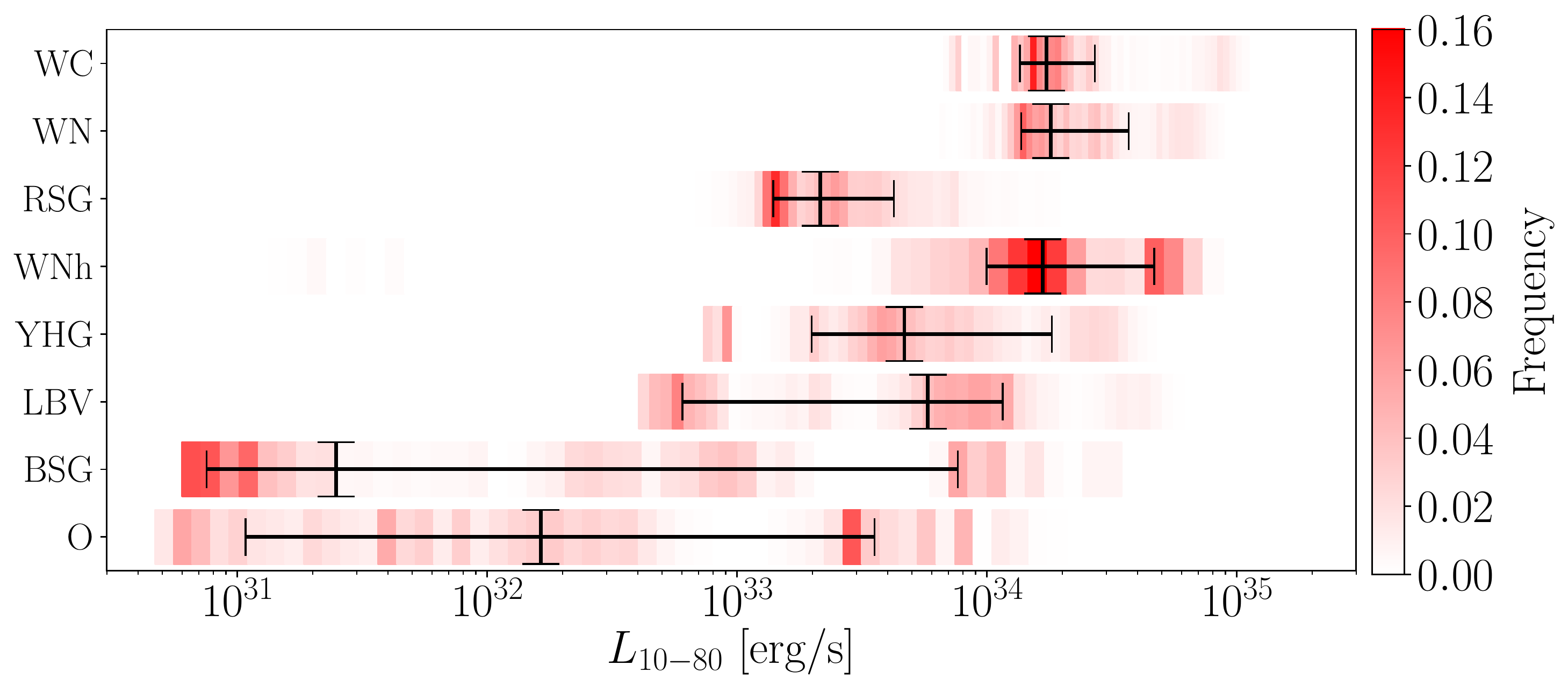}
\caption{(Upper Left) The Wd1 axion spectrum assuming $g_{a\gamma\gamma} = 10^{-12}$ GeV$^{-1}$ (black) plotted against the NuSTAR effective area (blue). The analysis range, from 15 - 80 keV, is shaded in gray. (Upper Right) The individual contributions of each stellar classification to the Wd1 axion spectrum. The analysis range is again shaded.  (Bottom) The 10-80 keV luminosity distribution assigned to each stellar classification in Wd1. In red we show the frequency with which each luminosity occurs, while the black error bars show the mean and 1$\sigma$ band.
\label{fig:Wd1_subtypes}
}
\end{center}
\end{figure}

We now show the axion luminosity and spectra that go into the right panel of Fig.~\ref{fig:all-maps-Wd1}. In the upper left panel of Fig.~\ref{fig:Wd1_subtypes}, we show the mean expected axion luminosity, as a function of energy, of the Wd1 cluster, assuming $g_{a\gamma\gamma} = 10^{-12}$ GeV$^{-1}$. In the upper right panel, we show the contribution of each spectral classification in Wd1 to this luminosity, summed over all stars with the given classification. For all energies of interest, the WN stars dominate the cluster luminosity, although the WC stars are important as well. As in Quintuplet, this is due to the fact that WR stars have the hottest cores, but in this case there are more WN stars than WC stars. In the bottom panel, we show the 10 - 80 keV luminosity distribution for each spectral classification, along with the 1$\sigma$ bands and the mean expectation. Again, the more evolved stars produce more axion flux, because their core temperatures increase with time. As in the case of Quintuplet, the O and BSG stars may be pre- or post-helium ignition. 
The luminous blue variable (LBV), yellow hypergiant (YHG), and cool red supergiant (RSG) stars are all post-helium ignition, although have generically cooler cores than the WR stars.  The WNh stars are entirely helium burning.

\begin{table*}[]
\centering
\begin{tabular}{|c||c|c|c|c|c|c|c|}
\hline
           & O & (B/R)SG/YHG   & LBV     & WNh   & WC/WN & tot & tot (10-80 keV)    \\ \hline
$N_{\rm star}$	
& $\begin{array}{c} 72 \end{array}$  
& $\begin{array}{c} 56 \end{array}$     
& $\begin{array}{c} 1 \end{array}$    
& $\begin{array}{c} 2 \end{array}$    
& $\begin{array}{c} 22 \end{array}$ 
& 153
& 153
\\ \hline
$\begin{array}{c}
z = 0.018 \\ \mu_{\rm rot} = 100 \, \,  {\rm km/s }
\end{array}$
& $\begin{array}{c}  1.6_{-0.6}^{+0.9} \times 10^{35} \end{array}$ 
& $2.4_{-0.8}^{+1.3} \times 10^{35}$        
& $1.4_{-1.3}^{+2.8} \times 10^{34}$      
& $2.2_{-1.5}^{+4.8} \times 10^{35}$        
& $4.3_{-1.4}^{+1.7} \times 10^{36}$   
& $5.2_{-1.4}^{+1.7} \times 10^{36}$
& $1.3_{-0.2}^{+0.2} \times 10^{36}$
\\ \hline
$\begin{array}{c}
z = 0.035 \\ \mu_{\rm rot} = 100 \, \,  {\rm km/s }
\end{array}$
& $\begin{array}{c}  2.6_{-1.1}^{+1.5} \times 10^{35} \end{array}$ 
& $3.9_{-1.5}^{+2.6} \times 10^{35}$        
& $7.1_{-6.5}^{+10} \times 10^{33}$      
& $8.7_{-4.5}^{+37} \times 10^{34}$        
& $2.0_{-0.7}^{+1.0} \times 10^{36}$   
& $3.1_{-0.9}^{+1.1} \times 10^{36}$
& $9.9_{-1.4}^{+1.4} \times 10^{35}$ \\
\hline 
$\begin{array}{c}
z = 0.035 \\ \mu_{\rm rot} = 150 \, \,  {\rm km/s  	}
\end{array}$
& $\begin{array}{c}  2.3_{-1.0}^{+1.3} \times 10^{35} \end{array}$ 
& $3.5_{-1.5}^{+2.6} \times 10^{35}$        
& $7.1_{-6.5}^{+9.0} \times 10^{33}$      
& $6.2_{-2.8}^{+31} \times 10^{34}$        
& $1.8_{-0.7}^{+1.0} \times 10^{36}$ 
& $2.6_{-0.8}^{+1.0} \times 10^{36}$
& $9.0_{-1.0}^{+1.0} \times 10^{35}$
\\ \hline
\end{tabular}
\caption{As in Tab.~\ref{tab:SSC-Quint} but for Wd1. 
}
\label{tab:SSC-Wd1}
\end{table*}

In Tab.~\ref{tab:SSC-Wd1} we provide detailed luminosities for each of the stellar sub-types for different choices of initial $Z$ and $\mu_{\rm rot}$ for Wd1, as we did in Tab.~\ref{tab:SSC-Quint}.  Note that we assume $Z = 0.035$ and $\mu_{\rm rot} = 150$ km/s for our fiducial analysis, even though it is likely that the initial $Z$ is closer to solar (in which case the luminosities would be enhanced, as seen in Tab.~\ref{tab:SSC-Wd1}).

\subsubsection{Systematics on the extracted spectrum}

\begin{figure}[htb]
\hspace{0pt}
\vspace{-0.2in}
\begin{center}
\includegraphics[width=0.55\textwidth]{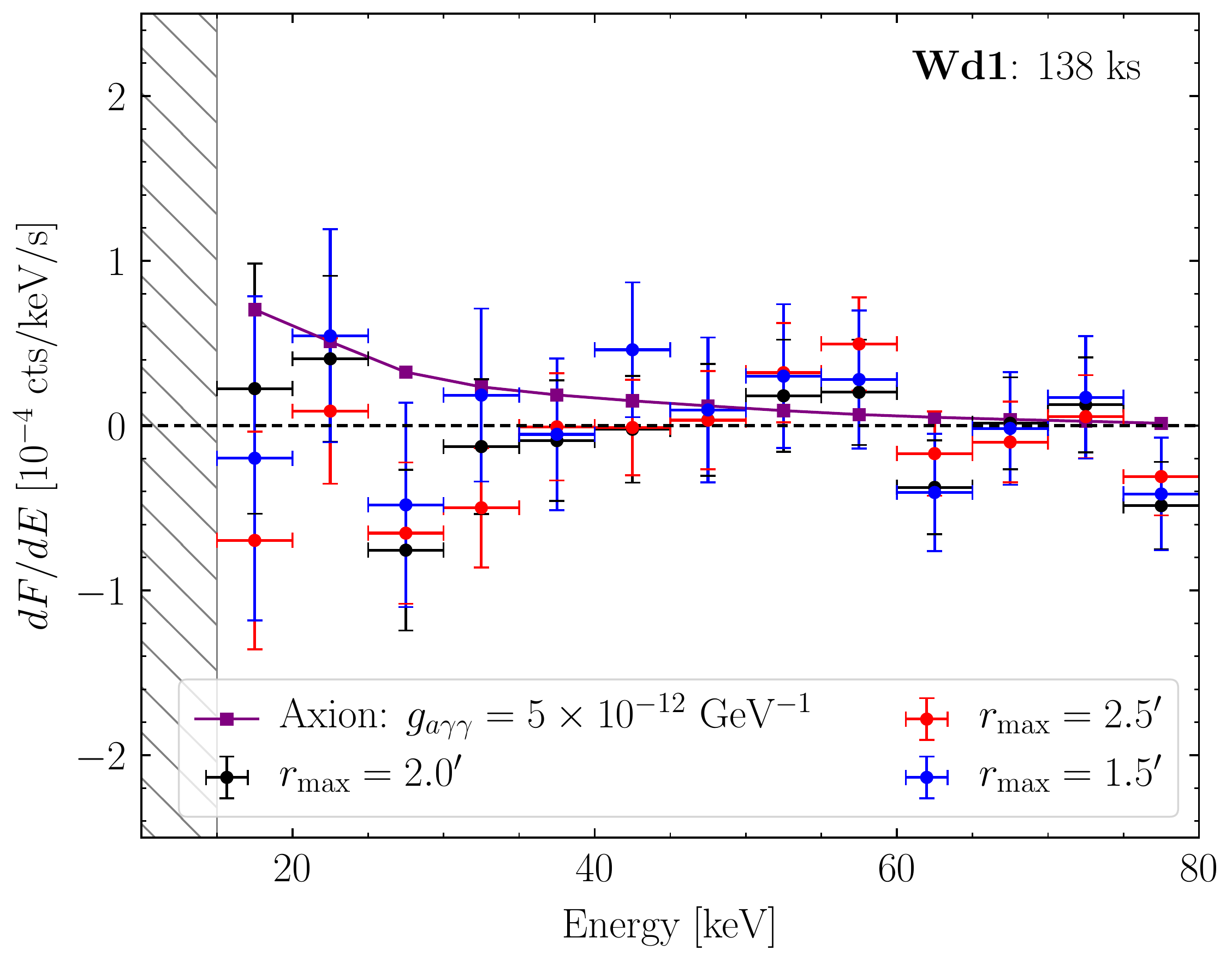}
\caption{As in Fig.~\ref{fig:flux_spectra-syst} but for the Wd1 analysis.  Note that we only include energies above 15 keV in our analysis because of ghost-ray contamination.  
\label{fig:flux_spectra-wd1}
}
\end{center}
\end{figure}

In analogy to the Quintuplet analysis we may profile over emission associated with the background template to measure the spectrum from 15 - 80 keV associated with the axion-induced signal template shown in Fig.~\ref{fig:all-maps-Wd1}.  That spectrum is reproduced in Fig.~\ref{fig:flux_spectra-wd1}.  For our default analysis we use the ROI with all pixels contained with $r_{\rm max} = 2.0'$ of the cluster center, except for those in the magnetar mask, as indicated in Fig.~\ref{fig:all-maps-Wd1}.  However, as a systematic test we also compute the spectrum associated with the axion-induced template for $r_{\rm max} = 2.5'$ and $1.5'$, as shown in Fig.~\ref{fig:flux_spectra-wd1}.   
We measure a consistent spectrum across ROIs at these energies. 

\subsection{Arches}

In this subsection we present results from the analysis of archival NuSTAR data for an axion-induced signal from the Arches cluster. The Arches cluster is at a similar location, $\sim$30 pc from the GC, as the Quintuplet cluster.  Arches hosts even younger and more extreme ({\it e.g.}, hotter and more massive) stars than the nearby Quintuplet cluster.  Indeed, it is estimated that all $\sim$105 spectroscopically classified stars within Arches may become core-collapse supernovae within the next $\sim$10 Myr~\cite{clark2018arches}.  \textit{A priori}, the Arches and Quintuplet clusters should have similar sensitivities to axions, though as we discuss below 
the axion prediction from Arches is less robust to uncertainties in the initial metallicity than the Quintuplet prediction.

\subsubsection{Axion Luminosity}

\begin{figure}[htb]  
\hspace{0pt}
\vspace{-0.2in}
\begin{center}
\includegraphics[width=0.495\textwidth]{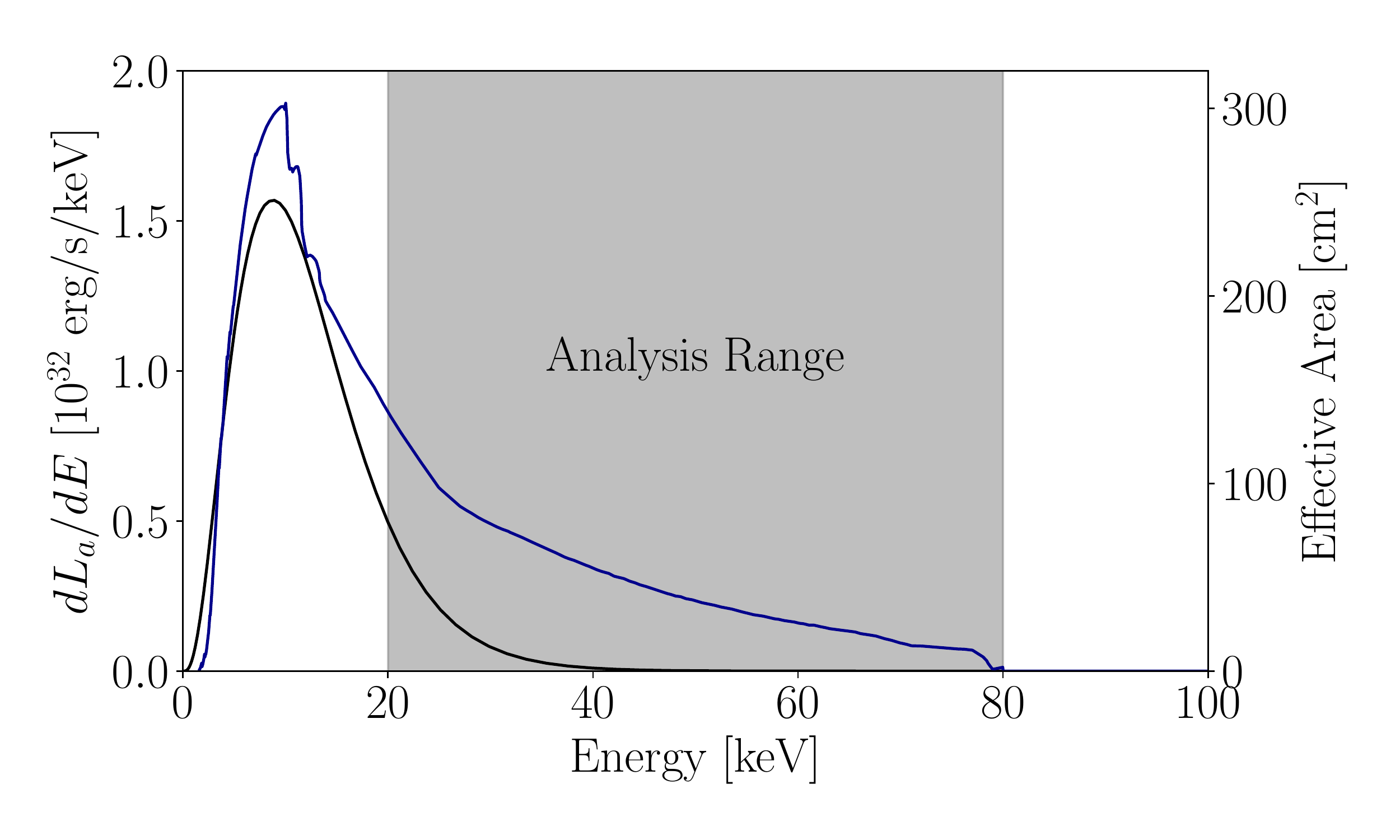}
\includegraphics[width=0.495\textwidth]{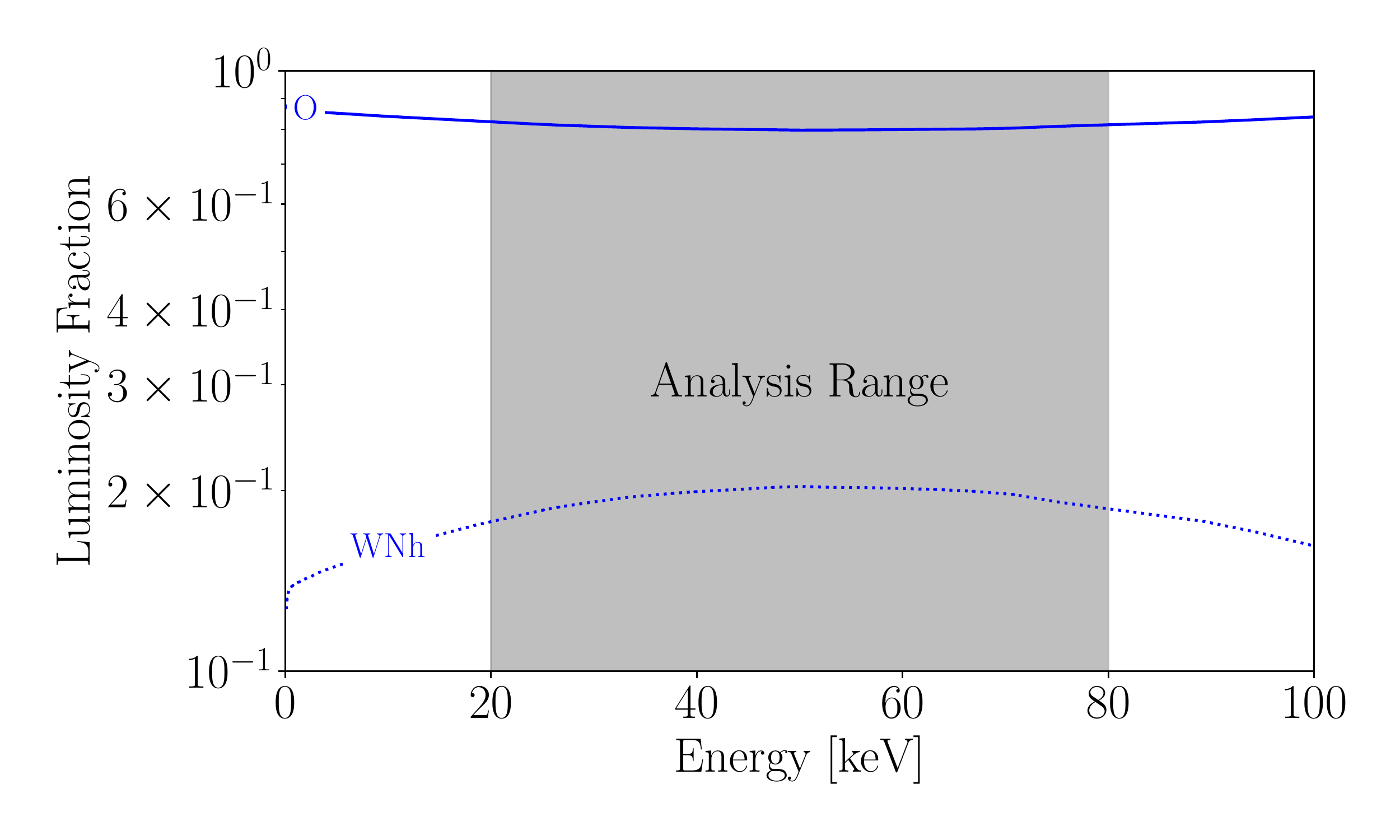}\\
\includegraphics[width=0.695\textwidth]{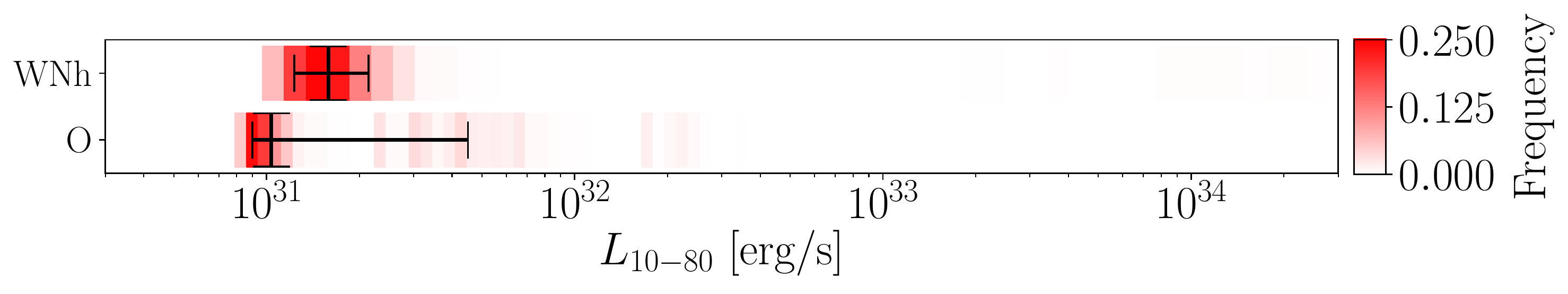}
\caption{(Upper Left) The Arches axion spectrum assuming $g_{a\gamma\gamma} = 10^{-12}$ GeV$^{-1}$ (black) plotted against the NuSTAR effective area (blue). The analysis range, from 20 - 80 keV, is shaded in gray. (Upper Right) The individual contributions of each stellar classification to the Arches axion spectrum. The analysis range is again shaded.  (Bottom) The 10-80 keV luminosity distribution assigned to each stellar classification in Arches. In red we show the frequency with which each luminosity occurs, while the black error bars show the mean and 1$\sigma$ band.
\label{fig:Arches_subtypes}
}
\end{center}
\end{figure}

We now describe the axion luminosity and spectra for Arches. In the upper left panel of Fig.~\ref{fig:Arches_subtypes}, we show the mean expected axion luminosity, as a function of energy, of the Arches cluster, assuming $g_{a\gamma\gamma} = 10^{-12}$ GeV$^{-1}$. The luminosity peaks at very low energies, although we could not analyze these energies due to contamination from the molecular cloud. As shown by the upper right panel, the Arches luminosity is dominated by the O stars, since the WNh stars are always hydrogen burning with our assumed metallicity of $Z = 0.035$ and there are many more O stars than WNh stars. In the bottom panel, we show the 10 - 80 keV luminosity distribution for the O and WNh stars, along with the 1$\sigma$ bands and the mean expectation. 

However, unlike for the Quintuplet and Wd1 clusters we find that the Arches luminosity is a strong function of the initial metallicity $Z$, as illustrated in Tab.~\ref{tab:SSC-Arches}.  As seen in that table, changing the metallicity from $Z = 0.035$ to $Z = 0.018$ increases the flux by over an order of magnitude.  This is because at the higher metallicity values the WNh stars are typically not in the He burning phase, while decreasing the initial metallicity slightly causes the WNh stars to enter the He burning phase.  Note that at solar initial metallicity ($Z = 0.02$, and also taking $\mu_{\rm rot} = 100$ km/s) we find that the 10-80 keV
flux is $8.7_{-5.6}^{+9.4} \times 10^{34}$ erg/s, comparable to but slightly larger than that found for $Z = 0.018$.  Thus, it is possible that the sensitivity of the Arches observations is comparable to that from Quintuplet, but given the larger uncertainties related to the stellar modeling of the Arches stars the limit is, at present, less robust.  We stress that the qualitative difference between Arches and Quintuplet that is responsible for this difference is that Quintuplet has a large cohort of WC and WN stars, which are robustly He burning, while Arches does not have any stars in these stellar classes.

\begin{table*}[]
\centering
\begin{tabular}{|c||c|c|c|c|c|c|c|}
\hline
           & O & (B/R)SG/YHG   & LBV     & WNh   & WC/WN & tot & tot (10-80 keV)    \\ \hline
$N_{\rm star}$	
& $\begin{array}{c} 96 \end{array}$  
& $\begin{array}{c} 0 \end{array}$     
& $\begin{array}{c} 0 \end{array}$    
& $\begin{array}{c} 13 \end{array}$    
& $\begin{array}{c} 0 \end{array}$ 
& 109
& 109
\\ \hline
$\begin{array}{c}
z = 0.018 \\ \mu_{\rm rot} = 100 \, \,  {\rm km/s }
\end{array}$
& $\begin{array}{c}  2.3_{-0.1}^{+0.2} \times 10^{33} \end{array}$ 
& 0        
& 0      
& $8.7_{-5.2}^{+6.5} \times 10^{34}$        
& 0   
& $8.9_{-5.2}^{+6.5} \times 10^{34}$
& $6.6_{-3.6}^{+5.6} \times 10^{34}$
\\ \hline
$\begin{array}{c}
z = 0.035 \\ \mu_{\rm rot} = 100 \, \,  {\rm km/s }
\end{array}$
& $\begin{array}{c} 
3.9_{-1.9}^{+1.8} \times 10^{35} \end{array}$ 
& 0        
& 0      
& $3.9_{-0.6}^{+217} \times 10^{32}$        
& 0   
& $7.2_{-4.9}^{+16} \times 10^{33}$
& $5.7_{-2.8}^{+23} \times 10^{33}$ \\
\hline 
$\begin{array}{c}
z = 0.035 \\ \mu_{\rm rot} = 150 \, \,  {\rm km/s } 	
\end{array}$
& $\begin{array}{c} 
3.5_{-1.6}^{+2.1} \times 10^{33} \end{array}$ 
& 0        
& 0      
& $3.6_{-0.3}^{+125} \times 10^{32}$
& 0 
& $4.7_{-2.2}^{+12} \times 10^{33}$
& $3.7_{-2.4}^{+13} \times 10^{33}$
\\ \hline
\end{tabular}
\caption{As in Tab.~\ref{tab:SSC-Quint} but for Arches. 
}
\label{tab:SSC-Arches}
\end{table*}

\subsubsection{Data analysis, results, and systematic tests}

We reduce and analyze  370 ks 
of archival NuSTAR data from  Arches.  
 The Arches observations (IDs \texttt{40010005001,
40101001004,
40101001002,
40202001002,
40010003001})
were performed as part of the same GC survey as the Quintuplet observations as well as for dedicated studies of the Arches cluster below 20 keV.  Note that we discard data from the Focal Plane Module B instrument for observations \texttt{40101001004}, \texttt{40101001002}, \texttt{40202001002}, and \texttt{40010003001} because of ghost-ray contamination.  We perform astrometric calibration using the low-energy data on the Arches cluster itself, which is a bright point source above 3 keV.

  In the Arches analysis it is known that there is a nearby molecular cloud that emits in hard $X$-rays~\cite{Krivonos:2016omp}.  We follow~\cite{Krivonos:2016omp} and model emission associated with this extended cloud as a 2D Gaussian centered at R.A.=$17^h45^m50.62^s$,  Dec.=$-28^\circ49'47.17''$ with a FWHM of $72.4''$.  The hard $X$-ray spectrum associated with the molecular cloud has been observed to extend to approximately 40 keV~\cite{Krivonos:2016omp}; indeed, we see that including the molecular cloud template, with a free normalization parameter, at energies below 40 keV affects the spectrum that we extract for the axion template, but it does not significantly affect the spectrum extraction above 40 keV.  The non-thermal flux associated with the molecular cloud is expected to be well described by a power-law with spectral index $\Gamma \approx 1.6$ and may arise from the collision of cosmic-ray ions generated within the star cluster with gas in the nearby molecular cloud~\cite{Krivonos:2013sqa}.  With this spectral index the molecular cloud should be a sub-dominant source of flux above $\sim$20 keV, and we thus exclude the 10-20 keV energy range from the Arches analysis, though {\it e.g.} including the 15-20 keV bin results in nearly identical results (as does excluding the 20 - 40 keV energy range).

\begin{figure}[htb]  
\hspace{0pt}
\vspace{-0.2in}
\begin{center}
\includegraphics[width=0.325\textwidth]{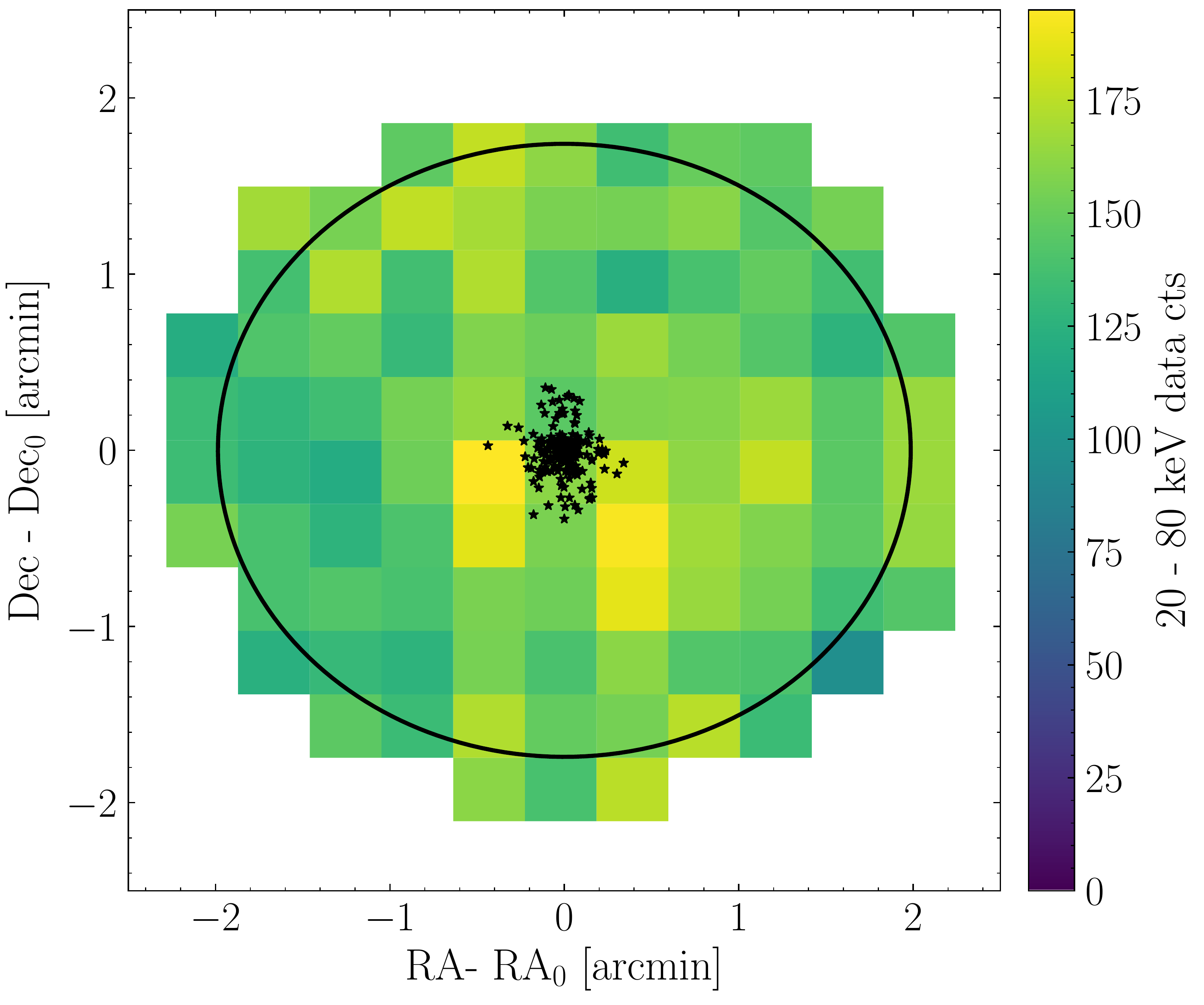}
\includegraphics[width=0.325\textwidth]{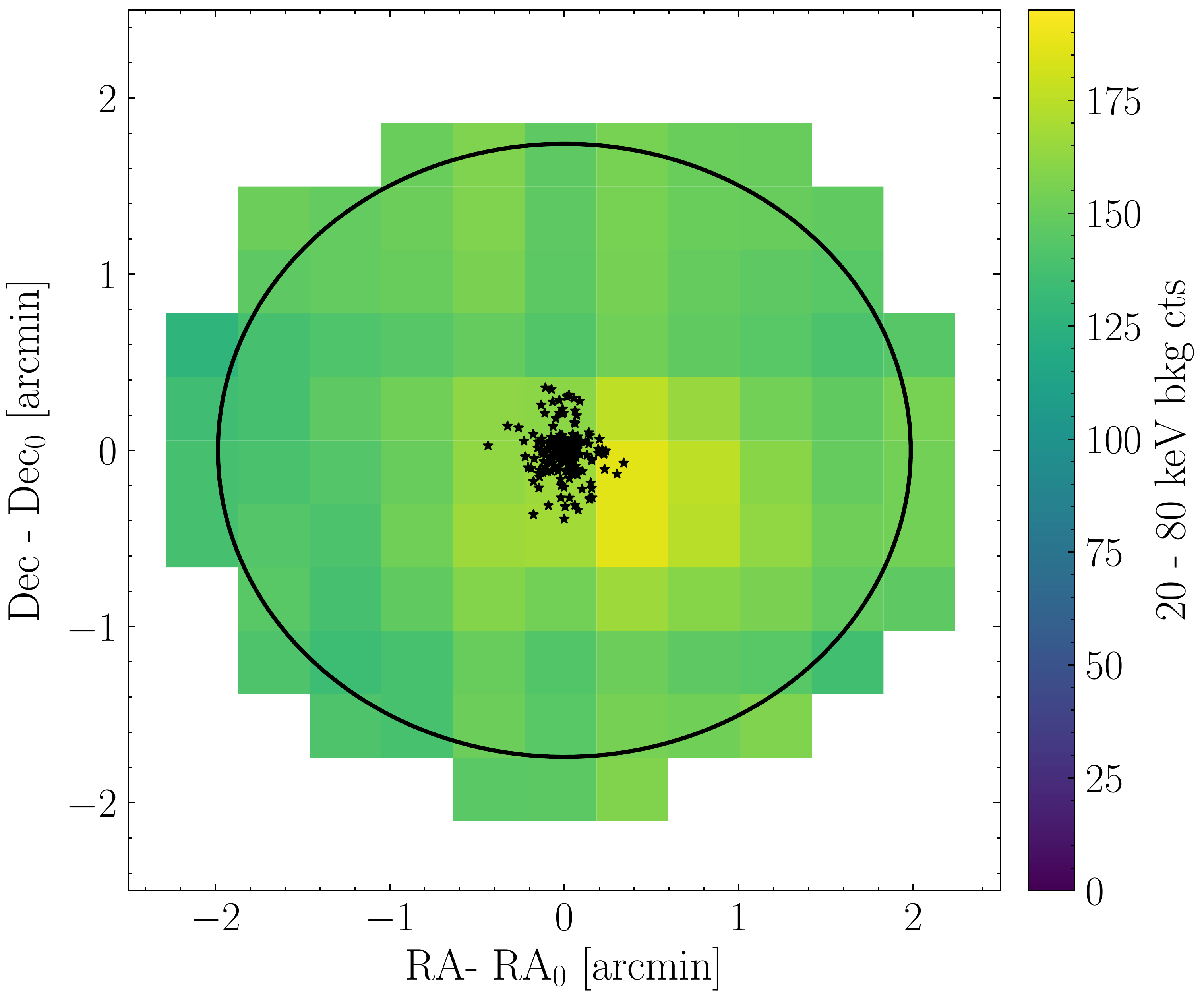}
\includegraphics[width=0.325\textwidth]{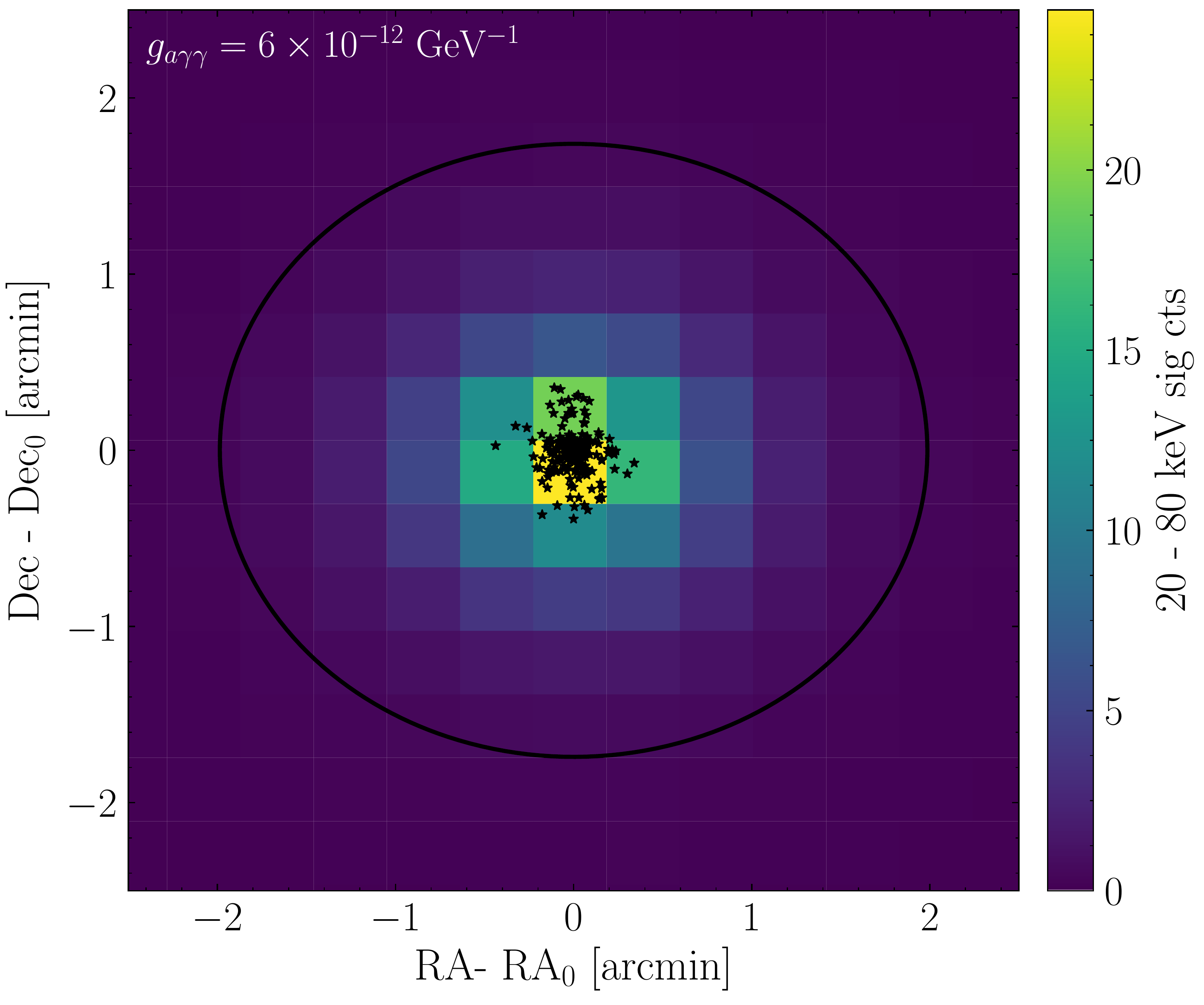}
\includegraphics[width=0.325\textwidth]{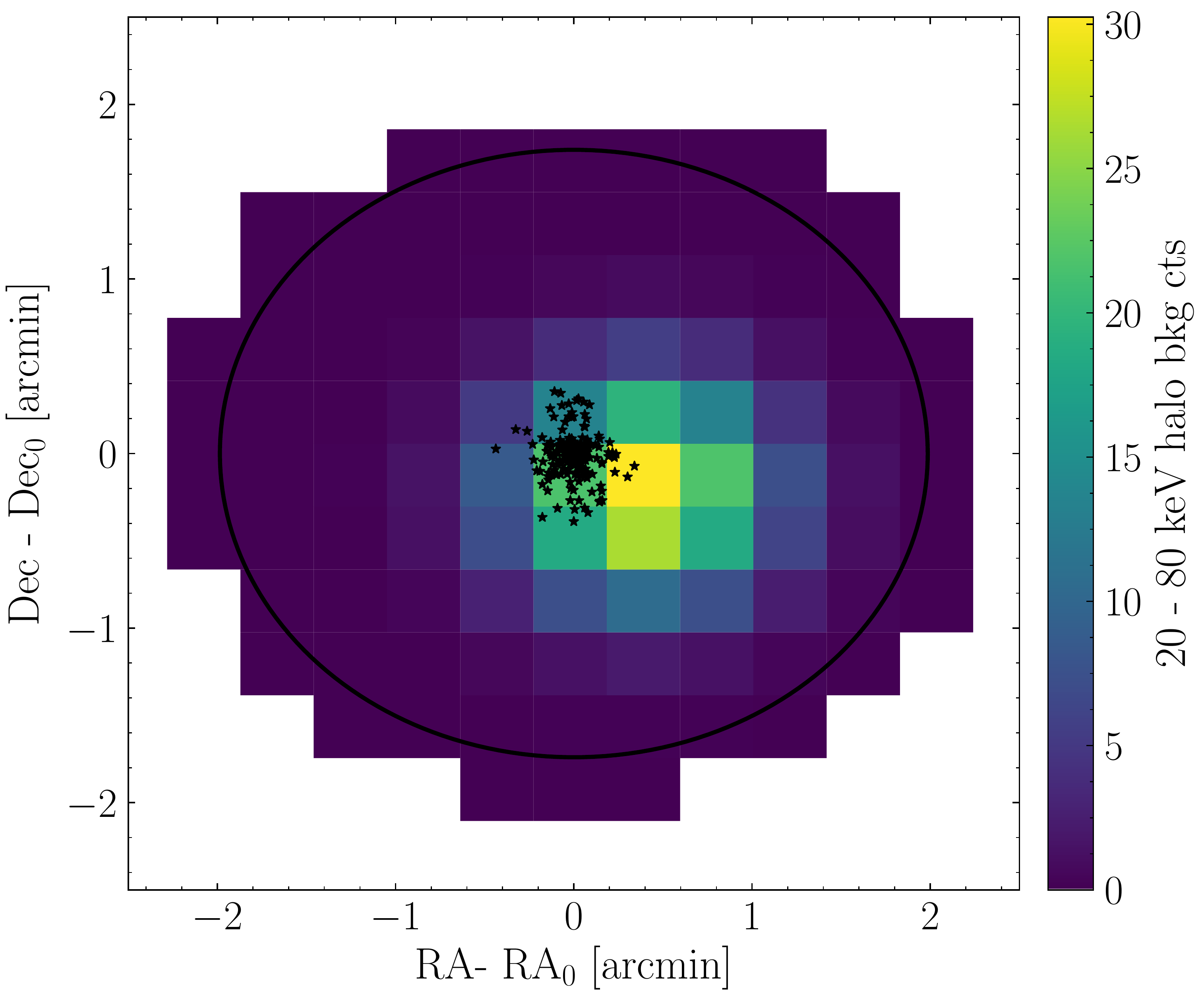}
\includegraphics[width=0.325\textwidth]{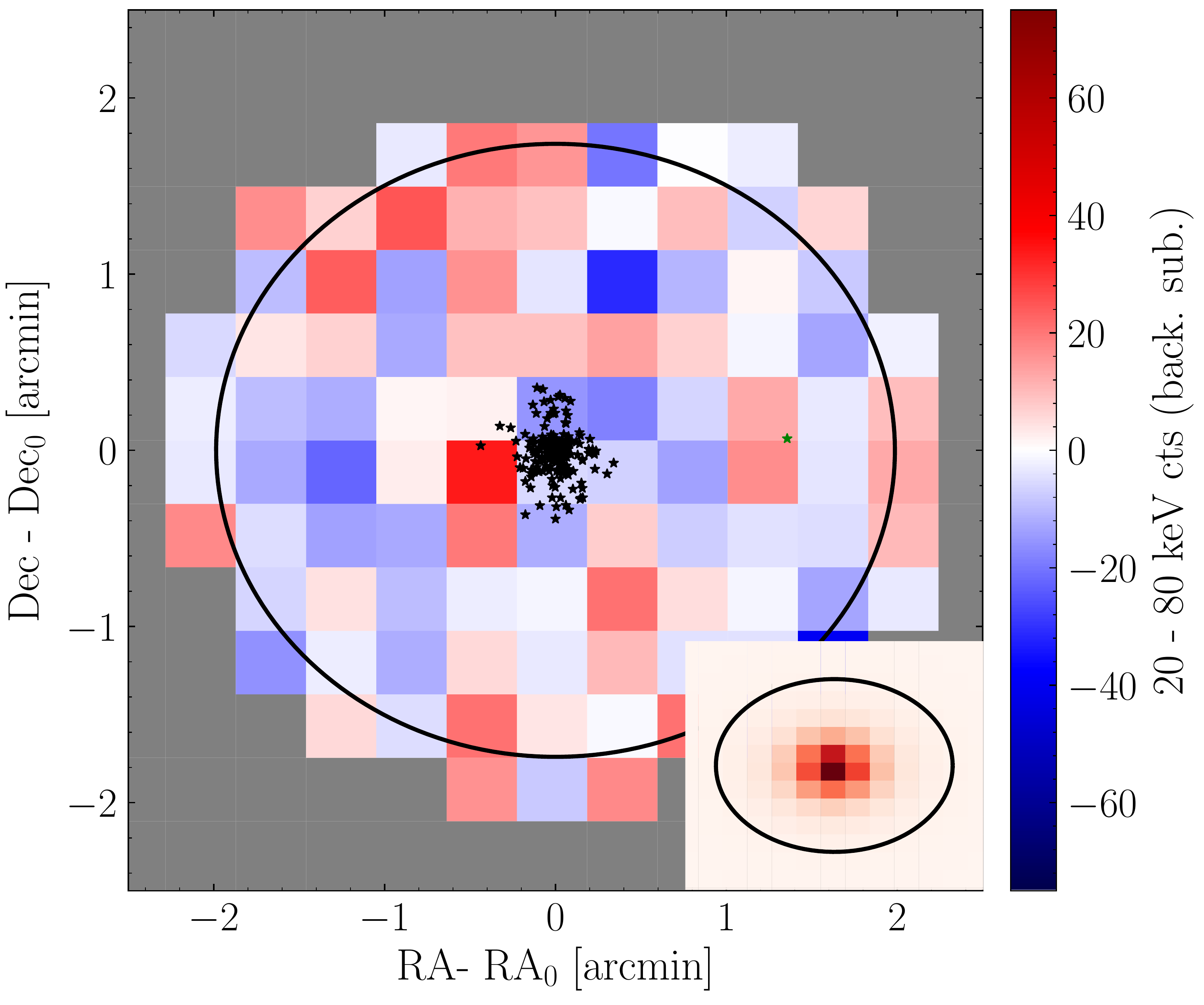}
\caption{
(Top Panel) As in Fig.~\ref{fig:all-maps-quint}, but for the Arches cluster. (Bottom left) We show the best-fit emission associated with the halo template that describes emission from the nearby molecular cloud. (Bottom right) As in in Fig.~\ref{fig:ill}, but for Arches.  
\label{fig:all-maps-arches}
}
\end{center}
\end{figure}

The molecular cloud template is illustrated in the bottom left panel of Fig.~\ref{fig:all-maps-arches}.  In that figure we also show the data, background templates, signal template, and background-subtracted counts, as in Fig.~\ref{fig:all-maps-quint} for the Quintuplet analysis.  Note that we profile over emission associated both the background template and with the halo template when constraining the flux in each energy bin associated with the signal template. 

\begin{figure}[htb]  
\hspace{0pt}
\vspace{-0.2in}
\begin{center}
\includegraphics[width=0.49\textwidth]{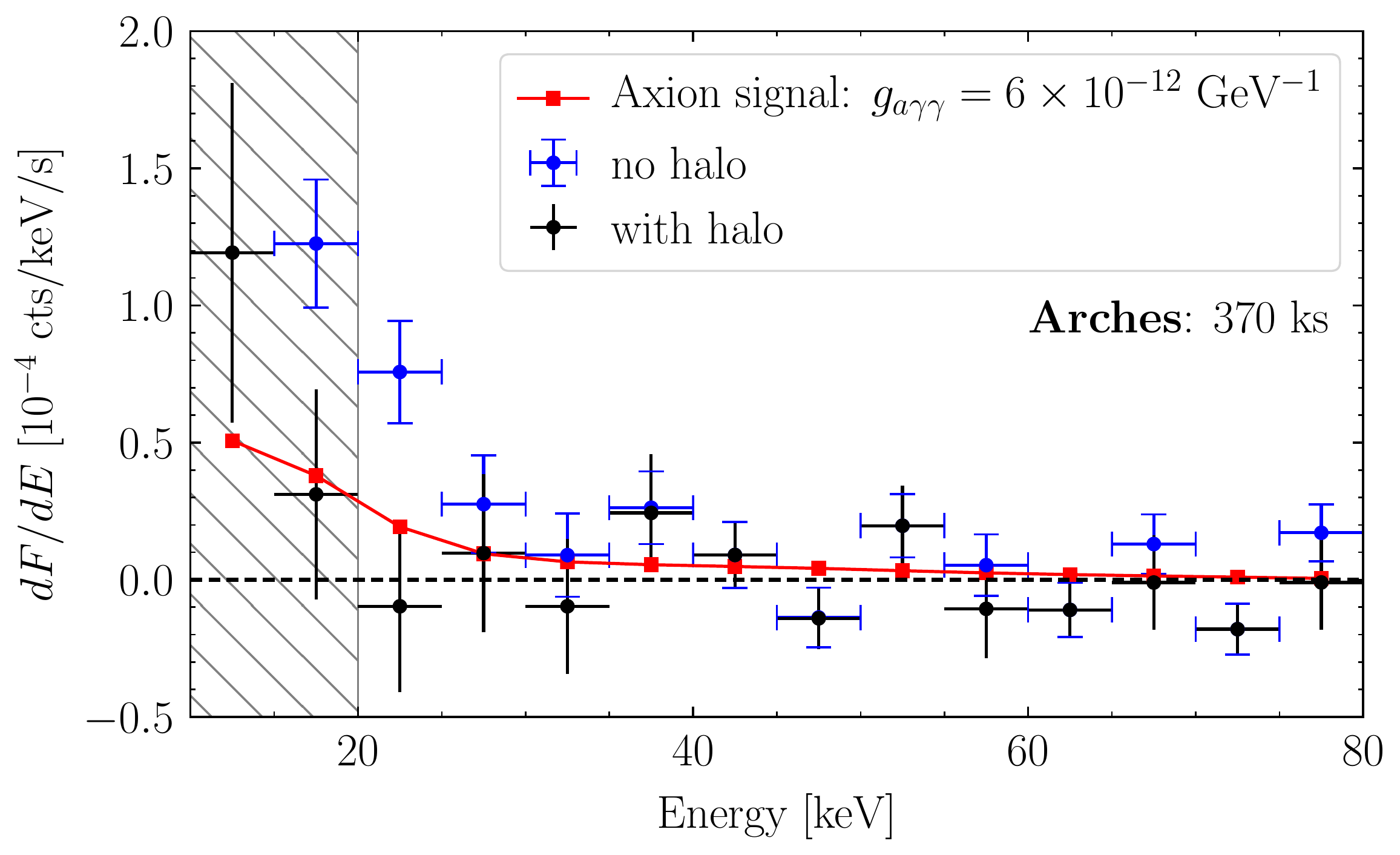}
\includegraphics[width=0.49\textwidth]{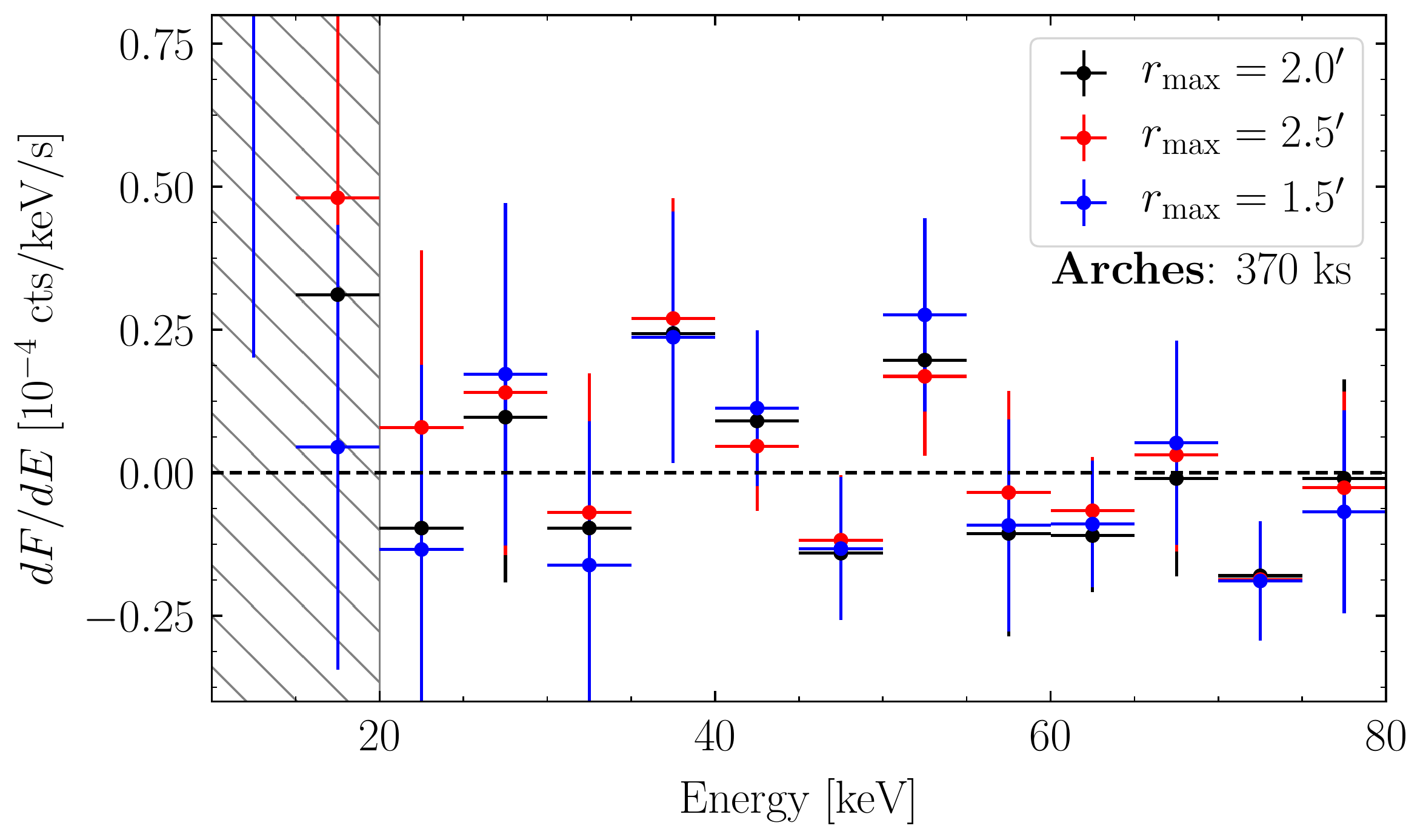}
\caption{
(Left) The Arches spectrum measured with and without the halo template.  Note that we use the spectrum with the halo template in our fiducial analysis, though the difference between the two results is relatively minor above $\sim$20 keV.  (Right) As in Fig.~\ref{fig:flux_spectra-syst} but for the Quintuplet analysis. Note that these spectra are computed while profiling over halo emission.  Above $\sim$20 keV the different ROIs produce consistent results.
\label{fig:Arches-no-halo}
}
\end{center}
\end{figure}

As a systematic test of our signal extraction procedure we show in Fig.~\ref{fig:Arches-no-halo} (left panel) the spectrum extracted for axion emission from the Arches cluster both with and without the halo template.  The two spectra diverge below $\sim$20 keV but give consistent results above this energy.  Similarly, we find that the spectrum is relatively insensitive to the ROI size for energies above $\sim$20 keV, as shown in the right panel of Fig.~\ref{fig:Arches-no-halo}, which is analogous to the Quintuplet Fig.~\ref{fig:flux_spectra-syst}.

\begin{figure}[htb]  
\hspace{0pt}
\vspace{-0.2in}
\begin{center}
\includegraphics[width=0.55\textwidth]{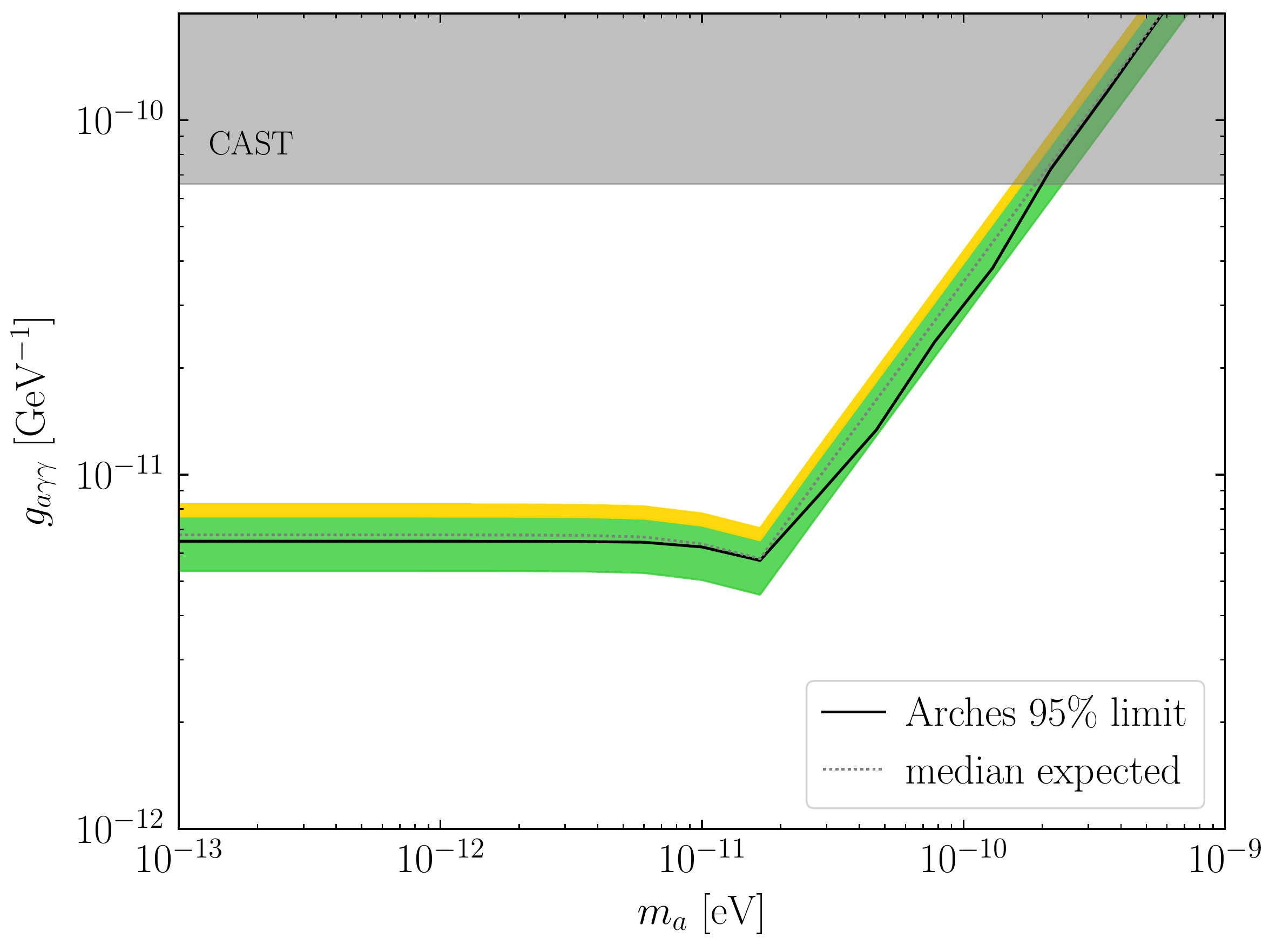}
\caption{
As in Fig.~\ref{fig:limits} but from the analysis towards the Arches SSC.  No evidence for axions is found from this search.
\label{fig:lim-Arches}
}
\end{center}
\end{figure}

In Fig.~\ref{fig:lim-Arches} we show the 95\% upper limit we obtain on $g_{a\gamma\gamma}$ from the Arches analysis, using the conservative modeling with $Z = 0.035$ and $\mu_{\rm rot} = 150$ km/s.  We find no evidence for an axion-induced signal from this search.  Note that, as in indicated in Fig.~\ref{fig:Arches-no-halo}, we do not include data below 20 keV in this analysis.

\section{Initial metallicity determination for Quintuplet and Arches}

\begin{figure}[htb]  
\hspace{0pt}
\vspace{-0.2in}
\begin{center}
\includegraphics[width=0.7\textwidth]{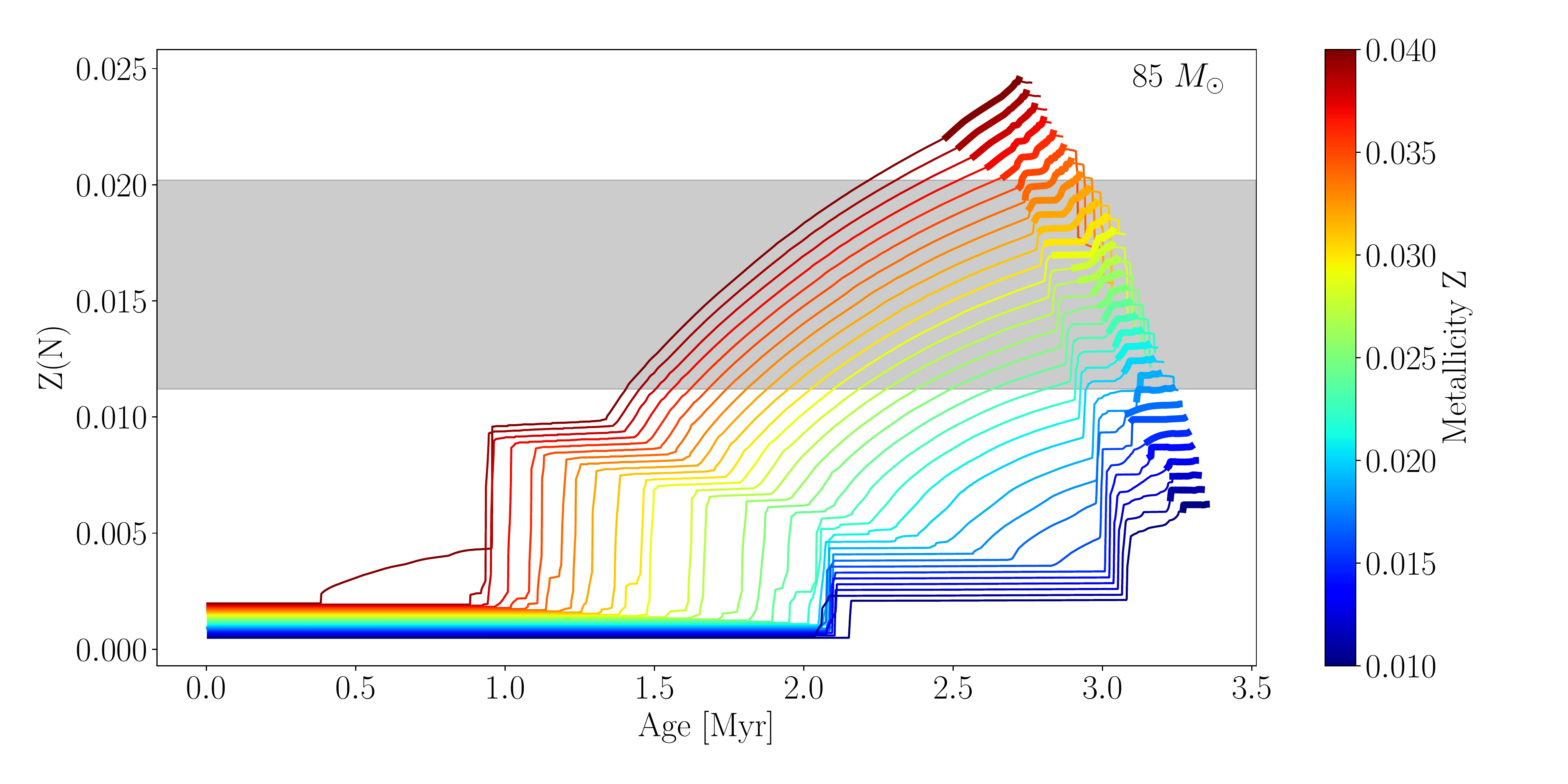}
\caption{
(Left) The evolution of the nitrogen abundance Z(N) over time from MESA simulations of a non-rotating 85 $M_\odot$ star with initial metallicity $Z = 0.01$ to $Z = 0.04$. The bolded sections of the lines correspond to the WNh phase. The gray shaded region indicates the measurements of nitrogren abundances of the Arches WNh stars from~\cite{Najarro:2004qm}.
\label{fig:metals}
}
\end{center}
\end{figure}

In our fiducial analysis we assumed the cluster metallicity was $Z = 0.035$, which we take as the highest allowed metallicity in the Quintuplet cluster. In this subsection we show how we arrived at this value. The cluster metallicity is an important parameter in that it affects the mass loss rates in the stellar winds, the lifetime of individual evolutionary stages, and the surface abundances. Here we use measurements of the nitrogren abundances of WNh stars in the Arches cluster to estimate the uncertainty on the cluster metallicities. The nitrogen abundance during the WNh phase reaches a maximum that depends only on the original CNO content, and as such is a direct tracer of stellar metallicity (and increases with increasing metallicity). Ref.~\cite{Najarro:2004qm} measured the nitrogen abundance in the WNh stars in the Arches cluster at present to be $0.0157 \pm 0.0045$. We run MESA simulations of the Arches WNh stars on a grid of metallicities from $Z = 0.01$ to $Z = 0.04$ and find this measurement implies that the Arches initial metallicity is between $Z = 0.018$ and $Z = 0.035$. The results are shown in Fig.~\ref{fig:metals}, where we see that the nitrogen abundance during the WNh phase intersects with the measurement only for the initial metallicities in that range. Although there are no measurements of the Quintuplet WNh nitrogren abundance, note that a similar abundance was found in the nearby GC SSC of $0.0143 \pm 0.0042$~\cite{Martins:2007kw}. Given the similarity of these two measurements, we assume the same metallicity range for Quintuplet as computed for Arches. 

\begin{figure}[htb]  
\hspace{0pt}
\vspace{-0.2in}
\begin{center}
\includegraphics[width=0.49\textwidth]{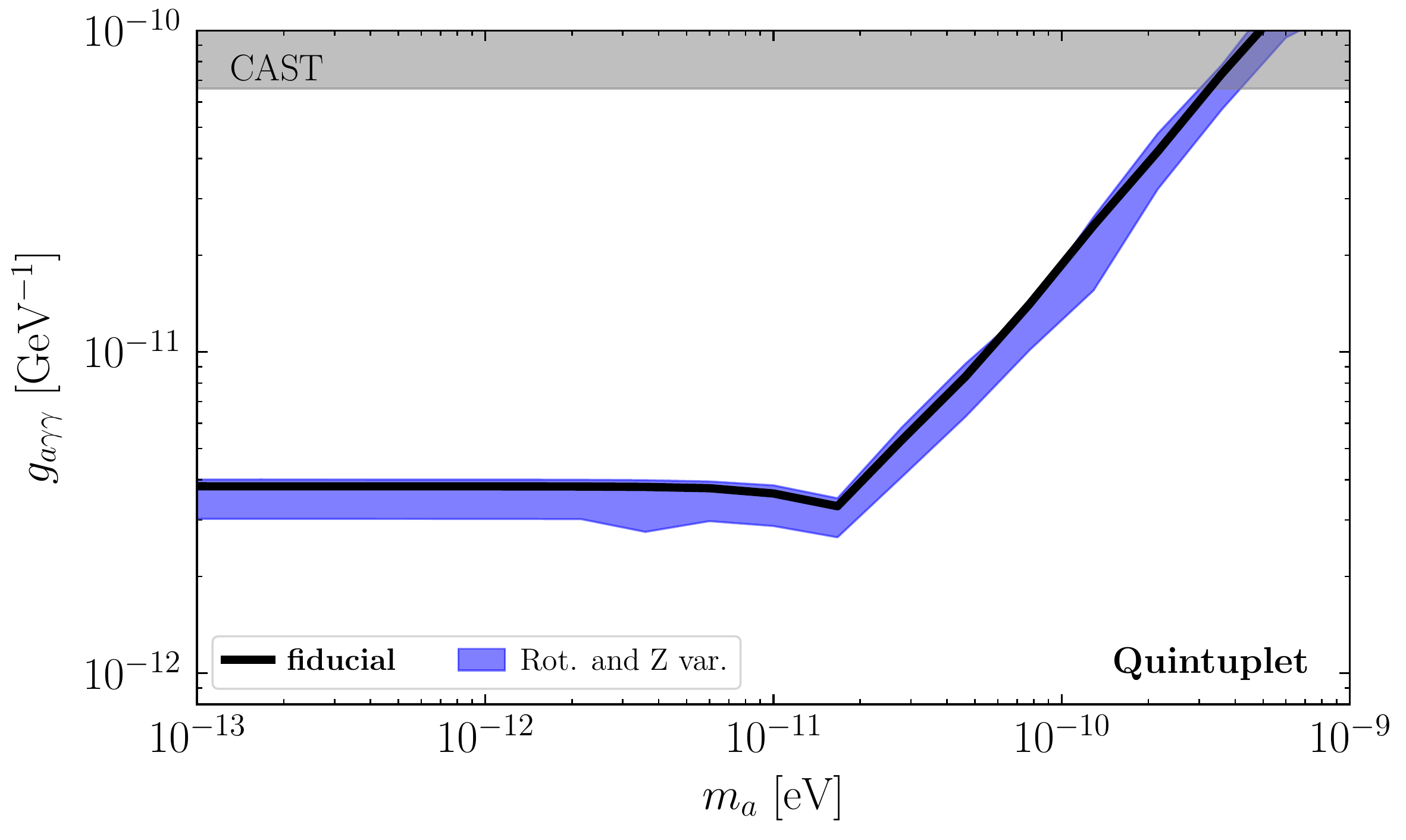}
\includegraphics[width=0.49\textwidth]{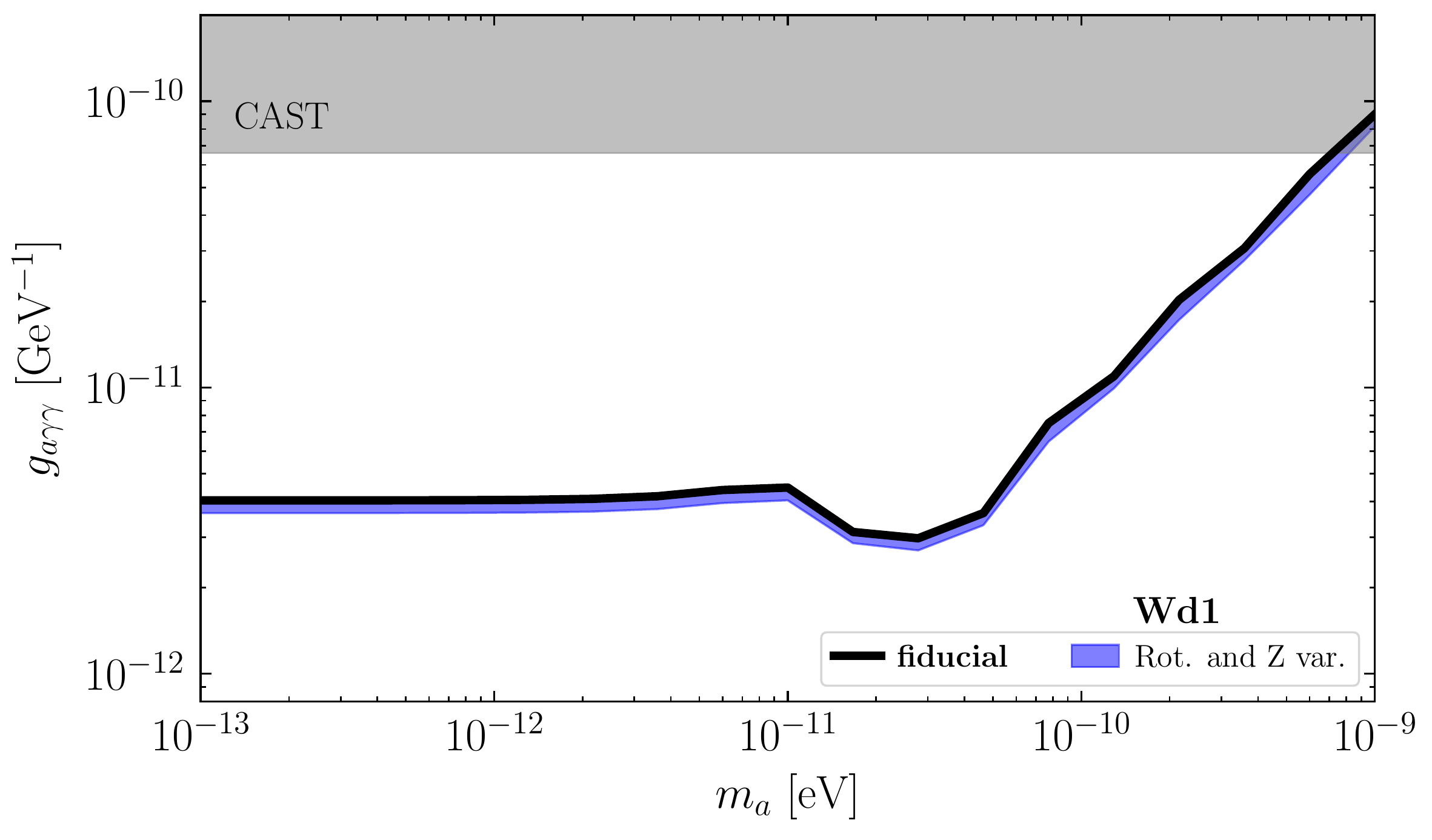}
\caption{
(Left) The variation to the 95\% upper limit found by varying the initial metallicity and rotation in the range  $Z \in (0.018, 0.035)$ and $\mu_{\rm rot} \in (50, 150)$ km/s for the Quintuplet analysis.  The blue region indicates the maximum and minimum limit found, while the black curve shows our fiducial limit. (Right) As in the left panel but for Wd1.
\label{fig:rot_and_Z}
}
\end{center}
\end{figure}

\new{
\section{Variation of upper limits with initial conditions}
}
In this section we show the variation in the upper limits as we vary over our initial conditions $Z \in (0.018, 0.035)$ and $\mu_{\rm rot} \in (50, 150)$ km/s.  These initial conditions represent the dominant uncertainties in our stellar modeling. Recall that in our fiducial analysis we assume the initial metallicity and rotation giving the most conservative upper limits: $Z = 0.035$ and $\mu_{\rm rot} = 150$ km/s. 
Fig.~\ref{fig:rot_and_Z} shows, for both Quintuplet and Wd1, how our 95\% upper limit varies as we scan over $Z$ and $\mu_{\rm rot}$.  In particular, the shaded blue regions show the minimum and maximum limit obtained when varying $Z$ and $\mu_{\rm rot}$.  Note that our fiducial limits, solid black, are the most conservative across most axion masses, though the effect of the $Z$ and $\mu_{\rm rot}$ is relatively minimal, especially for Wd1.

\end{document}